\documentclass[%
	reprint,
	amsmath,amssymb,
	aps,
	prb,
]{revtex4-2}

\usepackage{tikz}
\usepackage{pifont}
\usepackage{graphicx}
\usepackage{dcolumn}
\usepackage{bm}

\usepackage{mathrsfs}
\usepackage{hyperref}
\hypersetup{
	colorlinks = true,
	urlcolor = blue,
	linkcolor = blue,
	citecolor = green,
	filecolor = magenta,
}

\makeatletter
\DeclareRobustCommand{\loplus}{\mathbin{\mathpalette\dog@lsemi{+}}}
\DeclareRobustCommand{\lotimes}{\mathbin{\mathpalette\dog@lsemi{\times}}}
\DeclareRobustCommand{\roplus}{\mathbin{\mathpalette\dog@rsemi{+}}}
\DeclareRobustCommand{\rotimes}{\mathbin{\mathpalette\dog@rsemi{\times}}}

\newcommand{\dog@rsemi}[2]{\dog@semi{#1}{#2}{-90,90}}
\newcommand{\dog@lsemi}[2]{\dog@semi{#1}{#2}{270,90}}
\newcommand{\dog@semi}[3]{%
  \begingroup
  \sbox\z@{$\m@th#1#2$}%
  \setlength{\unitlength}{\dimexpr\ht\z@+\dp\z@\relax}%
  \makebox[\wd\z@]{\raisebox{-\dp\z@}{%
    \begin{picture}(1,1)
    \linethickness{\variable@rule{#1}}
    \roundcap
    \put(0.5,0.5){\makebox(0,0){\raisebox{\dp\z@}{$\m@th#1#2$}}}
    \put(0.5,0.5){\arc[#3]{0.5}}
    \end{picture}%
  }}%
  \endgroup
}
\newcommand{\variable@rule}[1]{%
  \fontdimen8  
  \ifx#1\displaystyle\textfont3\else
    \ifx#1\textstyle\textfont3\else
      \ifx#1\scriptstyle\scriptfont3\else
        \scriptscriptfont3\relax
  \fi\fi\fi
}
\makeatother

\newcommand{\arrowIn}{
\tikz \draw[-stealth] (-1pt,0) -- (1pt,0);
}

\newcommand{\ud}{\mathrm{d}}
\newcommand{\pd}{\partial}
\newcommand{\lie}{\mathscr L}
\newcommand{\order}[1]{\mathcal O\left(#1\right)}

\newcommand{\sD}{\mathscr{D}}

\newcommand{\scR}{{\mathscr R}}
\newcommand{\tsD}{\tilde{\mathscr{D}}}
\newcommand{\tdu}{{\tilde{u}}}

\newcommand{\tdvrho}{{\tilde{\varrho}}}
\newcommand{\tdvth}{{{\tilde{\vartheta}}}}

\newcommand{\ta}{{\tilde{A}}}

\newcommand{\tmu}{{{\mu}}}
\newcommand{\tnu}{{{\nu}}}
\newcommand{\trho}{{{\rho}}}

\newcommand{\hzero}{{\hat{0}}}
\newcommand{\hyi}{{\hat{1}}}
\newcommand{\her}{{\hat{2}}}
\newcommand{\hsan}{{\hat{3}}}
\newcommand{\hi}{{\hat{i}}}
\newcommand{\hj}{{\hat{j}}}
\newcommand{\hk}{{\hat{k}}}
\newcommand{\hl}{{\hat{l}}}

\newcommand{\mcI}{{\mathcal{I}}}

\newcommand{\mcG}{{\mathcal{G}}}
\newcommand{\mcT}{{\mathcal{T}}}
\newcommand{\mcW}{{\mathcal{W}}}
\newcommand{\hmcW}{{\mathcal{W}}}

\newcommand{\cZ}{{\mathcal{Z}}}

\newcommand{\cS}{{\mathcal{S}}}
\newcommand{\cV}{{\mathcal{V}}}

\newcommand{\jac}{\mathcal{J}}
\newcommand{\ijac}{\mathcal{L}}
\newcommand{\matk}{{\mathcal{K}}}

\newcommand{\rotm}{{\mathcal{R}}}

\newcommand{\mfk}{{\mathfrak{K}}}

\begin{document}


\title{Shaving off soft hairs and black hole image memory effect}

\author{Shaoqi Hou}
\email{hou.shaoqi@whu.edu.cn}
\affiliation{School of Physics and Technology, Wuhan University, Wuhan, Hubei 430072, China}
\author{Zong-Hong Zhu}
\email{zhuzh@whu.edu.cn}
\affiliation{School of Physics and Technology, Wuhan University, Wuhan, Hubei 430072, China}


\date{\today}

\begin{abstract}
    Soft hairs of black holes are the Noether charges associated with the generalized Bondi-Metzner-Sachs symmetries.
    In this work, the images of soft-haired Kerr black holes are studied.
    For an eternal black hole, the image is rotated, dilated, and drifting compared to that of the bald counterpart in the celestial plane.
    The rotation and the dilation are independent of time, while the drifting occurs at a constant speed and in a fixed direction.
    These effects all depend on angular directions.
    The soft hair of an astronomical black hole can change due to the emission of gravitational or electromagnetic waves from various physical processes occurring in the vicinity of the horizon.
    Then, the image roams in the observer's view, causing the image memory effect, the smoking gun for the existence of soft hair.
    The magnitude of the image memory effect of a huge, spinning black hole accompanied by a much smaller one is estimated.
    It turns out that this effect is proportional to the mass of the large black hole, increases with its spin, but decreases with the mass ratio.
    Due to the limited angular resolution of current and future detectors, this effect is hard to detect if the impact of cosmological expansion is ignored.
\end{abstract}

\maketitle


\section{Introduction}
\label{sec-int}

The no-hair theorem depicts a very simple picture of the zoo of black holes: all isolated, stationary, and charge-neutral black hole spacetimes are diffeomorphic to the Kerr spacetime \cite{Chrusciel:2012jk}.
It is generally believed that the diffeomorphism is just a pure gauge, and two black hole metrics are treated as physically equivalent to each other if they are related via a diffeomorphism \cite{mtw}.
However, this picture started to become a mosaic ever since the study of the infrared structure of the gravitating system \cite{Strominger:2014pwa,Strominger2014bms,He:2014laa,Strominger:2018inf}.
There exist some diffeomorphisms that transform a black hole spacetime into a physically distinct one.
These are the Bondi-Metzner-Sachs (BMS) coordinate transformations of the asymptotically flat spacetime \cite{Bondi:1962px,Sachs:1962wk,Sachs1962asgr}.
The Noether charges associated with these transformations are nontrivial \cite{Ashtekar:1981bq,Flanagan:2015pxa}, which generate the corresponding canonical transformations in the phase space.
More specifically, the supertranslation, a member of the BMS group, transforms a black hole into a distinct one carrying a different set of soft hairs \cite{Hawking:2016msc,Hawking:2016sgy}, which are the Noether charges \emph{induced} by the supertranslation.
These black holes are thus labeled by a function $\mcT(\vartheta,\varphi)$ of the angular coordinates that parameterizes the supertranslation.
Let us refer to these black holes as soft-haired.

Recently, the asymptotic symmetry group has been expanding.
In fact, the BMS group is the semidirect product of the Lorentz group and the supertranslation group \cite{Sachs1962asgr}.
If one replaces the Lorentz group with the Virasoro group, one gets the extended BMS group \cite{Barnich:2009se,Barnich:2010eb,Barnich:2011ct}.
Alternatively, if one generalizes the Lorentz group to the diffeomorphism group of the topological 2-sphere $\mathbb S$, the generalized BMS group is obtained \cite{Campiglia:2014yka,Campiglia:2015yka,Campiglia:2020qvc}.
One may call the Virasoro group or the diffeomorphism group of $\mathbb S$ the super-Lorentz group in both enlargement schemes.
Like the Lorentz group, the super-Lorentz group consists of super-boosts, super-rotations, and their combinations.
The final enlargement of the BMS group is the Weyl BMS group, which contains the supertranslation $\mcT$, the Weyl rescaling $\mcW$, and the diffeomorphism $\Upsilon^A$ transformations \cite{Freidel:2021fxf}.
Like $\mcT$, $\mcW$ and $\Upsilon^A$ depend on the angles $\vartheta^A=\vartheta,\varphi$.
In this group, $\mcW$ and $\Upsilon^A$ are independent of each other.
In contrast, the super-boost in the extended/generalized BMS group can be viewed as a special Weyl rescaling $\mcW$ associated with a specific $\Upsilon^A$, while the super-rotation is a transformation with $\mcW=0$ and $\Upsilon^A$ generating the conformal transformation of the metric on $\mathbb S$.

As in the case of the (standard) BMS group, a black hole spacetime may carry Noether charges induced by the supertranslation $\mcT$, the Weyl rescaling $\mcW$, and the diffeomorphism transformation $\Upsilon^A$.
This spacetime is labeled by $\mcT$, $\mcW$, and $\Upsilon^A$.
Thus, it is physically distinct from one with different $\mcT$, $\mcW$, and $\Upsilon^A$.
These are the degenerate gravitational vacua \cite{Compere:2018ylh,Freidel:2021fxf}.
These spacetimes are theorized to be physically distinguishable from each other, so there must be some observables that differentiate them.

Previously, people focused on the physical consequences of the supertranslated black holes.
Whether the soft hair impacts the Hawking radiation was the first topic discussed.
The information paradox \cite{Hawking:1976ra} may be resolved if there exists a supertranslated Vaidya black hole \cite{Vaidya:1951zza}, as the information might be stored in the soft hair \cite{Hawking:2016msc,Hawking:2016sgy}.
The tunneling method was used to study the Hawking radiation of this black hole, and it was found that the spectrum depends on the soft hair \cite{Chu:2018tzu}.
However, if a Schwarzschild black hole is supertranslated, the Hawking spectrum is not altered \cite{Iofa:2017ukq}.
There were works arguing that the soft hair is irrelevant to the information paradox as the soft and hard modes decouple dynamically \cite{Mirbabayi:2016axw,Bousso:2017dny,Bousso:2017rsx}.
Some researchers also found that the soft hair is too sparse to store enough information \cite{Marolf:2017jkr,Compere:2019rof}.

Recently, the images of the supermassive black holes M87 and Sagittarius A* taken by the Event Horizon Telescope (EHT) \cite{EventHorizonTelescope:2019dse,EventHorizonTelescope:2019uob,EventHorizonTelescope:2019jan,EventHorizonTelescope:2019ths,EventHorizonTelescope:2019pgp,EventHorizonTelescope:2019ggy,EventHorizonTelescope:2022wkp,EventHorizonTelescope:2022apq,EventHorizonTelescope:2022wok,EventHorizonTelescope:2022exc,EventHorizonTelescope:2022urf,EventHorizonTelescope:2022xqj} offered a new avenue for studying physical processes in the strong gravity regime \cite{Bambi:2019tjh,Vagnozzi:2022moj,Khodadi:2024ubi}.
Thus, the image of a supertranslated Schwarzschild black hole was also considered.
In Refs.~\cite{Lin:2022ksb,Zhu:2022shb}, it was found that the shadow of a supertranslated Schwarzschild black hole has the same shape and size as the standard one but is shifted in the viewing plane.
The photon sphere of a supertranslated Vaidya black hole exhibits special dynamics, which might help distinguish a supertranslated one from its bald counterpart \cite{Sarkar:2021djs}.
The image of a superrotated Schwarzschild black hole was later investigated and found to have a different shape when viewed by a near-zone observer \cite{Lin:2024xzo}.
However, the superrotation considered in Ref.~\cite{Lin:2024xzo} is not the one defined at null infinity but instead on the horizon \cite{Donnay:2015abr}.

Note that in all of these works, only the linearized supertranslation or superrotation was considered.
For an arbitrary transformation, the resultant black hole spacetime may possess new properties.
Up to now, there has been only one work studying the quasinormal mode of a finitely supertranslated Schwarzschild black hole \cite{Hou:2024odb}.
It was found that the supertranslation causes the quasinormal mode to have a different phase from that of a standard black hole.
The phase shift depends on $\mcT$ and is anisotropic.
Thus, to detect this phase shift and further infer the existence of the supertranslated black hole, one must observe the quasinormal mode using multiple interferometers at different angular directions.

In this work, we would like to consider the image of the black hole obtained by transforming the standard Kerr metric via some \emph{finite} generalized BMS transformation.
As discussed previously, the generalized BMS group is a subset of the Weyl BMS group.
The following derivation shows that the generalized BMS transformation can be viewed as the ordinary Poincar\'e transformation in the local inertial frame at a large distance.
It is easier to understand the impact of the generalized BMS transformation on the image of the black hole.
The generalization to the full Weyl BMS group is presented in Appendix~\ref{sec-g-wbms}.
We do not consider the extended BMS transformation in the current work, as it may involve singularities on the celestial sphere $\mathbb S$, which complicates the discussion.

The method for obtaining the desired image is to perform the appropriate coordinate transformation on the photon's 4-momentum at a large distance, as done in Refs.~\cite{Zhu:2022shb,Lin:2024xzo}.
This is because the geodesic equation is a tensor equation.
If it holds in one coordinate system, such as the familiar Boyer-Lindquist (BL) coordinates, it holds in any other coordinate system.
In this way, one does not have to solve the null geodesic equation, which can be complicated and time-consuming for a soft-haired black hole.
Moreover, this method works for all possible forms of the soft-haired black hole metrics.
It is also irrelevant to the properties of the light source, such as the accretion disk, as the coordinate transformation is carried out near the distant observer.
Therefore, this is a universal method, which results in the most general conclusions about the soft-haired black hole image.

Our derivation reveals that for an eternal black hole carrying soft hair, the image is rotated, dilated, and drifting at a constant velocity compared to that of the bald counterpart in the celestial plane.
The rotation, dilation, and drifting velocity depend on $\mcT$, $\mcW$, and $\Upsilon^A$, and therefore, are functions of the angular coordinates.
Nevertheless, the shape of the image remains the same as the bald one.
However, an astronomical black hole always comes with some sorts of matter and energy around it, and some energetic processes occur.
For example, a large black hole may be accompanied by a much smaller  one, forming a binary system and emitting gravitational waves.
This changes the soft hair, and thus the image.
It turns out that although the image is still rotated and dilated by the same amount during the emission, it is no longer drifting at a constant velocity.
Instead, it roams along some curved trajectory in the observer's view.
Once the smaller black hole is merged into the central one, and the binary system settles down to a single black hole, the image would again drift at the same constant velocity as before, but in a different straight line.
That is, before and after the emission of the gravitational wave, the image drifts at different straight line segments.
This is the image memory effect.
The image memory effect would be the smoking gun for the existence of the soft hair.

An important remark is in order.
The generalized BMS transformation that will be performed below shall be considered to be the \emph{active transformation}.
That is, the physical fields are transformed.
It is not simply the passive transformation, i.e., the change in the four numbers (the coordinates) that label a spacetime point.
Since we work in the framework of general relativity, which is a diffeomorphism invariant theory, the active transformation and the passive one are mathematically equivalent \cite{Wald:1984rg}.
This equivalence permits us to formally use the passive coordinate transformation in this work.
Also, in the following, we often talk about how the generalized BMS transformation affects the image, which shall be understood as the influence of the soft hair.

Another remark is that we will ignore the influence of the cosmological expansion on the black hole image.
This will be the topic of the future work.

So this work is organized in the following way.
In Section~\ref{sec-afs}, we will briefly review the basics of the asymptotically flat spacetime.
The finite Weyl BMS transformation and the vacuum structure of the asymptotically flat spacetime will be presented in Section~\ref{sec-inf-wbms}.
Then, the finite generalized BMS transformation will be summarized in Section~\ref{sec-f-gbms}.
Section~\ref{sec-def-cc} defines the adapted tetrads and the celestial coordinates, which are useful for describing the image of the black hole in the BS coordinates.
After that, the image of an eternal Kerr black hole is discussed in Section~\ref{sec-kerr-im}.
In this section, we first consider the computation of the image of a bald Kerr black hole in the BS coordinates in Section~\ref{sec-bk-im}, and then, obtain the image of a soft-haired black hole in Section~\ref{sec-sh-im}.
Next, we consider the image of an astronomical soft-haired black hole, and the image memory effect in Section~\ref{sec-ime}.
The magnitude of the image memory effect is also estimated in Section~\ref{sec-est-im}.
Finally, there is a conclusion.
Throughout, the geometric units is used with $G=c=1$.

\section{Asymptotically flat spacetimes}
\label{sec-afs}

It is very convenient to use the BS coordinates $(u,\varrho,\vartheta,\varphi)$ to describe asymptotically flat spacetimes \cite{Bondi:1962px,Sachs:1962wk}.
The spacetime metric is formally written as \cite{Freidel:2021fxf}
\begin{equation}
    \label{eq-afs-met-f}
    \begin{split}
        \ud s^2= & -2e^{2\varsigma}\ud u(\ud \varrho+F\ud u)                  \\
                 & +g_{AB}(\ud\vartheta^A-U^A\ud u)(\ud\vartheta^B-U^B\ud u),
    \end{split}
\end{equation}
where $\varsigma,F,U^A,g_{AB}$ are the metric functions, and $\vartheta^A=(\vartheta,\varphi)$.
In addition, one imposes the determinant condition,
\begin{equation}
    \label{eq-bs-dc}
    \frac{\pd\sqrt{\det(g_{AB}/\varrho^2)}}{\pd\rho}=0.
\end{equation}
By solving the vacuum Einstein equation, one finds that
\begin{subequations}
    \label{eq-afs-met-exp}
    \begin{gather}
        \varsigma=\frac{\varsigma_2}{\varrho^2}+\order{\frac{1}{\varrho^3}},\quad\varsigma_2\equiv-\frac{c_{AB}c^{AB}}{32},\\
        F=\frac{\mathscr R}{4}-\frac{M}{\varrho}+\order{\frac{1}{\varrho^2}},\\
        \begin{split}
            U^A= & -\frac{\sD^Bc_{AB}}{2\varrho^2}+\frac{1}{\varrho^3}\left(-\frac{2}{3}N^A-2\sD^A\varsigma_2\right. \\
                 & \left.+\frac{1}{2}c^{AB}\sD^Cc_{BC}\right)+\order{\frac{1}{\varrho^4}},
        \end{split}\\
        g_{AB}=\varrho^2h_{AB}+\varrho c_{AB}+\order{\varrho^0}.
    \end{gather}
\end{subequations}
Here, $M$ is called the Bondi mass aspect, $N_A$ is the angular momentum aspect, $h_{AB}$ is a metric on the 2-sphere $\mathbb S$ parameterized by $\vartheta^A$, $\sD_A$ is the covariant derivative of $h_{AB}$, and $\scR$ is its Ricci scalar.
In addition, $c_{AB}$ is named the asymptotic shear tensor, which contains the radiative degrees of freedom of the theory.
One uses $h_{AB}$ and its inverse $h^{AB}$ to lower or raise the capital Latin indices.
In the (standard) BMS case, one sets $h_{AB}=\gamma_{AB}$, the unit round metric on $\mathbb S$, and thus, $\scR=2$.
Here, one can relax this requirement by allowing $h_{AB}$ to depend on $\vartheta,\varphi$ arbitrarily, so that one is allowed to enlarge the asymptotic symmetry group.
The Einstein equation also leads to the following evolution equation,
\begin{gather}
    \pd_u M=-\frac{1}{8}N_{AB}N^{AB}+\frac{1}{4}\sD_A\sD_BN^{AB}+\frac{1}{8}\sD^2\scR,\label{eq-evo-m}
\end{gather}
where $N_{AB}=\pd_uc_{AB}$ is the news tensor, and $\sD^2=\sD_A\sD^A$.
The evolution of $N_A$ can also be obtained, but will not be used in this work.

For later reference, one can substitute Eq.~\eqref{eq-afs-met-exp} into Eq.~\eqref{eq-afs-met-f} to get the explicit form of the metric,
\begin{widetext}
    \begin{equation}
        \label{eq-afs-met}
        \begin{split}
            \ud s^2= & \left[-\frac{\mathscr R}{2}+\frac{2M}{\varrho}+\order{\frac{1}{\varrho^2}}\right]\ud u^2+2\left[-1+\frac{c_{AB}c^{AB}}{16\varrho^2}+\order{\frac{1}{\varrho^2}}\right]\ud u\ud\varrho \\
                     & +2\left[\frac{\sD^Bc_{AB}}{2}+\frac{1}{\varrho}\left(\frac{2}{3}N_A-\frac{1}{8}c_{BC}\sD_Ac^{BC}\right)+\order{\frac{1}{\varrho^2}}\right]\ud u\ud\vartheta^A                         \\
                     & + [\varrho^2 h_{AB}+\varrho c_{AB}+\order{\varrho^0}]\ud\vartheta^A\ud\vartheta^B.
        \end{split}
    \end{equation}
\end{widetext}
Therefore, to specify an asymptotically flat spacetime, one has to give $h_{AB}(\bm\vartheta)$, $c_{AB}(u,\bm\vartheta)$, $M(u,\bm\vartheta)$ and $N_A(u,\bm\vartheta)$ on a null hypersurface $u=u_0$, where we used $\bm\vartheta$ to represent $(\vartheta,\varphi)$ to avoid cluttered notation.
These quantities define the asymptotically flat spacetime up to the order shown in Eq.~\eqref{eq-afs-met}.

\subsection{Weyl BMS transformations and the vacuum structure}
\label{sec-inf-wbms}

There is the coordinate freedom to write down the metric.
It is easy to determine an infinitesimal coordinate transformation parameterized by a vector field $\xi^\mu$ that satisfies the following conditions,
\begin{equation}
    \lie_\xi g_{\varrho\varrho}=0,\quad \lie_\xi g_{\varrho A}=0,\quad \pd_\varrho(g^{AB}\lie_\xi g_{AB})=0.
\end{equation}
Furthermore, one also demands the transformed metric to respect the same falloff behavior of the original one, i.e., Eq.~\eqref{eq-afs-met}.
Thus, the components of $\xi^\mu$ can be easily obtained \cite{Freidel:2021fxf},
\begin{subequations}
    \label{eq-ass-sol}
    \begin{gather}
        \xi^u=\tau\equiv T+uW,\\ \xi^A=Y^A-\frac{\sD^A\tau}{\varrho}+\order{\frac{1}{\varrho^2}},\\ \xi^\varrho=-\varrho W+\frac{1}{2}\sD^2\tau+\order{\frac{1}{\varrho}},
    \end{gather}
\end{subequations}
where $T$, $W$ and $Y^A$ are arbitrary functions of $\bm\vartheta$.
Therefore, $\xi^\mu$ is determined by $T$, $W$ and $Y^A$.
Such a vector field satisfying Eq.~\eqref{eq-ass-sol} is said to generate the Weyl BMS symmetry, and $\xi^\mu$ is thus called the BMSW vector field \cite{Freidel:2021fxf}.
$T$ generates the so-called supertranslation, $W$ corresponds to the Weyl rescaling of the celestial sphere $\mathbb S$, and $Y^A$ is associated with the diffeomorphism of $\mathbb S$.

It can also be shown that under the infinitesimal coordinate transformation \cite{Freidel:2021fxf},
\begin{subequations}
    \label{eq-bmsw-hc}
    \begin{gather}
        \delta h_{AB}=(\lie_Y-2W)h_{AB},\label{eq-dh-ab}\\
        \delta c_{AB}=(\tau\pd_u+\lie_Y-W)c_{AB}-2\sD_{\langle A}\sD_{B\rangle}\tau.
    \end{gather}
\end{subequations}
For the transformation rules of $M$ and $N_A$, please refer to Ref.~\cite{Freidel:2021fxf}.
So the transformation of $c_{AB}$ is not linear.
This may cause some trouble when one attempts to define a vacuum state of the theory.
That is, it seems natural to define the vacuum state to be the one with $c_{AB}=0$.
However, the coordinate freedom suggests this is not the unique configuration.
Instead, since
\begin{equation}
    \label{eq-dotn-tfr}
    \delta \dot N_{AB}=(\tau\pd_u+\lie_Y+W)\dot N_{AB},
\end{equation}
which is linear,
one may require that in the vacuum state,
\begin{equation}
    \label{eq-def-gv-1}
    \dot N_{AB}=0,
\end{equation}
which is invariant under the Weyl BMS transformation.

In order to completely specify the vacuum state, one also has to impose a certain condition on $h_{AB}$.
Although its transformation rule is linear, one cannot simply set it to zero for the vacuum state, otherwise $h_{AB}$ is singular and the area cannot be defined at the null infinity.
The way out is to introduce,
\begin{equation}
    \label{eq-def-mm}
    \mathcal M\equiv M+\frac{1}{8}c_{AB}N^{AB},\; M^A\equiv\frac{1}{2}\sD_BN^{AB}+\frac{1}{4}\sD^A\scR,
\end{equation}
which transform according to \cite{Freidel:2021fxf}
\begin{subequations}
    \begin{gather}
        \delta \mathcal M=(\tau\pd_u+\lie_Y+3W)\mathcal M+M^A\pd_A\tau,\label{eq-dotm-tfr}\\
        \delta M^A=(\tau\pd_u+\lie_Y+4W)M^A+\frac{1}{2}\dot N^{AB}\pd_B\tau.
    \end{gather}
\end{subequations}
So the second equation implies that if in the vacuum state, one demands $M^A=0$, it remains vanishing.

Therefore, the gravitational vacuum state is proposed to be described by \cite{Freidel:2021fxf}
\begin{equation}
    \label{eq-gv-cond}
    \dot N_{AB}^\mathsf{vac}=0,\quad M^A_\mathsf{vac}=0.
\end{equation}
Under the Weyl BMS transformation, these conditions are invariant, as illustrated above.
\textcolor{black}{It is worth to mention that these conditions are equivalent to the vanishing of the leading order components the Newman-Penrose variables $\Psi_3$ and $\Psi_4$ \cite{Newman:1961qr,Newman:1962cia,Newman:1968uj,Adamo:2009vu,Freidel:2021qpz}, if one of the null tetrad is in the radial direction.
    These conditions can also be expressed in terms of the vanishing of the Cotton tensor $\mathcal C_{AB}\equiv \epsilon_{A}{}^{CD}\sD_{C}(\scR_{BD}-\scR h_{BD}/4)$ with $\epsilon_{ABC}$ the volume element of the null infinity \cite{Campoleoni:2023fug}.}
By Eq.~\eqref{eq-dotm-tfr}, if $\mathcal M=0$ further, the Weyl BMS transformation does not alter it, either.
The very vacuum state with $\mathcal M=0$ is called the flat vacuum, which includes the familiar Minkowski spacetime as a special case.
More nontrivial vacuum states are Schwarzschild \cite{Schwarzschild:1916uq}, Kerr \cite{Kerr:1963ud},  and supertranslated Schwarzschild \cite{Compere:2016hzt} spacetimes, to name a few.
From the definition of the vacuum state, and the transformation laws Eqs.~\eqref{eq-dotn-tfr} and \eqref{eq-dotm-tfr}, it follows that given a vacuum state, one can obtain an infinitely many new vacuum states by applying the Weyl BMS transformation to the given vacuum.
So if one works out all possible finite Weyl BMS transformations, one can obtain all possible vacuum states connected to the given one.

It shall be emphasized that although these vacuum states are connected by the Weyl BMS transformations, they are physically distinctive from each other, as their Noether charges are different, in general.
Indeed, it was found out that the Noether charges associated with the supertranslation $T$, the Weyl rescaling $W$ and the diffeomorphism $Y^A$ on $\mathbb S$ are \cite{Compere:2018ylh,Compere:2020lrt,Freidel:2021fxf},
\begin{subequations}
    \label{eq-def-ncs}
    \begin{gather}
        Q_T=\int_{\mathbb S}\ud^2\vartheta\sqrt hT\left(M+\frac{1}{4}\sD_A\sD_Bc^{AB}\right),\label{eq-def-qt}\\
        Q_W=Q_{T=uW}+4\int_{\mathbb S}\ud^2\vartheta\sqrt hW\varsigma_2\\
        Q_Y=\int_{\mathbb S}\ud^2\vartheta\sqrt hY^A\left(N_A+4\sD_A\varsigma_2-\frac{1}{4}c_{AB}\sD_Cc^{BC}\right),
    \end{gather}
\end{subequations}
respectively.
For a generic Weyl BMS generator $\xi^\mu(T,W,Y)$, its charge is $Q_\xi=Q_T+Q_W+Q_Y$.
So the vacuum states are labeled by different Noether charges, and are believed to be physically distinguishable from each other.
\textcolor{black}{Note that the vacua carrying different values of Noether charges are still degenerate, as Eq.~\eqref{eq-gv-cond} hold for them.
    Being physically distinguishable means that some phenomena may look different for the \emph{same} observer, as to be illustrated in Sec.~\ref{sec-kerr-im}.}

Now, the basics of the Weyl BMS transformation has been reviewed.
The restriction to the generalized BMS transformation can be straightforwardly done.
In the Weyl BMS group, $W$ is completely independent of $Y^A$, while in the generalized BMS group, these are connected.
That is, $W=\sD_AY^A/2$ in the generalized BMS algebra.
So there are two kinds of generators in the generalized BMS case, $T$ and $Y^A$.
In this work, the transformation generated by $Y^A$ will be called the super-Lorentz transformation.
In particular, if $W=\sD_AY^A/2\ne0$, the transformation will be called the super-boost.
If $W=\sD_AY^A/2=0$, one has the super-rotation.
As discussed in the Introduction, we will focus on the generalized BMS transformation in the main text, since its influence on the black hole image can be more easily understood.

\subsection{The finite generalized BMS transformation}
\label{sec-f-gbms}

A finite transformation can be obtained by exponentiating Eq.~\eqref{eq-ass-sol}.
Since these transformations generally do not commute with each other, it is necessary to specify the sequence in which to perform the transformation.
One can first execute a super-Lorentz transformation ($\Upsilon^A$) to transform the old coordinates $(u,\varrho,\bm\vartheta)$ to the intermediate ones $(u',\varrho',\bm\vartheta')$, and then a supertranslation ($\mcT$) to arrive at the final coordinates $(\tdu,\tdvrho,\bm\tdvth)$.
The composite transformation rule is \cite{Chrusciel:1993hx,Freidel:2021fxf,Flanagan:2023jio}
\begin{subequations}
    \label{eq-f-gbms-vac}
    \begin{gather}
        \tdu=\mcG-\frac{e^{-\hmcW}}{2\varrho}\sD_A\mcG \sD^A\mcG+\order{\frac{1}{\varrho^2}},\\
        \tdvrho=e^{-\hmcW}\varrho+\frac{e^{-2\hmcW}}{2}\sD^2\mcG+\order{\frac{1}{\varrho}},\\
        \tdvth^A=\Upsilon^A(\bm\vartheta)-\frac{e^{-\hmcW}}{\varrho}\jac_B^A\sD^B\mcG+\order{\frac{1}{\varrho^2}},
    \end{gather}
    where $\sD_A$ acts on $\bm\vartheta$, and
    $\mcG\equiv\mcT+ue^\hmcW$
    with both $\mcT$ and $\hmcW$ viewed as functions of $\bm\Upsilon(\bm\vartheta)$.
    One also has \cite{Flanagan:2023jio}
    \begin{equation}
        \label{eq-det-dif}
        e^{2\hmcW}=|\det(\jac_A^B)|\sqrt{\frac{\tilde h(\bm\Upsilon(\bm\vartheta))}{h(\bm\vartheta)}},\quad \jac_A^B\equiv\frac{\pd\Upsilon^B}{\pd\vartheta^A},
    \end{equation}
    where $h=\det(h_{AB})$ and $\tilde h=\det(\tilde h_{AB})$.
\end{subequations}
Here, different symbols ($\mcT,\mcW,\Upsilon^A$) are used in order to emphasize the difference between the infinitesimal and the finite transformations.
The particular transformation with $\mcT\ne0$, $\mcW=0$ and $\Upsilon^A=0$ is the finite supertranslation, which was used in the previous work \cite{Hou:2024odb}.
If $\mcT=0$, one has the finite super-Lorentz transformation.
In particular, if $\mcW\ne0$ and is related to $\Upsilon^A$ via Eq.~\eqref{eq-det-dif}, one gets the finite super-boost transformation.
If $\mcW=0$, and $\Upsilon^A$ satisfies Eq.~\eqref{eq-det-dif}, one obtains the finite super-rotation transformation.

Based on the finite generalized BMS transformations, one can work out how the gravitational vacuum transforms.
As long as one finds a particular vacuum state, one can perform a series of the basic generalized BMS transformations to get any vacuum state.
One of the special vacuum state is given by,
\begin{equation}
    \label{eq-def-fvac}
    h_{AB}=\gamma_{AB},\quad c_{AB}=0,
\end{equation}
which includes the Kerr spacetime to be shown below.
Let us construct a general vacuum from this particular one.
One can show that after the transformation~\eqref{eq-f-gbms-vac}, one has\footnote{For other ways of parameterizing the vacuum, please refer to Refs.~\cite{Compere:2018ylh,Freidel:2021fxf,Flanagan:2023jio}.}
\begin{subequations}
    \label{eq-sh-h-sh}
    \begin{gather}
        \tilde h_{AB}(\bm\tdvth)=e^{2\hmcW}\ijac_A^C\ijac_B^D\gamma_{CD}(\bm\vartheta),\label{eq-h-af-nc}\\
        \tilde c_{AB}=2\tsD_{\langle A}\tsD_{B\rangle}\mcT-2(\tdu-\mcT)e^\hmcW\tsD_{\langle A}\tsD_{B\rangle}e^{-\hmcW},\label{eq-shear-af-nc}
    \end{gather}
\end{subequations}
where $\ijac_A^C\jac_C^B=\delta_A^B$, $\gamma_{AB}$ is the unit round metric on $\mathbb S$, and $\langle\cdot\rangle$, enclosing the indices, means to take the traceless part with respect to $\tilde h_{AB}$.
One can compare Eq.~\eqref{eq-sh-h-sh} with Eqs.~(3.8) and (3.10) in Ref.~\cite{Compere:2018ylh}, and recognizes that $-\mcT$ is the supertranslation field and $-2\hmcW$ is the super-boost field.
Of course, a complete description of the vacuum state requires the specification of $M$ and $N_A$ at the same time.
This will be done for the soft-haired Kerr black hole later\footnote{For the complete specification of a nontrivial flat vacuum, please refer to Refs.~\cite{Compere:2016jwb,Compere:2018ylh}.}.

Now, let us examine how the various specific generalized BMS transformations affect the vacuum.
First, if the Poincar\'e transformation is performed\footnote{Since one starts with a particular vacuum~\eqref{eq-def-fvac}, the Poincar\'e transformation can be well defined.}, $\tilde h_{AB}=\gamma_{AB}$ and $\tilde c_{AB}=0$, so the vacuum stays intact, and the black hole is still bald.
Second, suppose that one performs a proper supertranslation, i.e., the one not belonging to the Poincar\'e group, then $\tilde h_{AB}=\gamma_{AB}$ and $\tilde c_{AB}=2\tsD_{\langle A}\tsD_{B\rangle}\mcT$.
So the shear changes, and the vacuum transition occurs.
The black hole becomes soft-haired, as it carries more charges, as shown later for the soft-haired Kerr black hole.
Finally, consider the proper super-Lorentz transformation.
Then, $\tilde h_{AB}=e^{2\hmcW}\ijac_A^C\ijac_B^D\gamma_{CD}$ and $\tilde c_{AB}=-2\tdu e^\hmcW\tsD_{\langle A}\tsD_{B\rangle}e^{-\hmcW}$.
Again, the vacuum state changes, and the black hole also carries the soft hair.
Therefore, a \emph{proper} generalized BMS transformation manifests as the vacuum transition \cite{Strominger:2014pwa,Freidel:2021fxf}.

\subsection{Adapted tetrads and celestial coordinates}
\label{sec-def-cc}

In the following sections, we will obtain the image of the soft-haired Kerr black hole.
The image is basically the brightness pattern of the light viewed by an observer.
It is related to the direction of the photon received by the observer, and also the intensity.
So it is necessary to set up a local tetrad basis to quantify the direction.

The tetrads adapted to Eq.~\eqref{eq-afs-met-f} can be chosen to be
\begin{subequations}
    \label{eq-def-tetrad}
    \begin{gather}
        \sigma^\hzero=\frac{1}{2}(2F+e^{2\varsigma})\ud u+\ud\varrho,\\ \sigma^\hyi=\ud\varrho+\frac{1}{2}(2F-e^{2\varsigma})\ud u,\\ \sigma^\hi=\varrho E^\hi(\ud\vartheta^A-U^A\ud u),
    \end{gather}
\end{subequations}
where $E^\hi_A$ ($\hi=\her,\hsan$) represents the dyad on the sphere $\mathbb S$ with $g_{AB}=\varrho^2E^\hi_AE^\hj_B\delta_{\hi\hj}$.
The dual basis is also useful, and given by
\begin{subequations}
    \begin{gather}
        \sigma_\hzero=e^{-2\varsigma}\left[\pd_u-\frac{1}{2}(2F-e^{2\varsigma})\pd_\varrho+U^A\pd_A\right],\\
        \sigma_\hyi=-e^{-2\varsigma}\left[\pd_u-\frac{1}{2}(2F+e^{2\varsigma})\pd_\varrho+U^A\pd_A\right],\\
        \sigma_\hi=\frac{E_\hi^A}{\varrho}\pd_A,
    \end{gather}
\end{subequations}
with $E^\hi_AE_\hj^A=\delta^\hi_\hj$.
These tetrads are normalized and orthogonal to each other, so that the spacetime metric has the diagonal form $\text{diag}(-1,1,1,1)$ in this basis.

Near the null infinity, one can show that
\begin{subequations}
    \label{eq-tet-nl}
    \begin{gather}
        \sigma^\hzero=\frac{1}{2}\left(\frac{\scR}{2}+1\right)\ud u+\ud\varrho+\order{\frac{1}{\varrho^2}},\\ \sigma^\hyi=\frac{1}{2}\left(\frac{\scR}{2}-1\right)\ud u+\ud\varrho+\order{\frac{1}{\varrho}}, \\ \sigma^\hi=\varrho\matk^\hi_A\ud\vartheta^A+\order{\varrho^0},
    \end{gather}
\end{subequations}
where $\matk^\hi_A$ satisfies $h_{AB}=\matk^\hi_A\matk^\hj_B\delta_{\hi\hj}$.
In fact, $\matk^\hi_A$ is the leading order term of $E^\hi_A$.
Its exact form is not important for now.
The series expansions of the dual tetrads are
\begin{subequations}
    \label{eq-tet-nl-vec}
    \begin{gather}
        \sigma_\hzero=\pd_u-\frac{1}{2}\left(\frac{\scR}{2}-1\right)\pd_\varrho+\order{\frac{1}{\varrho}},\\ \sigma_\hyi=-\pd_u+\frac{1}{2}\left(\frac{\scR}{2}+1\right)\pd_\varrho+\order{\frac{1}{\varrho}},\\
        \sigma_\hi=\frac{\matk_\hi^A}{\varrho}\pd_A+\order{\frac{1}{\varrho^2}},
    \end{gather}
\end{subequations}
with $\matk^\hi_A\matk^A_\hj=\delta^\hi_\hj$.
These adapted bases are associated with the preferred observer with the 4-velocity $\sigma_\hzero$.

Let the photon received by the observer have the 4-momentum $p^\mu$, where the subscript $o$ indicates to evaluate quantities at the observer.
Then, the observer can define the following celestial coordinates,
\begin{equation}
    \label{eq-def-sc-abs}
    x^\hi=\frac{p^\hi}{p^\hzero},
\end{equation}
where $p^{\hat\mu}$'s are the components of $p^\mu$ in the observer's frame, that is, $p^{\hat\mu}=p^\nu(\sigma^{\hat\mu})_\nu$.
We will show that this definition agrees with those used in the literature \cite{Grenzebach:2014fha,Gralla:2017ufe,Perlick:2021aok}.
Indeed, we will obtain the exact same expression for the bald Kerr black hole.
Then, since the soft-haired black hole can be got by the generalized BMS transformation, the basic task of the following is to determine the transformation rule of $x^\hi$ under the generalized BMS transformation.

\section{The eternal black hole image}
\label{sec-kerr-im}

Now, it is ready to determine the image of an eternal black hole.
We will first derive the celestial coordinates of the photons running in the bald Kerr black hole in the BS coordinates in Section~\ref{sec-bk-im}.
So in this section, after reviewing the basics of the Kerr metric and the photon's momentum in the BL coordinates, we will obtain the BS coordinates, and determine the celestial coordinates of the photon.
Then, we will carry out the generalized BMS transformation to arrive at  the soft-haired black hole image in Section~\ref{sec-sh-im}.
There, we will start with the properties of the soft-haired Kerr metric and its conserved charges, then, derive the transformation rule of the adapted tetrad basis, and finally, get the celestial coordinates for the soft-haired black hole.
The impacts of the soft hair on the black hole image will be discussed.

\subsection{The bald Kerr black hole image}
\label{sec-bk-im}

In the  BL coordinates ($t, r,\theta,\phi$), the Kerr metric takes the following form \cite{mtw,Wald:1984rg},
\begin{subequations}
    \begin{equation}
        \label{eq-kerr-met-bl}
        \ud s^2=-\frac{\Delta\Sigma}{\Xi}\ud t^2+\frac{\Sigma}{\Delta}\ud r^2+\Sigma\ud\theta^2+\frac{\Xi\sin^2\theta}{\Sigma}(\ud\phi-\omega\ud t)^2,
    \end{equation}
    where
    \label{eq-kerr-met-bl}
    \begin{gather}
        \omega=\frac{2a\mathring M r}{\Xi},\\ \Delta= r^2-2\mathring M r+a^2,\\ \Sigma= r^2+a^2\cos^2\theta,\\ \Xi=( r^2+a^2)^2-\Delta a^2\sin^2\theta.
    \end{gather}
\end{subequations}
This metric possesses two Killing vector fields, $\mfk_t\equiv\pd_t$ and $\mfk_\phi\equiv\pd_\phi$, as well as a Killing-Yano tensor \cite{Walker:1970un}.
So a null geodesic, or the photon's 4-momentum $p^\mu$, is labeled by three conserved quantities, i.e., the energy $E$, the angular momentum $\lambda E$, and the Carter constant $q E^2$ \cite{Carter:1968rr}.
More explicitly, the photon's 4-momentum $p^\mu$ is given by \cite{Gralla:2017ufe},
\begin{subequations}
    \label{eq-p-4}
    \begin{gather}
        p^ r=\pm\frac{E}{\Sigma}\sqrt{\mathcal R( r)},\\
        p^\theta=\pm\frac{E}{\Sigma}\sqrt{\Theta(\theta)},\\
        p^\phi=\frac{E}{\Sigma}\left[\frac{a}{\Delta}( r^2+a^2-a\lambda)-a+\frac{\lambda}{\sin^2\theta}\right],\\
        p^ t=\frac{E}{\Sigma}\left[\frac{ r^2+a^2}{\Delta}( r^2+a^2-a\lambda)-a(a\sin^2\theta-\lambda)\right].
    \end{gather}
\end{subequations}
Here, $\mathcal R( r)\equiv( r^2+a^2-a\lambda)^2-[q+(a-\lambda)^2]\Delta$ and $\Theta(\theta)\equiv q+a^2\cos^2\theta-\lambda^2\cot^2\theta$ are the radial and angular potentials, respectively.
At certain values of $ r$ and $\theta$, they vanish, so $p^ r$ and $p^\theta$ change signs there.
These are the turning points of the null geodesics, so there are $\pm$'s in the above expressions.
As one can see, the energy $E$ shows up as an overall factor, so its value only determines the parameterization of the null curve but does not affect the ``shape'' of the geodesic.
Therefore, the null geodesic is basically determined by $\lambda$ and $q$.
Of course, $E$ is also related to the intensity of the light, but in this work, we will not consider the brightness of the image.
The momentum $p^\mu$ can be expanded at the large $r$,
\begin{subequations}
    \label{eq-p-4-lr}
    \begin{gather}
        p^t=E\left(1+\frac{2\mathring M}{r}\right)+\order{\frac{1}{r^2}},\\
        p^r=E+\order{\frac{1}{r^2}},\\
        p^\theta=\pm \frac{E}{r^2}\sqrt{q+a^2\cos^2\theta-\lambda^2\cot^2\theta}+\order{\frac{1}{r^3}},\\
        p^\phi=\frac{E\lambda \csc^2\theta}{r^2}+\order{\frac{1}{r^3}}.
    \end{gather}
\end{subequations}
This asymptotic behavior actually determines the image of the bald Kerr black hole.
To show this, let us first express everything in the BS coordinates.

There were several works on the transformation of the Kerr metric to the BS coordinates \cite{Fletcher2003bs,Bishop:2006kerr,Bai:2007rs,Wu:2008yi}.
We will follow Ref.~\cite{Bai:2007rs} mainly.
The basic idea of obtaining the BS form of the Kerr metric is to first determine the solution $u$ to the eikonal equation $g^{\tmu\tnu}\pd_\tmu u\pd_\tnu u=0$, which is given by
\begin{equation}
    \label{eq-tdu-eis}
    u= t- r-\mathring M\ln\frac{ r}{2\mathring M}+\order{\frac{1}{ r}},
\end{equation}
where the higher order terms depend on both $\mathring M$ and $a$, but are useless for the current discussion.
Then, choosing a suitable set of Newman-Penrose tetrads \cite{Newman:1961qr}, one can determine the Newman-Unti coordinates $(u,\trho,\vartheta,\varphi)$ \cite{Newman:1962cia}, where $\trho$ is the affine parameter for the null generator of the light cone $u=\text{const.}$ with the tangent $g^{\tmu\tnu}\pd_{\tnu}u$.
Finally, making a further coordinate transformation, $\varrho=\trho+\order{1/\trho^5}$, one obtains the desired BS coordinates.
It is easy to write the BL coordinates in terms of the BS coordinates
\begin{subequations}
    \label{eq-bl-bs}
    \begin{gather}
        t= u+\varrho+\mathring M\ln\frac{\varrho}{2\mathring M}+\order{\frac{1}{\varrho}},\\
        r=\varrho-\frac{a^2\sin^2\vartheta}{2\varrho}+\order{\frac{1}{\varrho^2}},\\
        \theta=\vartheta-\frac{a^2\sin2\vartheta}{4\varrho^2}+\order{\frac{1}{\varrho^4}},\\
        \phi=\varphi-\frac{\mathring M a}{\varrho^2}+\order{\frac{1}{\varrho^3}}.
    \end{gather}
\end{subequations}
The Kerr metric now takes the following form,
\begin{widetext}
    \begin{equation}
        \label{eq-kem-bs}
        \begin{split}
            \ud s^2= & -\left[1-\frac{2\mathring M}{\varrho}+\order{\frac{1}{\varrho^3}}\right]\ud u^2-2\left[1+\order{\frac{1}{\varrho^4}}\right]\ud u\ud\varrho \\
                     & -\left[\frac{4\mathring M a\sin^2\vartheta}{\varrho}+\order{\frac{1}{\varrho^2}}\right]\ud u\ud\varphi
            +\left[\varrho^2+\order{\frac{1}{\varrho}}\right](\ud\vartheta^2+\sin^2\vartheta\ud\varphi^2)                                                         \\
                     & +\order{\frac{1}{\varrho^2}}\ud u\ud\vartheta+\order{\frac{1}{\varrho^3}}\ud\vartheta\ud\varphi.
        \end{split}
    \end{equation}
\end{widetext}
By comparing this metric with Eq.~\eqref{eq-afs-met}, one can find out that $h_{AB}$ is the unit round metric $\gamma_{AB}$ (the last round brackets in the second line), and the shear tensor $c_{AB}=0$.
So this metric describes a trivial vacuum state.
The Bondi mass aspect is $M=\mathring M$, and the angular momentum aspect is
\begin{equation}
    \label{eq-amas-k}
    N_A=-\mathring Ma\sin^2\vartheta \delta_A^\varphi.
\end{equation}
It is difficult to write down a closed form for the Kerr metric in the BS coordinates\footnote{There is a closed form \cite{PhysRevD.73.084023}. Unfortunately, that is too complicated for our purpose.}.
But this is not a big problem for our purpose, as the image of a black hole is observed at a large distance, and the coordinate transformation is known there, such as Eq.~\eqref{eq-bl-bs}.

The adapted tetrad basis is given by
\begin{subequations}
    \label{eq-tb-bkerr}
    \begin{gather}
        \sigma^\hzero=\ud u+\ud\varrho+\order{\frac{1}{\varrho^2}},\\ \sigma^\hyi=\ud\varrho+\order{\frac{1}{\varrho}},\\ \sigma^\hi=\varrho\matk^\hi_A\ud\vartheta^A+\order{\varrho^0},
    \end{gather}
\end{subequations}
where for convenience, set $\matk^\hi_A=\text{diag}(1,\csc\vartheta)$.
The dual basis is explicitly
\begin{subequations}
    \label{eq-tbd-bk}
    \begin{gather}
        \sigma_\hzero=\pd_u+\order{\frac{1}{\varrho}},\\ \sigma_\hyi=-\pd_u+\pd_\varrho+\order{\frac{1}{\varrho}},\\ \sigma_\hi=\frac{\matk^A_\hi}{\varrho}\pd_A+\order{\frac{1}{\varrho^2}}.
    \end{gather}
\end{subequations}
So asymptotically, the preferred observer associated with this set of tetrads has $u$ as its proper time.
$\sigma_\hyi$ points in the radial direction, and $\sigma_\hi$'s are in the angular directions.
The choice of $\matk^\hi_A$ makes $\sigma_\her$ point in the $\vartheta$-direction, while $\sigma_\hsan$ in the $\varphi$-direction.
With the chosen tetrad basis and the definition of the celestial coordinates~\eqref{eq-def-sc-abs} introduced in Section~\ref{sec-def-cc}, one can compute the celestial coordinates of the photon received by the observer, which are
\begin{subequations}
    \label{eq-cc-bk-bs-1}
    \begin{gather}
        x^\her=\pm\frac{\sqrt{q+a^2\cos^2\vartheta-\lambda^2\cot^2\vartheta}}{\varrho}+\order{\frac{1}{\varrho^2}},\\ x^\hsan=\frac{\lambda}{\varrho\sin\vartheta}+\order{\frac{1}{\varrho^2}},
    \end{gather}
\end{subequations}
based on Eqs.~\eqref{eq-p-4-lr} and \eqref{eq-bl-bs}.
These agree with the results in previous literature \cite{Grenzebach:2014fha,Gralla:2017ufe,Perlick:2021aok}.
As one can see, these coordinates are determined by $\lambda$ and $q$, irrelevant to the energy $E$ of the photon.

Before moving on, let us calculate the conserved charges of the bald Kerr black hole.
With Eq.~\eqref{eq-def-ncs}, one can verify that the only nonvanishing Noether charges are the total energy $\mathring M$ and the spin $a$.
These are associated with the Killing vectors $\mfk_t$ and $\mfk_\phi$, respectively.
The two Killing vectors approach
\begin{equation}
    \mfk_t=\pd_u+\order{\frac{1}{\varrho}}, \quad \mfk_\phi=\pd_\varphi+\order{\frac{1}{\varrho^2}},
\end{equation}
near the null infinity in the BS coordinates.
More precisely, Eq.~\eqref{eq-def-ncs} gives
\begin{equation}
    \label{eq-cal-qtp}
    Q_t=4\pi\mathring M,\quad Q_\phi=-4\pi\mathring Ma.
\end{equation}
Both $\mfk_t$ and $\mfk_\phi$ belong to the Poincar\'e algebra, and neither $\mathring M$ nor $a$ represents the soft hair, which is the Noether charge of the proper generalized BMS transformation.
Therefore, the Kerr black hole described by Eq.~\eqref{eq-kerr-met-bl} or \eqref{eq-kem-bs} is bald.

\subsection{The soft-haired black hole image}
\label{sec-sh-im}

Now, one tries to determine the image of the soft-haired Kerr black hole.
The soft-haired Kerr metric can be computed using the transformation rule given in Eq.~\eqref{eq-f-gbms-vac}, so we will use $(\tdu,\tdvrho,\tdvth^A)$ as the BS coordinates.
The metric will be formally given by Eq.~\eqref{eq-afs-met} with the appropriate expansion coefficients.
That is, replace $h_{AB}$ and $c_{AB}$ by $\tilde h_{AB}$ and $\tilde c_{AB}$ given in Eq.~\eqref{eq-sh-h-sh}, respectively.
$\sD_A$ shall be substituted with $\tsD_A$, the compatible derivative for $\tilde h_{AB}$, and $\scR$ shall be $\tilde\scR\equiv2(e^{-2\hmcW}+\tsD^2\hmcW)$, the Ricci scalar of $\tilde h_{AB}$.
At the same time, it can be shown that the Bondi mass aspect $M$ in Eq.~\eqref{eq-afs-met} shall be given by
\begin{equation}
    \label{eq-bma-shk}
    \tilde M=e^{-3\hmcW}\mathring M-\frac{1}{8}\tilde c_{AB}\tilde N^{AB},\quad \tilde N_{AB}=-2e^\hmcW\tsD_{\langle A}\tsD_{B\rangle}e^{-\hmcW},
\end{equation}
while the angular momentum aspect $N_A$ is now
\begin{equation}
    \label{eq-baa-shk}
    \tilde N_A=-e^{-2\hmcW}\ijac^\varphi_A\mathring Ma\sin^2\vartheta,
\end{equation}
where $\vartheta$ is now viewed as a function of $\bm\tdvth$ via the relation $\tdvth^A=\Upsilon^A(\bm\vartheta)$.

One shall realize that although the new spacetime is described by a more complex metric, it still possesses two Killing vectors $\mfk_t$ and $\mfk_\phi$, which are now approximated by
\begin{subequations}
    \begin{gather}
        \mfk_t=e^\mcW\pd_\tdu+\frac{1}{2}\tsD^2e^\mcW\pd_\tdvrho+\order{\frac{1}{\tdvrho}},\\
        \mfk_\phi=\tau_\varphi\pd_\tdu-\left(\tdvrho\pd_\varphi\mcW-\frac{1}{2}\tsD^2\tau_\varphi\right)\pd_\tdvrho+\jac_\varphi^A\pd_\ta+\order{\frac{1}{\tdvrho}},\label{eq-phi-asy}
    \end{gather}
\end{subequations}
respectively, where $\tau_\varphi\equiv\pd_\varphi\mcT+(\tdu-\mcT)\pd_\varphi\mcW$.
So $\mfk_t$ is a supertranslation generator, and $\mfk_\phi$ represents a generalized BMS generator, near the null infinity in the new spacetime.
The conserved charges are
\begin{subequations}
    \label{eq-tqt-qp}
    \begin{gather}
        \tilde Q_t=4\pi\mathring M+\int\ud^2\tdvth\sqrt{\tilde h}\left(\frac{1}{4}\tsD_A\tsD_B\tilde c^{AB}-\frac{1}{8}\tilde c_{AB}\tilde N^{AB}\right),\\
        \tilde Q_\phi=-4\pi\mathring Ma+\int\ud^2\tdvth\sqrt{\tilde h}\left\{\tau_\varphi\left(\tilde M+\frac{1}{4}\tsD_A\tsD_B\tilde c^{AB}\right)\right.\nonumber\\
        \left.+4\tilde\varsigma_2\pd_\varphi\hmcW+\jac_\varphi^A\left(4\pd_A\tilde\varsigma_2-\frac{1}{4}\tilde c_{AB}\tsD_C\tilde c^{BC}\right)\right\},
    \end{gather}
\end{subequations}
with $\tilde\varsigma_2=-\tilde c_{AB}\tilde c^{AB}/32$.
These charges differ from $Q_t$ and $Q_\phi$ in Eq.~\eqref{eq-cal-qtp}.
The new spacetime carries more nonzero charges.
For example, Eq.~\eqref{eq-def-qt} becomes
\begin{equation}
    \label{eq-qt-sh}
    \tilde Q_T=\int_{\mathbb S}\ud^2\tdvth\sqrt {\tilde h}T\left(\tilde M+\frac{1}{4}\tsD_A\tsD_B\tilde c^{AB}\right),
\end{equation}
which is generally nonvanishing for a generic $T$.
Therefore, the new spacetime is physically different from the bald Kerr spacetime discussed in the previous subsection, and is called the soft-haired.

The image of the soft-haired black hole is expected to be different from that of the bald counterpart.
Again, it is not necessary to write down the full metric of the soft-haired Kerr black hole, as we are actually able to determine the celestial coordinates using the generalized BMS transformation rule~\eqref{eq-f-gbms-vac}.
Indeed, as mentioned at the beginning, the image of the soft-haired Kerr black hole will be obtained by applying the finite generalized BMS transformation to that of the bald Kerr black hole.
This demands that one shall study how the adapted tetrads transform.
Due to Eq.~\eqref{eq-f-gbms-vac}, the new basis $\{\tilde\sigma^{\hat\mu}\}$ is related to the old $\{\sigma^{\hat\mu}\}$ in the following way,
\begin{subequations}
    \label{eq-tet-tf}
    \begin{gather}
        \begin{split}
            \tilde\sigma^\hzero= & \frac{1}{2}\left(\tsD^2+1+\frac{\tilde\scR}{2}\right)e^\hmcW(\sigma^\hzero-\sigma^\hyi)+e^{-\hmcW}\sigma^\hyi \\
                                 & +\sigma^\hi\matk_\hi^A\pd_Ae^{-\hmcW}+\order{\frac{1}{\varrho}},
        \end{split}\\
        \begin{split}
            \tilde \sigma^\hyi= & \frac{1}{2}\left(\tsD^2-1+\frac{\tilde\scR}{2}\right)e^\hmcW(\sigma^\hzero-\sigma^\hyi)+e^{-\hmcW}\sigma^\hyi \\
                                & +\sigma^\hi\matk_\hi^A\pd_Ae^{-\hmcW}+\order{\frac{1}{\varrho}},
        \end{split}\\
        \begin{split}
            \tilde \sigma^\hi= & \rotm^\hi_\hj \sigma^\hj-\tilde\matk^\hi_A\tsD^Ae^\hmcW(\sigma^\hzero-\sigma^\hyi)
            +\cS^\hi\sigma^\hyi                                                                                     \\
                               & +\cZ^\hi+\order{\frac{1}{\varrho^2}},
        \end{split}
    \end{gather}
    where $\cZ^\hi$ represents a complicated but useless contribution, and the matrix $\rotm^\hi_\hj$ and the vector $\cS^\hi$ are
    \begin{gather}
        \label{eq-def-rotm}
        \rotm^\hi_\hj(\bm{\tilde\vartheta})\equiv e^{-\hmcW}\tilde{\matk}^\hi_A\jac^A_B\matk^B_\hj,\\
        \cS^\hi(\tdu,\bm{\tdvth})\equiv\frac{1}{\tdvrho}\tilde\matk^\hi_A[\tsD^A\mcT+(\tdu-\mcT)\tsD^A\hmcW],
    \end{gather}
    with $\tilde\matk^\hi_A\tilde\matk^\hj_B\tilde h^{AB}=\delta^{\hi\hj}$.
\end{subequations}
$\tilde\matk^\hi_A$ defines the dyads associated with $\tilde h_{AB}$ of the soft-haired spacetime.
As in the previous subsection, we choose $\tilde\matk^\hsan_A\ud\tdvth^A$ to be parallel to $\ud\tilde\varphi$ direction.
This tetrad basis~\eqref{eq-tet-tf} is associated with the preferred observer in the soft-haired black hole.
It is interesting to study how the 4-velocity of the preferred observer changes,
\begin{equation}
    \label{eq-4v-obs-tf}
    \begin{split}
        \sigma_\hzero= & \frac{1}{2}\tilde\sigma_\hzero\left(\tsD^2+1+\frac{\tilde\scR}{2}\right)e^{\hmcW}+\frac{1}{2}\tilde\sigma_\hyi\left(\tsD^2-1+\frac{\tilde\scR}{2}\right)e^{\hmcW} \\
                       & -\tilde\sigma_\hi\tilde\matk^\hi_A\tsD^Ae^{\hmcW}
        +\order{\frac{1}{\tdvrho}}.
    \end{split}
\end{equation}
Therefore, the preferred observer in the bald Kerr spacetime is moving relative to the one in the haired spacetime, as long as $\hmcW\ne0$.
This also explains why the super-Lorentz transformation with $\hmcW\ne0$ can be named the super-boost.
Let us also consider how the tetrads in the angular directions transform,
\begin{equation}
    \label{eq-4v-obs-tf-a}
    \sigma_\hi=(\tilde\sigma_\hzero+\tilde\sigma_\hyi)\tilde\matk^A_\hi\tsD_Ae^{-\hmcW}+\rotm^\hj_\hi\tilde\sigma_\hj+\order{\frac{1}{\tdvrho}}.
\end{equation}
The straightforward computation shows that $\rotm^\hi_\hj$ is actually a rotation matrix,
\begin{equation*}
    \label{eq-rotm-pro}
    \begin{split}
        \rotm^\hk_\hi\rotm^\hl_\hj\delta_{\hk\hl}= & e^{-2\hmcW}\tilde\matk^\hk_A\tilde\matk_B^\hl\delta_{\hk\hl}\jac^A_C\jac^B_D\matk^C_\hi\matk^D_\hj           \\
        =                                          & e^{-2\hmcW}\tilde h_{AB}\jac^A_C\jac^B_D\matk^C_\hi\matk^D_\hj=h_{AB}\matk^A_\hi\matk^B_\hj=\delta_{\hi\hj},
    \end{split}
\end{equation*}
where Eq.~\eqref{eq-det-dif} has been applied.
So when $\hmcW=0$, $\sigma_\hi$'s are just rotated relative to $\tilde\sigma_\hi$'s.
Thus, according to Eqs.~\eqref{eq-det-dif} and \eqref{eq-def-rotm}, $\Upsilon^A$ determines the super-rotation.
This discussion also shows that the super-boost transformation looks like the ordinary boost in the local inertial frame $\{\tilde\sigma^{\hat\mu}\}$, and similarly, the super-rotation can be viewed as the ordinary rotation locally.
In this sense, super-boosts and super-rotations are \emph{local} boosts and rotations, respectively.
They explicitly depend on the angular coordinates, $\hmcW=\hmcW(\bm{\tdvth})$ and $\rotm^\hi_\hj=\rotm^\hi_\hj(\bm{\tdvth})$.
As one can check, the supertranslation $\mcT$ does not show up in the leading order of the transformation of the tetrads, but it affects the celestial coordinates, to be derived immediately.

Now, consider the celestial coordinates of the very photon still labeled by $(\lambda,q)$.
This is meaningful, as the soft-haired Kerr black hole still has the Killing vector fields $\mfk_t$ and $\mfk_\phi$, and also the Killing-Yano tensor, so the photon can still be described by $E,\,\lambda$, and $q$.
The photon has the following new momentum components,
\begin{subequations}
    \begin{gather}
        \tilde p^\hzero=e^{-\hmcW}p^\hyi+\order{\frac{1}{\varrho}}\approx \tilde p^\hyi,\\
        \tilde p^\hi=\rotm^\hi_\hj p^\hj+e^{-\hmcW}\cS^\hi p^\hyi+\order{\frac{1}{\varrho^2}},
    \end{gather}
\end{subequations}
in terms of the momentum $p^{\hat\mu}$ in the old tetrad basis.
Now, it is easy to work out the transformation rule of the celestial coordinates,
\begin{equation}
    \label{eq-cc-deg}
    \tilde x^\hi=\frac{\tilde p^\hi}{\tilde p^\hzero}=e^{\hmcW}\rotm^\hi_\hj(\bm{\tilde\vartheta})x^\hj+\cS^\hi(\tdu,\bm{\tdvth})+\order{\frac{1}{\varrho^2}}.
\end{equation}
Note that according to Eq.~\eqref{eq-cc-bk-bs-1}, $x^\hi$ are of the order $1/\varrho$.
So this equation tells us that the image of a soft-haired Kerr black hole can be obtained from the one of the bald counterpart by rotating it by $\rotm^\hi_\hj(\bm{\tdvth})$, dilating it by $e^{\hmcW}$, and shifting it by $\cS^\hi(\tdu,\bm\tdvth)$.
Moreover, by Eq.~\eqref{eq-def-rotm}, the shifting consists of two ``steps'', i.e., the image is first displaced by
\begin{equation}
    \label{eq-def-mci}
    \mcI^\hi\equiv\left.\frac{1}{\tdvrho}\tilde\matk^\hi_A(\tsD^A\mcT-\mcT\tsD^A\hmcW)\right.,
\end{equation}
and then drifting at a constant speed
\begin{equation}
    \label{eq-def-ve}
    \cV^\hi\equiv\dot \cS^\hi=\left.\frac{1}{\tdvrho}\tilde\matk^\hi_A\tsD^A\hmcW\right.,
\end{equation}
in the celestial plane.
Since $\rotm^\hi_\hj$ is a rotation matrix, the images of the soft-haired and bald black holes are similar in shape.

\begin{widetext}
    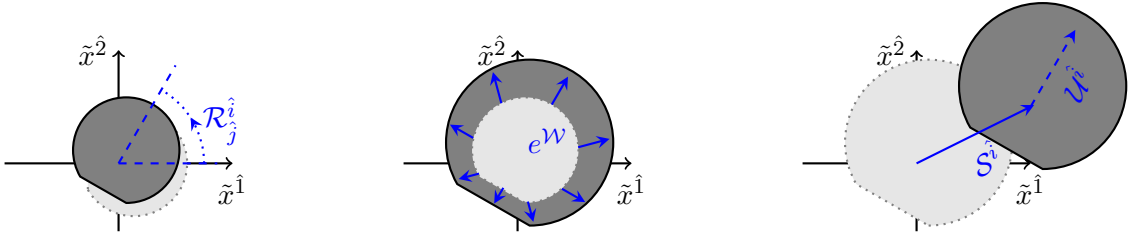
\begin{figure*}[ht]
        \centering
        \tikzset{
            dshape/.pic={
                    \fill [fill=gray,draw=black,thick]({-1/sqrt(3)},-0.5) arc (-150:150:1) -- ({-1/sqrt(3)},0.5) -- cycle;
                }
        }
        \tikzset{
            dshape2/.pic={
                    \fill [fill=gray!20,draw=black!50,dotted,thick]({-1/sqrt(3)},-0.5) arc (-150:150:1) -- ({-1/sqrt(3)},0.5) -- cycle;
                }
        }
        \begin{tikzpicture}[scale=0.75]
            \draw[->,thick] (-2,0) -- (2,0) node[anchor=north]{$\tilde x^\hyi$};
            \draw[->,thick] (0,-1.2) -- (0,2) node[anchor=east]{$\tilde x^\her$};
            \pic[scale=0.7] at (0,0) {dshape2};
            \pic[scale=0.7,rotate=60] at (0,0) {dshape};
            \draw[blue,thick,dashed] (0,0) -- (1.,{1.*sqrt(3)});
            \draw[blue,thick,dashed] (0,0) -- ({sqrt(3)},0);
            \draw [blue,thick,dotted] (1.5,0) arc [start angle=0,end angle=60,radius=1.5] node[sloped,pos=0.5,allow upside down,scale=1.5]{\arrowIn};
            \node (rotm) at  ({1.5*cos(30)},{1.5*sin(30)}) [right,blue] {$\rotm^\hi_\hj$};
            \draw[->,thick] (5,0) -- (9,0) node[anchor=north]{$\tilde x^\hyi$};
            \draw[->,thick] (7,-1.2) -- (7,2) node[anchor=east]{$\tilde x^\her$};
            \pic[scale=1.1,rotate=60] at (7,0) {dshape};
            \pic[scale=0.7,rotate=60] at (7,0) {dshape2};
            \draw[thick,blue]({7+1.2*cos(60)},{1.2*sin(60)}) -- ({7+1.7*cos(60)},{1.7*sin(60)}) node[sloped,scale=1.5,rotate=60]{\arrowIn};
            \draw[thick,blue]({7+1.1*cos(105)},{1.1*sin(105)}) -- ({7+1.6*cos(105)},{1.6*sin(105)}) node[sloped,scale=1.5,rotate=105]{\arrowIn};
            \draw[thick,blue]({7+0.9*cos(150)},{0.9*sin(150)}) -- ({7+1.3*cos(150)},{1.3*sin(150)}) node[sloped,scale=1.5,rotate=150]{\arrowIn};
            \draw[thick,blue]({7+0.7*cos(195)},{0.7*sin(195)}) -- ({7+1.0*cos(195)},{1.0*sin(195)}) node[sloped,scale=1.5,rotate=195]{\arrowIn};
            \draw[thick,blue]({7+0.7*cos(285)},{0.7*sin(285)}) -- ({7+1.0*cos(285)},{1.0*sin(285)}) node[sloped,scale=1.5,rotate=285]{\arrowIn};
            \draw[thick,blue]({7+1.1*cos(15)},{1.1*sin(15)}) -- ({7+1.6*cos(15)},{1.6*sin(15)}) node[sloped,scale=1.5,rotate=15]{\arrowIn};
            \draw[thick,blue]({7+0.9*cos(-30)},{0.9*sin(-30)}) -- ({7+1.3*cos(-30)},{1.3*sin(-30)}) node[sloped,scale=1.5,rotate=-30]{\arrowIn};
            \draw[thick,blue]({7+0.5*cos(-120)},{0.5*sin(-120)}) -- ({7+0.75*cos(-120)},{0.75*sin(-120)}) node[sloped,scale=1.5,rotate=-120]{\arrowIn};
            \node (sct) at (7,0) [above right,blue] {$e^{\hmcW}$};
            \draw[->,thick] (12,0) -- (16,0) node[anchor=north]{$\tilde x^\hyi$};
            \draw[->,thick] (14,-1.2) -- (14,2) node[anchor=east]{$\tilde x^\her$};
            \pic[scale=1.1,rotate=60] at (14,0) {dshape2};
            \pic[scale=1.1,rotate=60] at (16,1) {dshape};
            \draw[thick,blue] (14.0,0) -- node [sloped,below]{$\cS^\hi$} (16.0,1) node[sloped,scale=1.5,rotate={atan(0.5)}]{\arrowIn};
            \draw[thick,blue,dashed](16,1) -- node [sloped,below]{$\cV^\hi$} ({16+1.5*cos(60)},{1+1.5*sin(60)}) node[sloped,scale=1.5,rotate=60]{\arrowIn};
        \end{tikzpicture}
        \caption{
            The impacts of the soft hair on the image of an eternal black hole, represented by the changes in the shadow of a highly spinning black hole, illuminated by a large planar light source.
            Left: the rotation  $\rotm^\hi_\hj$ of the image;
            middle: the dilation $e^\hmcW$ of the image;
            right: the shifting $\cS^\hi$ of the image at the constant velocity $\cV^\hi$.
        }
        \label{fig-bhigbmset}
    \end{figure*}
\end{widetext}

Figure~\ref{fig-bhigbmset} shows the transformation in the shadow, if the black hole is backlit by a huge planar light source.
The left panel displays the effect of the rotation, where the gray region represents the shadow of the bald black hole, and the black region is for the soft-haired one.
We use $\rotm^\hi_\hj$ to schematically represent the amount of the rotation.
In this panel, in order to show the effect of the rotation more clearly, we assume that the bald black hole has a very large spin, so that its image has the ``D'' shape  when viewed near the equatorial plane \cite{Chandrasekhar:1985kt}.
The middle panel displays the effect of the subsequent dilation  $e^\hmcW$, in which the gray region is the shadow after the rotation, and the black region is the result of the dilation.
And finally, the right panel demonstrates the shifting of the shadow.
Here, the gray area is the shadow at an initial time, say $\tdu=0$, and later, it shifts to the black region.
The solid arrow stands for the total displacement $\cS^\hi=\mcI^\hi+\tdu\cV^\hi$, while the dashed arrow indicates the velocity $\cV^\hi$.

Let us also consider how each specific transformation changes the image.
First, the supertranslation $\mcT$ appears only in Eq.~\eqref{eq-def-mci}, so it causes a constant displacement of the image by $\mcI^\hi$.
The image, neither rotated nor dilated, is just displaced to a new position, and stays there.
Therefore, the supertranslation can be viewed as an ordinary translation locally.
This result agrees with the linearized analysis \cite{Lin:2022ksb,Zhu:2022shb}.
Second, the super-rotation transformation $\Upsilon^A$ with $\hmcW=0$ merely rotates the image by $\rotm^\hi_\hj$.
Finally, the super-boost transformation with $\hmcW$ related to $\Upsilon^A$ via Eq.~\eqref{eq-det-dif} modifies the image in the most complex way.
Basically, it leads to all of the effects of a generic generalized BMS transformation, except that its displacement $\mcI^\hi=0$.

\textcolor{black}{These results also show that although the bald Kerr black hole in Sec.~\ref{sec-bk-im} and the soft-haired one in this subsection are the degenerate vacuum states, their images are different, measured by a single observer, if this observer happens to take the pictures of the two black holes simultaneously.
    In this sense, the gravitational vacua with different Noether charges are physically distinct.}

Together with the discussion on the transformation rule of the tetrads, one finds out that supertranslations, super-boosts, and super-rotations are \emph{local} translations, boosts, and rotations, respectively, at least in the phenomenology of the black hole image.
Being local implies that the effects induced by these transformations on the image can be erased by suitably choosing a different local inertial frame, i.e., a new tetrad basis not being the adapted one.
For example, since the soft-haired spacetime also has the Killing vector $\mfk_\phi$, and it has an angular component asymptotically according to Eq.~\eqref{eq-phi-asy}, one can make the following choice,
\begin{equation}
    \tilde\sigma_\hsan=\frac{\tilde\matk^A_\hsan}{\tdvrho}\pd_\ta+\order{\frac{1}{\tdvrho^2}}\parallel \jac_\varphi^A\pd_\ta+\order{\frac{1}{\tdvrho}}.
\end{equation}
This implies that
\begin{equation}
    \tilde\matk^A_\hi=e^{-\hmcW}\jac^A_B\matk^B_\hi,
\end{equation}
and thus,
\begin{equation}
    \rotm^\hi_\hj=\delta^\hi_\hj,
\end{equation}
that is, the rotation effect can be canceled.
If the black hole is only supertranslated, one can change the origin of the celestial coordinates to eliminate the supertranslation effect \cite{Zhu:2022shb}.
The effects of the super-boost are harder to be erased, as this requires to (locally) boost the detector, although it is still possible in principle.
However, one shall keep in mind that the elimination of these effects is only local.
That is, one may negate the impacts of the generalized BMS transformation at a specific time $\tdu$ and angular position $\bm{\tdvth}$, but not at other times and positions.
So if one observes these effects caused by the soft hair, one shall make more measurements at different times and positions to confirm that they are not just local artifacts of the choice of the tetrad basis.

Now, let us comment on the consequences of the super-boost on the image.
Intuitively, one may expect that the super-boost will cause the dilation and the drifting of the image, but not the rotation.
This is true for some particular super-boosts.
For instance, if a super-boost can be treated as a local boost in $\tilde\sigma_{\hyi}$ direction, the image would not be rotated.
However, in the previous derivation, the super-boost is a general one, corresponding to a generic local boost in an arbitrary direction.
Indeed, Eq.~\eqref{eq-4v-obs-tf} shows that the relative velocity of the two preferred observers has three spatial components.
A general boost results in the rotation formally, as shown in Ref.~\cite{Jackson:1998nia}.

\section{The black hole image memory effect}
\label{sec-ime}

In the previous section, the focus was on the image of an eternal black hole.
However, a black hole in a more realistic scenario is always accompanied by various forms of matter and energy,
and some energetic processes occur, e.g.,
turbulent disk accretion \cite{1991ApJ...376..214B,RevModPhys.70.1}, relativistic jets \cite{1918PLicO..13....9C,2015ASSL..414.....C},
tidal disruption events \cite{1975Natur.254..295H,Rees:1988bf},
the Kozai-Lidov mechanism \cite{Kozai:1962zz,Lidov:1962wjn}, etc.
A large black hole may also have a secondary star or a smaller black hole \cite{1975Natur.254..295H,Rees:1988bf}, or even a binary system forming a hierarchical triple system \cite{Will:2020tri,Conway:2024azg,Conway:2025ixt}.
So the spacetime is slightly perturbed, and the energy can be released in the form of electromagnetic waves or gravitational waves \cite{Gupta:2019unn,Toscani:2021bzr,Pfister:2021ton,Yuan:2025fde}.
The emission of radiation will change the soft hair of the black hole \cite{Flanagan:2015pxa,Freidel:2021fxf}, and thus, the image.
This leads to the so-called image memory effect.

To explain this effect, we shall make use of the evolution equation~\eqref{eq-evo-m}, and the fact that $\tilde h_{AB}$ is independent of $\tdu$ \cite{Freidel:2021fxf}.
To be more specific, let us consider a binary system consisting of a central large black hole and a much smaller one.
Suppose at the beginning of the time $\tdu_0$, the central black hole is very far away from the smaller one, so the spacetime is basically a vacuum.
This means that initially, $\tilde h_{AB}(\tdu_0)$ and $\tilde c_{AB}(\tdu_0)$ are given by Eq.~\eqref{eq-sh-h-sh} with $\mcT$ labeled by a subscript ``0''.
$\mcT_0$ represents the initial value of the supertranslation.
It is not necessary to label $\hmcW$ or $\Upsilon^A$ (implicitly contained in $\ijac^A_B$ and $\tsD_A$), as $\tilde h_{AB}$ is constant in $\tdu$.
Then, the smaller black hole starts to inspiral into the central black hole, and the gravitational wave is emitted.
Eventually, after time $\tdu_f$, the black holes merge together, and the spacetime is again in a vacuum state.
At this time, $\tilde h_{AB}(\tdu_f)$ is still given by Eq.~\eqref{eq-h-af-nc}, but $\tilde c_{AB}(\tdu_f)$ is given by Eq.~\eqref{eq-shear-af-nc} with $\mcT$ replaced by $\mcT_f$.
Therefore, there is a total change in $\mcT$,
\begin{equation}
    \Delta\mcT=\mcT_f-\mcT_0.
\end{equation}
This can be computed by integrating Eq.~\eqref{eq-evo-m}, i.e.,
\begin{equation}
    \label{eq-int-evom}
    \begin{split}
        \tsD^2(\tsD^2+\tilde\scR)\Delta\mcT & -2\tsD_A\tsD_B(\tilde N^{AB}_f\Delta\mcT)                                 \\
        =                                   & 8\Delta \tilde M+\int_{\tdu_0}^{\tdu_f}\tilde N_{AB}\tilde N^{AB}\ud\tdu,
    \end{split}
\end{equation}
where $\tilde N_f^{AB}=-2e^\hmcW\tsD^{\langle A}\tsD^{B\rangle}e^{-\hmcW}$, and $\Delta\tilde M=\tilde M(\tdu_f)-\tilde M(\tdu_0)$.
Therefore, this equation implies that as long as there is gravitational wave emitted, and also the change in $\tilde M$, $\Delta\mcT\ne0$ generally.

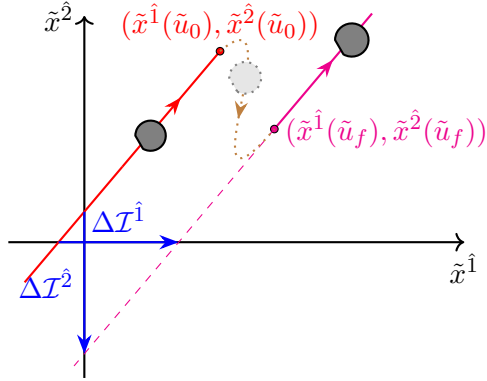
\begin{figure}[h]
    \centering
    \usetikzlibrary {arrows.meta,bending,positioning,fit,decorations.markings,decorations.pathmorphing, decorations.pathreplacing,decorations.shapes}
    \tikzset{->-/.style={decoration={
            markings,
            mark=at position #1 with {\arrow{{Stealth}}}},postaction={decorate}}}
    \tikzset{
        dshape2/.pic={
                \fill [fill=gray!20,draw=black!50,dotted,thick]({-1/sqrt(3)},-0.5) arc (-150:150:1) -- ({-1/sqrt(3)},0.5) -- cycle;
            }
    }
    \tikzset{
        dshape/.pic={
                \fill [fill=gray,draw=black,thick]({-1/sqrt(3)},-0.5) arc (-150:150:1) -- ({-1/sqrt(3)},0.5) -- cycle;
            }
    }
    \begin{tikzpicture}[scale=1]
        \draw[->,thick] (-1.0,0) -- (5,0) node[anchor=north]{$\tilde x^\hyi$};
        \draw[->,thick] (0,-1.8) -- (0,3) node[anchor=east]{$\tilde x^\her$};
        \coordinate (l1p1) at (0.5,1.0);
        \coordinate (l1p2) at ({0.5+2.0*cos(50)},{1.0+2.0*sin(50)});
        \draw[->-={0.6},thick,red] (l1p1) -- (l1p2) node[above]{$(\tilde x^\hyi(\tdu_0),\tilde x^\her(\tdu_0))$};
        \draw[fill=red](l1p2) circle (0.05);
        \pic[scale=0.2,rotate=30] at ({0.5+0.5*cos(50)},{1.0+0.5*sin(50)}) {dshape};
        \coordinate (l1p3) at ({0.5-2.0*cos(50)},{1.0-2.0*sin(50)});
        \draw[thick,red] (l1p1) -- (l1p3) ;
        \coordinate (l2p1) at (2.5,1.5);
        \coordinate (l2p2) at ({2.5+2.0*cos(50)},{1.5+2.0*sin(50)});
        \draw[->-={0.6},magenta,thick] (l2p1)  node[right]{$(\tilde x^\hyi(\tdu_f),\tilde x^\her(\tdu_f))$} -- (l2p2);
        \draw[fill=magenta](l2p1) circle (0.05);
        \pic[scale=0.22,rotate=30] at ({2.5+1.5*cos(50)},{1.5+1.5*sin(50)}) {dshape};
        \coordinate (l2p3) at ({2.5-4.1*cos(50)},{1.5-4.1*sin(50)});
        \draw[dashed,magenta] (l2p1) -- (l2p3);
        \coordinate (l1inty) at (0,{-0.5*tan(50)+1.0});
        \coordinate (l1intx) at ({0.5-1.0*cot(50)},0);
        \coordinate (l2inty) at (0,{-2.5*tan(50)+1.5});
        \coordinate (l2intx) at ({2.5-1.5*cot(50)},0);
        \draw[arrows = {-Stealth[scale=1.1]},blue,thick] (l1intx) --  node[sloped,above,align=center]{$\Delta \mcI^\hyi$}(l2intx);
        \draw[arrows = {-Stealth[scale=1.1]},blue,thick] (l1inty) --  node[left,align=center]{$\Delta \mcI^\her$}(l2inty);
        \coordinate (forcontrol1) at ({1.2+2.5*cos(50)},{1.5+2.5*sin(50)});
        \draw[->-={0.55},brown,dotted,thick] (l1p2) .. controls (forcontrol1) and (l2intx) .. (l2p1);
        \pic[scale=0.21,rotate=30] at ({1.1+1.5*cos(50)},{1.0+1.5*sin(50)}) {dshape2};
    \end{tikzpicture}
    \caption{The image memory effect.
        The red line on the left represents the trajectory before the emission of the radiation,
        while the magenta line on the right represents the trajectory after the emission.
        During the emission, the image accelerates, and its path is schematically represented by the brown dotted curve.
        The "D"-shaped areas stand for the black hole shadows at different locations.
        The blue arrows represent the components $\Delta \mcI^\hi$ of the image memory effect.
    }
    \label{fig-im}
\end{figure}

The change in $\mcT$ due to the emission of the gravitational wave results in the change in the image of the central black hole.
Let us first consider the image observed before $\tdu_0$ and after $\tdu_f$, which can be represented by the celestial coordinates of a particular photon,
\begin{subequations}
    \begin{gather}
        \tilde x^\hi(\tdu<\tdu_0)=e^{\hmcW}\rotm^\hi_\hj x^\hj+{\cS^\hi(\tdu<\tdu_0)}+\order{\frac{1}{\varrho^2}},\\
        \tilde x^\hi(\tdu>\tdu_f)=e^{\hmcW}\rotm^\hi_\hj x^\hj+{\cS^\hi(\tdu>\tdu_f)}+\order{\frac{1}{\varrho^2}},
    \end{gather}
\end{subequations}
where the dependence of $\bm{\tdvth}$ has been suppressed.
Therefore, the two images observed in different time periods differ in the shift $\cS^\hi$.
This means that the two images would drift in two different straight half-lines in the observer's view.
This is \emph{the image memory effect}, which is schematically shown in Fig.~\ref{fig-im}.
This figure displays the motion of the black hole shadow, the ``D'' shaped black and gray areas, due to the emission of the gravitational wave.
The red solid line represents the trajectory before $\tdu_0$, and the magenta stands for the trajectory after $\tdu_f$.
During the emission, the trajectory is no longer straight, given by the brown dotted curve.
Due to Eq.~\eqref{eq-def-ve}, these two half-lines share the same slope, as $\cV^\hi$ does not depend on $\tdu$.
So they are parallel to each other.
To further investigate the relation, it is useful to write down their equations.
The half-line before $\tdu_0$ can be given by
\begin{equation}
    \tilde X^\hi(\tdu<\tdu_0)=\tdu\cV^\hi+\tilde x^\hi_0,
\end{equation}
where $\tilde X^\hi_0$ is a constant, representing the interception with the $\tilde x^\hi$ axis.
After $\tdu_f$, it can be shown that the half-line is described by
\begin{equation}
    \tilde X^\hi(\tdu>\tdu_f)=\tdu_0\cV^\hi+\tilde X^\hi_0+\Delta \cS^\hi+(\tdu-\tdu_f)\cV^\hi,
\end{equation}
where $\Delta\cS^\hi\equiv\cS^\hi(\tdu_f)-\cS^\hi(\tdu_0)$.
One can extend these half-lines, and compute their distance,
\begin{equation}
    \Delta\mcI^\hi=\frac{1}{\tdvrho}\tilde\matk^\hi_A(\tsD^A\Delta\mcT-\Delta\mcT\tsD^A\hmcW).
\end{equation}
This quantity measures the image memory effect, and is represented by the two blue arrows in Fig.~\ref{fig-im}.
During the emission of the gravitational wave, $\Delta\mcT$ is a function of $\tdu$ according to Eq.~\eqref{eq-int-evom}.
Since we assumed that the secondary black hole is much smaller than the central one, the change in $\mcT$ is expected to be very slow.
So Eq.~\eqref{eq-cc-deg} still holds approximately.
Then, the trajectory of the image becomes curved, which explains the brown dotted curve in Fig.~\ref{fig-im}.

Therefore, the image memory effect is basically the change in the half-lines in which the image drifts before and after the emission of the gravitational wave.
This change is permanent.
This effect is very similar to the gravitational wave memory effect \cite{Zeldovich:1974gvh,Braginsky:1986ia,Christodoulou1991,Thorne:1992sdb,DeLuca:2024cjl,DeLuca:2024asq} that might be detected by the gravitational wave interferometers \cite{Hubner:2019sly,Grant:2022bla,Gasparotto:2023fcg,Inchauspe:2024ibs,Hou:2024rgo}.
Like the gravitational wave memory effect, the image memory effect is also related to the vacuum transition, as before and after the emission, $\mcT$ changes \cite{Strominger:2014pwa,Strominger:2018inf,Hou:2020tnd,Tahura:2020vsa,Seraj:2021qja,Hou:2021oxe,Hou:2023pfz,Hou:2024xbv}.

\subsection{Estimation of the image memory effect}
\label{sec-est-im}

As mentioned above, the gravitational wave memory effect might be measured by the ground- and space-based interferometers in the future \cite{Hubner:2019sly,Grant:2022bla,Gasparotto:2023fcg,Inchauspe:2024ibs,Hou:2024rgo}.
In this subsection, we would like to estimate the image memory effect, and evaluate whether it can be observed by the current and future very long baseline interferometries (VLBI's).

To simplify the computation, one can assume that before $\tdu_0$, the central black hole carries little soft hair.
This requires that $\hmcW$ be small, and $\Upsilon^A$ generate a diffeomorphism on $\mathbb S^2$ that is very close to the identity.
So,
\begin{equation}
    \label{eq-happ}
    \tilde h_{AB}\approx \gamma_{AB},\quad \tsD_A\approx\sD_A,
\end{equation}
i.e., $\mathbb S^2$ is approximately a unit 2-sphere.
Therefore, such a black hole can also be characterized by the total mass $\mathring M$ and the spin $a$, of the corresponding bald counterpart.
This is because Eqs.~\eqref{eq-tqt-qp} imply that
\begin{equation}
    \tilde Q_t\approx4\pi\mathring M,\quad \tilde Q_\phi\approx-4\pi\mathring Ma.
\end{equation}
Of course, this black hole still carries a lot of soft hairs, but they are much smaller.
The emission of the gravitational wave is expected not to be strong, so $\Delta\mcT$ is small, too.
Then, Eq.~\eqref{eq-int-evom} is simplified,
\begin{equation}\label{eq-dmcT}
    \sD^2(\sD^2+2)\Delta\mcT \approx 8\Delta \mathring M+\int_{\tdu_0}^{\tdu_f}\ud \tdu\tilde N_{AB}\tilde N^{AB}.
\end{equation}
One can check that this is exactly one of the key equations used to compute the gravitational wave memory effect, by comparing it with, e.g., Eq.~(35) in Ref.~\cite{Mitman:2020pbt}.
The image memory effect is now approximated  by
\begin{equation}
    \label{eq-im-app}
    \Delta\mcI^\hi\approx \frac{\matk^\hi_A\sD^A\Delta\mcT}{\tdvrho},
\end{equation}
where $\matk^\hi_a$ can be taken to be the matrix $\text{diag}(1,\csc\tdvth)$.

As for the gravitational wave memory effect, one may call the contribution from $\Delta\mathring M$ the linear image memory effect, and the one from the energy flux [the second term on the right-hand side of Eq.~\eqref{eq-dmcT}] the null image memory effect \cite{Bieri:2013ada}.
Usually, the null memory effect dominates \cite{Mitman:2020pbt}, so we will ignore the linear memory in the following.
With the standard treatment \cite{Mitman:2020pbt,Grant:2022bla,Hou:2024rgo}, one first rewrites the shear tensor in terms of the complex strain $h=h_+-ih_\times$,
\begin{equation}
    \tilde c_{AB}=\frac{\tdvrho}{2}(\gamma_A\gamma_Bh+\bar \gamma_A\bar \gamma_B\bar h),
\end{equation}
where $\gamma_{A}=-(1,i\sin\tdvth)$ and the overhead bar means to take the complex conjugate.
Then, one can decompose $h=\sum_{\ell m}{}_{-2}Y_{\ell m}h_{\ell m}$ and $\Delta\mcT=\sum_{\ell m}Y_{\ell m}\Delta\mcT_{\ell m}$
where ${}_{-2}Y_{\ell m}$ is the spin-weighted spherical harmonics and $Y_{\ell m}$ the familiar spherical harmonics.
Using Eq.~\eqref{eq-im-app}, one obtains the expansion coefficients,
\begin{subequations}
    \begin{equation}
        \label{eq-phi-lm}
        \begin{split}
            \Delta\mcT_{\ell m}= & 2\tdvrho^2\frac{(\ell-2)!}{(\ell+2)!}\widehat\sum\mathcal C_\ell(-2,\ell_1,m_1;2,\ell_2,-m_2) \\
                                 & \times(-1)^{m_2}\int_{\tdu_0}^{\tdu}\dot h_{\ell_1m_1}\dot{\bar h}_{\ell_2m_2}\ud \tdu',
        \end{split}
    \end{equation}
    where the sum $\widehat\sum$ is over $\ell_1,m_1$ and $\ell_2,m_2$,
    and
    \begin{equation}
        \label{eq-def-cl}
        \begin{split}
                   & \mathcal C_\ell(s_1,\ell_1,m_1;s_2,\ell_2,m_2)                                                            \\
            \equiv & \int\ud^2\tdvth\sqrt\gamma({}_{s_1}Y_{\ell_1m_1})({}_{s_2}Y_{\ell_2m_2})({}_{s}\bar Y_{\ell m})|_\Lambda,
        \end{split}
    \end{equation}
\end{subequations}
with $\Lambda$ meaning to evaluate the integral under the following condition: $s=s_1+s_2$, $m=m_1+m_2$, and $\text{max}\{|\ell_1-\ell_2|,|m_1+m_2|,|s_1+s_2|\}\le\ell\le \ell_1+\ell_2$.

The image memory effect $\Delta\mcI^\hi$ can also be decomposed as
\begin{equation}
    \label{eq-im-dec}
    \Delta\mcI^\hi=\sum_{\ell m}Y_{\ell m}\Delta\mcI^\hi_{\ell m}.
\end{equation}
By Eq.~\eqref{eq-im-app}, one knows that
\begin{equation}
    \label{eq-im-dec-t}
    \Delta\mcI^\hi_{\ell m}\approx \frac{\Delta\mcT_{\ell m}}{\tdvrho}\matk^\hi_A\sD^AY_{\ell m}.
\end{equation}
Due to the overall factor in Eq.~\eqref{eq-phi-lm}, $\Delta\mcT_{\ell m}$ and so $\Delta\mcI^\hi_{\ell m}$, with a small $\ell$, contribute to the image memory effect the most.

\begin{figure}[ht]
    \centering
    \includegraphics[width=0.45\textwidth]{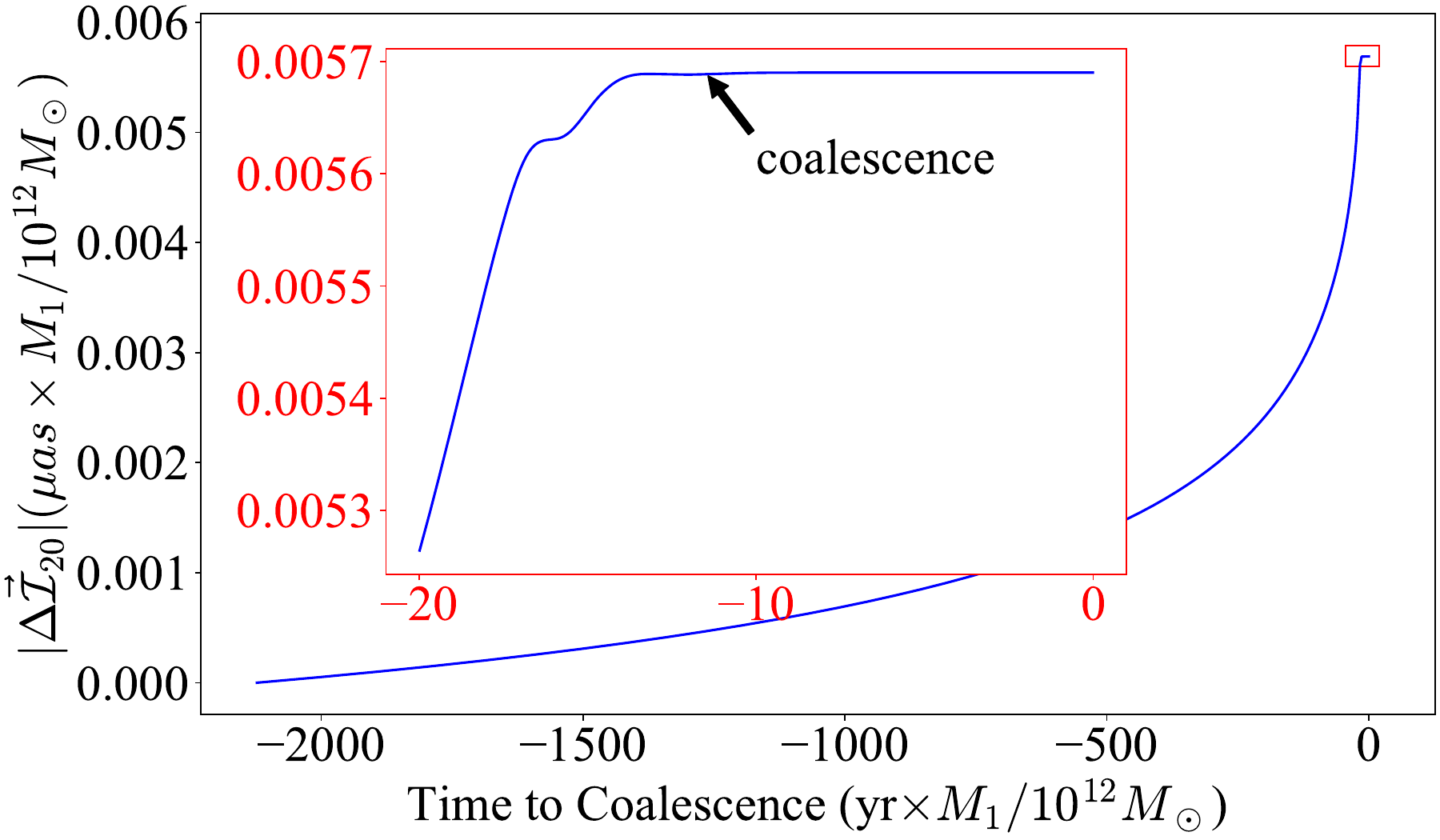}
    \caption{
        The estimation of the image memory effect for an intermediate mass ratio inspiral.
        The central black hole has a spin $a=0.8$, and the mass ratio is fixed at $q=M_1/M_2=100$.
        They are at a distance of $1\,\text{Gpc}$ from the Earth.
        The inset displays the details of the image memory effect during the last $20 \frac{M_1}{10^{12}M_\odot}$ years, corresponding to the small red rectangle at the top right corner.
    }
    \label{fig-imesim}
\end{figure}
Now, one can simulate the image memory effect.
We consider a huge, spinning central black hole with the mass $M_1=10^{12}M_\odot$ and the spin $a=0.8$.
Its companion is a spinless black hole of mass $M_2=10^{10}M_\odot$.
They are assumed to be at $\tdvrho=1$ Gpc away from the earth.
These two form an intermediate mass ratio inspiral, whose gravitational waveform can be computed with \verb+BHPTNRSurrogate+ \cite{islam2022surrogate}.
More specifically, the model \verb+BHPTNRSur2dq1e3+ was actually used.
This model allows $-0.8\le a\le 0.8$, and the mass ratio $q=M_1/M_2$ to be in the range $3\le q\le 1000$.
We used this model to generate all available spherical modes $h_{\ell m}$ \cite{Rink:2024swg}.
These modes $h_{\ell m}$ were then substituted into Eq.~\eqref{eq-phi-lm}, and it turned out that $\Delta\mcT_{20}$ is the largest, which is also true for other choices of $M_1$, $M_2$, and $a$.
According to Eq.~\eqref{eq-im-dec-t}, the vector $\Delta\mcI^\hi_{20}$ is expected to be the biggest, and its magnitude $|\Delta\bm\mcI_{20}|$ can be estimated to be
\begin{equation}
    \label{eq-est-im20}
    |\Delta\bm\mcI_{20}|\sim\frac{\Delta\mcT_{20}}{\tdvrho}.
\end{equation}
Varying $M_1$ while keeping $a$ and $q=M_1/M_2$ fixed, one finds out that $|\Delta\bm\mcI_{20}|$ and the evolution time are both proportional to $M_1$.
So we plot the evolution of $|\Delta\bm\mcI_{20}|$ in Fig.~\ref{fig-imesim} with the units of the axes scaled by $M_1$.
As one can see, the image memory effect is very small.
The total $|\Delta\bm\mcI^\text{tot}_{20}|$, i.e., the final value at the time 0, is about $5.7\times10^{-3}\frac{M_1}{10^{12}M_\odot}\mu$as, and it is accumulated over a long time, nearly $2000 \frac{M_1}{10^{12}M_\odot}$ yrs.
Within the last $10 \frac{M_1}{10^{12}M_\odot}$ yrs before the coalescence, the partial image memory effect $|\Delta\bm\mcI^\text{part}_{20}|$ is about $4.0\times10^{-4} \frac{M_1}{10^{12}M_\odot}\mu$as.
As one can see, the slope of the curve increases in the early time, and decreases later.
Near the coalescence, the slope changes in a more complicated way, as shown in the inset.
One may check that the slope is the largest at about $-20 \frac{M_1}{10^{12}M_\odot}$ yrs, i.e., the beginning of the curve in the inset.
This observation is useful, as for a very large central black hole, say $M_1=10^{14}M_\odot$ in the mass range of the speculated stupendously large black hole \cite{Carr:2020erq,Carr:2023tpt}, although $|\Delta\bm\mcI^\text{part}_{20}|\sim 0.02\mu$as, it is accumulated in 1000 yrs.
It is unlikely to have an observation plan of that long.
Since the slope of the curve in Fig.~\ref{fig-imesim} has a maximum, then if it happens to start to take data at about $-20 \frac{M_1}{10^{12}M_\odot}$ yrs, within 10 yrs, the image memory effect will be about
\begin{equation}
    \label{eq-ime-10y}
    |\Delta\bm\mcI^\text{10y}_{20}|\sim4\times10^{-4}\mu \text{as},
\end{equation}
for any $M_1$.

\begin{widetext}
    \begin{figure*}[ht]
        \centering
        \includegraphics[width=0.9\textwidth]{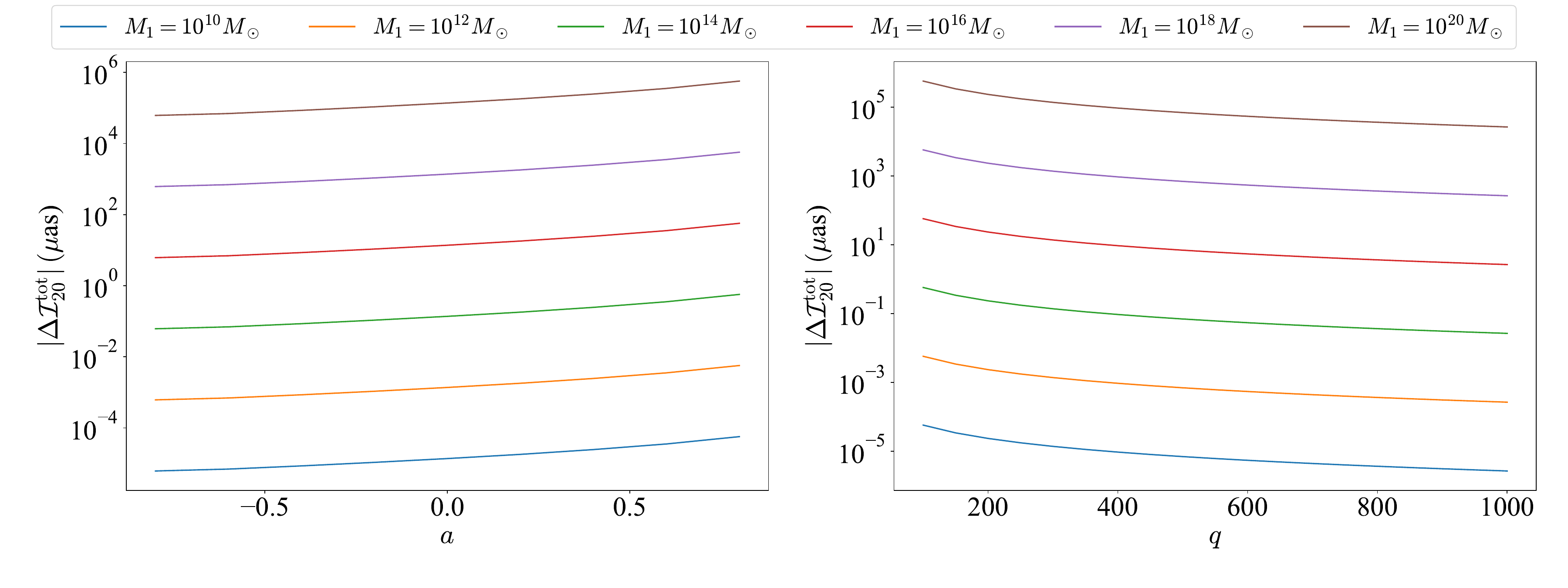}
        \caption{
            The total $|\Delta\bm\mcI^\text{tot}_{20}|$ as functions of $a$ and $q$ for different values of $M_1$.
            The left panel shows $|\Delta\bm\mcI^\text{tot}_{20}|\text{ vs. }a$ at $q=100$, while the right panel displays $|\Delta\bm\mcI^\text{tot}_{20}|\text{ vs. }q$ at $a=0.8$.
        }
        \label{fig-totimevsqs1}
    \end{figure*}
\end{widetext}

One may also vary the spin $a$ of the central black hole and the mass ratio $q$ to examine how the image memory effect changes.
Then, one can get Fig.~\ref{fig-totimevsqs1}.
In this figure, the left panel shows $|\Delta\bm\mcI^\text{tot}_{20}|\text{ v.s. }a$ at $q=100$, while the right panel displays $|\Delta\bm\mcI^\text{tot}_{20}|\text{ v.s. }q$ at $a=0.8$.
As one can see,  $|\Delta\bm\mcI^\text{tot}_{20}|$ increases with $a$, but decreases with $q$.
$M_1$ affects $|\Delta\bm\mcI^\text{tot}_{20}|$ the most, while $a$ and $q$ change the order of the magnitude of $|\Delta\bm\mcI^\text{tot}_{20}|$ approximately by 1 at most.

As estimated in Eq.~\eqref{eq-ime-10y}, during the 10 yrs in which the image memory effect accumulates the fastest, the image moves at most about $4\times10^{-4}\mu$as, which is much smaller than the angular resolution of the current and the next-generation EHT \cite{EventHorizonTelescope:2019dse,Ayzenberg:2023hfw}.
The higher angular resolution could be achieved with the space-borne VLBI, which has the longer baseline \cite{2021MNRAS.500.4866A,Gurvits:2022wgm,Hudson:2023tuy,2024SPIE13092E....C}.
The Moon-Earth VLBI projects were envisioned \cite{2019MNRAS.486..815K,2022MNRAS.511..668L,2025AcAau.232..564H,2025SCPMA..6919511H}, and the simulation with OmniUV showed that it is possible to resolve an image of the size of $0.01\mu$as \cite{2022AJ....164...67L}.
The estimation~\eqref{eq-ime-10y} is still smaller, even if one assumes the central black hole is at about 100 Mpc away from the earth.

However, this does not imply the image memory effect cannot be observed definitely.
Here, we did not consider the impact of the cosmological expansion on the black hole image, although we placed the black hole at 1 Gpc, which is quite distant.
In fact, the expansion of the universe might cause the size of the image to grow with the redshift $z$ of the central black hole if $z$ is large enough \cite{Perlick:2018iye,Bisnovatyi-Kogan:2018vxl,Tsupko:2019mfo,Li:2020drn}.
Although the image memory effect is not the size of the image, but the change in the position of the image, by the key idea of the derivations in Refs.~\cite{Perlick:2018iye,Bisnovatyi-Kogan:2018vxl,Tsupko:2019mfo,Li:2020drn}, i.e., the use of the angular diameter redshift relation \cite{Mukhanov:2005sc}, the image memory effect shall also increase with the redshift.
The influence of the cosmological expansion on the image memory effect will be the topic in the future.

\section{Conclusion}
\label{sec-con}

In this work, we considered the influence of soft hair on the black hole image.
The derivation shows that although the shape of the image remains intact, it is rotated, dilated, and drifting at a constant velocity in the observer's view if an eternal black hole has soft hairs.
This result was obtained via the suitable generalized BMS transformation.
Once the transformation is performed, the bald black hole metric becomes a new one carrying soft hair.
The change in the image can subsequently be determined.
This is a mathematical trick that greatly simplifies the derivation, and one does not have to worry about how the light source, such as the accretion disk, changes due to the soft hair.
This method gives the most general result.
Since the generalized BMS transformation is a diffeomorphism, the rotation, dilation, and drifting of the image are all local effects.
They can be canceled if a suitable local reference frame is chosen at a particular time and angular location, but they cannot be eliminated globally.

In a more realistic scenario, the image of the soft-haired black hole would change in a more interesting way.
That is, the emission of radiation, either gravitational waves or electromagnetic waves, would modify the soft hair, and thus, the image would no longer drift at a constant velocity.
Instead, it roams in the observer's view.
Specifically, if the gravitational wave is generated due to the inspiral of a much smaller black hole into a huge central black hole, there exists the image memory effect, the permanent change in the straight line along which the image drifts before and after the emission.
The primary simulation done in this work shows that this effect is very weak and cannot be detected by current and future VLBI's.
However, in the current derivation, the spacetime was taken to be asymptotically flat, so the impact of cosmological expansion was not considered.
It is expected that once the expansion of the universe is taken into account, the image memory effect would become more significant as long as the black hole is at a large enough redshift.
This will be the topic of future work.

\begin{acknowledgments}
    We were grateful for the helpful discussions with Yungui Gong, Pujian Mao, Wu-zhong Guo, Chen Yuan, Chao Zhang, Tieguang Zi, Shuxun Tian, and Zi-Chao Lin.
    This work was supported by the National Natural Science Foundation of China under grant Nos.~12205222, 12433001, and 12021003.
\end{acknowledgments}

%
\appendix
\section{Generalization to the Weyl BMS case}
\label{sec-g-wbms}

In the main text, the symmetry considered was the generalized BMS transformation, which belongs to the so-called Weyl BMS transformation \cite{Freidel:2021fxf,Flanagan:2023jio}.
One can modify the results in the main text easily to determine what one would observe in the Weyl BMS case.

The Weyl BMS transformation generalizes the generalized BMS transformation.
At the infinitesimal level, the generator is still formally given by Eq.~\eqref{eq-ass-sol} with $\sD_AY^A$ replaced by $2W(\bm\vartheta)$, independent of $Y^A(\bm\vartheta)$ and $T(\bm\vartheta)$.
In the Weyl BMS group, $T$ still parameterizes the supertranslation, but $Y^A$ parameterizes the arbitrary diffeomorphism on $\mathbb S$.
$W$ generates the so-called Weyl rescaling transformation.
The generalized BMS transformation is a special case of the Weyl BMS with $W$ determined by $Y^A$.

The finite Weyl BMS transformation is also given by Eq.~\eqref{eq-f-gbms-vac} without the condition~\eqref{eq-det-dif}, so that $\hmcW$ is independent of $\bm\Upsilon$.
Thus, it is easy to expect that Eq.~\eqref{eq-cc-deg} still describes how the image of the soft-haired BH changes under a generic Weyl BMS transformation.
More specifically, a purely finite supertranslation $\mcT$ causes a constant shift by $\tilde\matk^\hi_A\tsD^A\mcT/\tdvrho$; a pure diffeomorphism $\bm\Upsilon$ on $\mathbb S$ rotates the image by $\rotm^\hi_\hj$; and a pure Weyl rescaling transformation $\hmcW$ dilates and drifts the image, without rotating it.
As in the generalized BMS case, these effects are angle-dependent.
So if one observes the image of an eternal black hole, one would have to take the pictures at various angular coordinates in order to confirm whether the central black hole carries soft hairs.

The image memory effect in the Weyl BMS case also occurs, and is very similar to that discussed in the main text.
Its magnitude shall be weak, too.



\bibliography{bhigbms_v6.bbl}

\begin{thebibliography}{134}%
\makeatletter
\providecommand \@ifxundefined [1]{%
 \@ifx{#1\undefined}
}%
\providecommand \@ifnum [1]{%
 \ifnum #1\expandafter \@firstoftwo
 \else \expandafter \@secondoftwo
 \fi
}%
\providecommand \@ifx [1]{%
 \ifx #1\expandafter \@firstoftwo
 \else \expandafter \@secondoftwo
 \fi
}%
\providecommand \natexlab [1]{#1}%
\providecommand \enquote  [1]{``#1''}%
\providecommand \bibnamefont  [1]{#1}%
\providecommand \bibfnamefont [1]{#1}%
\providecommand \citenamefont [1]{#1}%
\providecommand \href@noop [0]{\@secondoftwo}%
\providecommand \href [0]{\begingroup \@sanitize@url \@href}%
\providecommand \@href[1]{\@@startlink{#1}\@@href}%
\providecommand \@@href[1]{\endgroup#1\@@endlink}%
\providecommand \@sanitize@url [0]{\catcode `\\12\catcode `\$12\catcode `\&12\catcode `\#12\catcode `\^12\catcode `\_12\catcode `\%12\relax}%
\providecommand \@@startlink[1]{}%
\providecommand \@@endlink[0]{}%
\providecommand \url  [0]{\begingroup\@sanitize@url \@url }%
\providecommand \@url [1]{\endgroup\@href {#1}{\urlprefix }}%
\providecommand \urlprefix  [0]{URL }%
\providecommand \Eprint [0]{\href }%
\providecommand \doibase [0]{https://doi.org/}%
\providecommand \selectlanguage [0]{\@gobble}%
\providecommand \bibinfo  [0]{\@secondoftwo}%
\providecommand \bibfield  [0]{\@secondoftwo}%
\providecommand \translation [1]{[#1]}%
\providecommand \BibitemOpen [0]{}%
\providecommand \bibitemStop [0]{}%
\providecommand \bibitemNoStop [0]{.\EOS\space}%
\providecommand \EOS [0]{\spacefactor3000\relax}%
\providecommand \BibitemShut  [1]{\csname bibitem#1\endcsname}%
\let\auto@bib@innerbib\@empty
\bibitem [{\citenamefont {Chrusciel}\ \emph {et~al.}(2012)\citenamefont {Chrusciel}, \citenamefont {Lopes~Costa},\ and\ \citenamefont {Heusler}}]{Chrusciel:2012jk}%
  \BibitemOpen
  \bibfield  {author} {\bibinfo {author} {\bibfnamefont {P.~T.}\ \bibnamefont {Chrusciel}}, \bibinfo {author} {\bibfnamefont {J.}~\bibnamefont {Lopes~Costa}},\ and\ \bibinfo {author} {\bibfnamefont {M.}~\bibnamefont {Heusler}},\ }\bibfield  {title} {\bibinfo {title} {{Stationary Black Holes: Uniqueness and Beyond}},\ }\href {https://doi.org/10.12942/lrr-2012-7} {\bibfield  {journal} {\bibinfo  {journal} {Living Rev. Rel.}\ }\textbf {\bibinfo {volume} {15}},\ \bibinfo {pages} {7} (\bibinfo {year} {2012})},\ \Eprint {https://arxiv.org/abs/1205.6112} {arXiv:1205.6112 [gr-qc]} \BibitemShut {NoStop}%
\bibitem [{\citenamefont {Misner}\ \emph {et~al.}(1973)\citenamefont {Misner}, \citenamefont {Thorne},\ and\ \citenamefont {Wheeler}}]{mtw}%
  \BibitemOpen
  \bibfield  {author} {\bibinfo {author} {\bibfnamefont {C.~W.}\ \bibnamefont {Misner}}, \bibinfo {author} {\bibfnamefont {K.~S.}\ \bibnamefont {Thorne}},\ and\ \bibinfo {author} {\bibfnamefont {J.~A.}\ \bibnamefont {Wheeler}},\ }\href@noop {} {\emph {\bibinfo {title} {{Gravitation}}}}\ (\bibinfo  {publisher} {W. H. Freeman},\ \bibinfo {address} {San Francisco},\ \bibinfo {year} {1973})\BibitemShut {NoStop}%
\bibitem [{\citenamefont {Strominger}\ and\ \citenamefont {Zhiboedov}(2016)}]{Strominger:2014pwa}%
  \BibitemOpen
  \bibfield  {author} {\bibinfo {author} {\bibfnamefont {A.}~\bibnamefont {Strominger}}\ and\ \bibinfo {author} {\bibfnamefont {A.}~\bibnamefont {Zhiboedov}},\ }\bibfield  {title} {\bibinfo {title} {{Gravitational Memory, BMS Supertranslations and Soft Theorems}},\ }\href {https://doi.org/10.1007/JHEP01(2016)086} {\bibfield  {journal} {\bibinfo  {journal} {JHEP}\ }\textbf {\bibinfo {volume} {01}},\ \bibinfo {pages} {086}},\ \Eprint {https://arxiv.org/abs/1411.5745} {arXiv:1411.5745 [hep-th]} \BibitemShut {NoStop}%
\bibitem [{\citenamefont {{Strominger}}(2014)}]{Strominger2014bms}%
  \BibitemOpen
  \bibfield  {author} {\bibinfo {author} {\bibfnamefont {A.}~\bibnamefont {{Strominger}}},\ }\bibfield  {title} {\bibinfo {title} {{On BMS invariance of gravitational scattering}},\ }\href {https://doi.org/10.1007/JHEP07(2014)152} {\bibfield  {journal} {\bibinfo  {journal} {JHEP}\ }\textbf {\bibinfo {volume} {07}},\ \bibinfo {eid} {152}},\ \Eprint {https://arxiv.org/abs/1312.2229} {arXiv:1312.2229 [hep-th]} \BibitemShut {NoStop}%
\bibitem [{\citenamefont {He}\ \emph {et~al.}(2015)\citenamefont {He}, \citenamefont {Lysov}, \citenamefont {Mitra},\ and\ \citenamefont {Strominger}}]{He:2014laa}%
  \BibitemOpen
  \bibfield  {author} {\bibinfo {author} {\bibfnamefont {T.}~\bibnamefont {He}}, \bibinfo {author} {\bibfnamefont {V.}~\bibnamefont {Lysov}}, \bibinfo {author} {\bibfnamefont {P.}~\bibnamefont {Mitra}},\ and\ \bibinfo {author} {\bibfnamefont {A.}~\bibnamefont {Strominger}},\ }\bibfield  {title} {\bibinfo {title} {{BMS supertranslations and Weinberg's soft graviton theorem}},\ }\href {https://doi.org/10.1007/JHEP05(2015)151} {\bibfield  {journal} {\bibinfo  {journal} {JHEP}\ }\textbf {\bibinfo {volume} {05}},\ \bibinfo {pages} {151}},\ \Eprint {https://arxiv.org/abs/1401.7026} {arXiv:1401.7026 [hep-th]} \BibitemShut {NoStop}%
\bibitem [{\citenamefont {Strominger}(2018)}]{Strominger:2018inf}%
  \BibitemOpen
  \bibfield  {author} {\bibinfo {author} {\bibfnamefont {A.}~\bibnamefont {Strominger}},\ }\href {https://press.princeton.edu/books/hardcover/9780691179506/lectures-on-the-infrared-structure-of-gravity-and-gauge-theory} {\emph {\bibinfo {title} {Lectures on the Infrared Structure of Gravity and Gauge Theory}}}\ (\bibinfo  {publisher} {Princeton University Press},\ \bibinfo {year} {2018})\ \Eprint {https://arxiv.org/abs/1703.05448} {arXiv:1703.05448} \BibitemShut {NoStop}%
\bibitem [{\citenamefont {Bondi}\ \emph {et~al.}(1962)\citenamefont {Bondi}, \citenamefont {van~der Burg},\ and\ \citenamefont {Metzner}}]{Bondi:1962px}%
  \BibitemOpen
  \bibfield  {author} {\bibinfo {author} {\bibfnamefont {H.}~\bibnamefont {Bondi}}, \bibinfo {author} {\bibfnamefont {M.~G.~J.}\ \bibnamefont {van~der Burg}},\ and\ \bibinfo {author} {\bibfnamefont {A.~W.~K.}\ \bibnamefont {Metzner}},\ }\bibfield  {title} {\bibinfo {title} {{Gravitational waves in general relativity. 7. Waves from axisymmetric isolated systems}},\ }\href {https://doi.org/10.1098/rspa.1962.0161} {\bibfield  {journal} {\bibinfo  {journal} {Proc. Roy. Soc. Lond. A}\ }\textbf {\bibinfo {volume} {269}},\ \bibinfo {pages} {21} (\bibinfo {year} {1962})}\BibitemShut {NoStop}%
\bibitem [{\citenamefont {Sachs}(1962{\natexlab{a}})}]{Sachs:1962wk}%
  \BibitemOpen
  \bibfield  {author} {\bibinfo {author} {\bibfnamefont {R.~K.}\ \bibnamefont {Sachs}},\ }\bibfield  {title} {\bibinfo {title} {{Gravitational waves in general relativity. 8. Waves in asymptotically flat space-times}},\ }\href {https://doi.org/10.1098/rspa.1962.0206} {\bibfield  {journal} {\bibinfo  {journal} {Proc. Roy. Soc. Lond. A}\ }\textbf {\bibinfo {volume} {270}},\ \bibinfo {pages} {103} (\bibinfo {year} {1962}{\natexlab{a}})}\BibitemShut {NoStop}%
\bibitem [{\citenamefont {Sachs}(1962{\natexlab{b}})}]{Sachs1962asgr}%
  \BibitemOpen
  \bibfield  {author} {\bibinfo {author} {\bibfnamefont {R.}~\bibnamefont {Sachs}},\ }\bibfield  {title} {\bibinfo {title} {Asymptotic symmetries in gravitational theory},\ }\href {http://link.aps.org/doi/10.1103/PhysRev.128.2851} {\bibfield  {journal} {\bibinfo  {journal} {Phys. Rev.}\ }\textbf {\bibinfo {volume} {128}},\ \bibinfo {pages} {2851} (\bibinfo {year} {1962}{\natexlab{b}})}\BibitemShut {NoStop}%
\bibitem [{\citenamefont {Ashtekar}\ and\ \citenamefont {Streubel}(1981)}]{Ashtekar:1981bq}%
  \BibitemOpen
  \bibfield  {author} {\bibinfo {author} {\bibfnamefont {A.}~\bibnamefont {Ashtekar}}\ and\ \bibinfo {author} {\bibfnamefont {M.}~\bibnamefont {Streubel}},\ }\bibfield  {title} {\bibinfo {title} {{Symplectic Geometry of Radiative Modes and Conserved Quantities at Null Infinity}},\ }\href {https://doi.org/10.1098/rspa.1981.0109} {\bibfield  {journal} {\bibinfo  {journal} {Proc. Roy. Soc. Lond. A}\ }\textbf {\bibinfo {volume} {376}},\ \bibinfo {pages} {585} (\bibinfo {year} {1981})}\BibitemShut {NoStop}%
\bibitem [{\citenamefont {Flanagan}\ and\ \citenamefont {Nichols}(2017)}]{Flanagan:2015pxa}%
  \BibitemOpen
  \bibfield  {author} {\bibinfo {author} {\bibfnamefont {E.~E.}\ \bibnamefont {Flanagan}}\ and\ \bibinfo {author} {\bibfnamefont {D.~A.}\ \bibnamefont {Nichols}},\ }\bibfield  {title} {\bibinfo {title} {{Conserved charges of the extended Bondi-Metzner-Sachs algebra}},\ }\href {https://doi.org/10.1103/PhysRevD.95.044002} {\bibfield  {journal} {\bibinfo  {journal} {Phys. Rev. D}\ }\textbf {\bibinfo {volume} {95}},\ \bibinfo {pages} {044002} (\bibinfo {year} {2017})},\ \Eprint {https://arxiv.org/abs/1510.03386} {arXiv:1510.03386 [hep-th]} \BibitemShut {NoStop}%
\bibitem [{\citenamefont {Hawking}\ \emph {et~al.}(2016)\citenamefont {Hawking}, \citenamefont {Perry},\ and\ \citenamefont {Strominger}}]{Hawking:2016msc}%
  \BibitemOpen
  \bibfield  {author} {\bibinfo {author} {\bibfnamefont {S.~W.}\ \bibnamefont {Hawking}}, \bibinfo {author} {\bibfnamefont {M.~J.}\ \bibnamefont {Perry}},\ and\ \bibinfo {author} {\bibfnamefont {A.}~\bibnamefont {Strominger}},\ }\bibfield  {title} {\bibinfo {title} {{Soft Hair on Black Holes}},\ }\href {https://doi.org/10.1103/PhysRevLett.116.231301} {\bibfield  {journal} {\bibinfo  {journal} {Phys. Rev. Lett.}\ }\textbf {\bibinfo {volume} {116}},\ \bibinfo {pages} {231301} (\bibinfo {year} {2016})},\ \Eprint {https://arxiv.org/abs/1601.00921} {arXiv:1601.00921 [hep-th]} \BibitemShut {NoStop}%
\bibitem [{\citenamefont {Hawking}\ \emph {et~al.}(2017)\citenamefont {Hawking}, \citenamefont {Perry},\ and\ \citenamefont {Strominger}}]{Hawking:2016sgy}%
  \BibitemOpen
  \bibfield  {author} {\bibinfo {author} {\bibfnamefont {S.~W.}\ \bibnamefont {Hawking}}, \bibinfo {author} {\bibfnamefont {M.~J.}\ \bibnamefont {Perry}},\ and\ \bibinfo {author} {\bibfnamefont {A.}~\bibnamefont {Strominger}},\ }\bibfield  {title} {\bibinfo {title} {{Superrotation Charge and Supertranslation Hair on Black Holes}},\ }\href {https://doi.org/10.1007/JHEP05(2017)161} {\bibfield  {journal} {\bibinfo  {journal} {JHEP}\ }\textbf {\bibinfo {volume} {05}},\ \bibinfo {pages} {161}},\ \Eprint {https://arxiv.org/abs/1611.09175} {arXiv:1611.09175 [hep-th]} \BibitemShut {NoStop}%
\bibitem [{\citenamefont {Barnich}\ and\ \citenamefont {Troessaert}(2010{\natexlab{a}})}]{Barnich:2009se}%
  \BibitemOpen
  \bibfield  {author} {\bibinfo {author} {\bibfnamefont {G.}~\bibnamefont {Barnich}}\ and\ \bibinfo {author} {\bibfnamefont {C.}~\bibnamefont {Troessaert}},\ }\bibfield  {title} {\bibinfo {title} {{Symmetries of asymptotically flat 4 dimensional spacetimes at null infinity revisited}},\ }\href {https://doi.org/10.1103/PhysRevLett.105.111103} {\bibfield  {journal} {\bibinfo  {journal} {Phys. Rev. Lett.}\ }\textbf {\bibinfo {volume} {105}},\ \bibinfo {pages} {111103} (\bibinfo {year} {2010}{\natexlab{a}})},\ \Eprint {https://arxiv.org/abs/0909.2617} {arXiv:0909.2617 [gr-qc]} \BibitemShut {NoStop}%
\bibitem [{\citenamefont {Barnich}\ and\ \citenamefont {Troessaert}(2010{\natexlab{b}})}]{Barnich:2010eb}%
  \BibitemOpen
  \bibfield  {author} {\bibinfo {author} {\bibfnamefont {G.}~\bibnamefont {Barnich}}\ and\ \bibinfo {author} {\bibfnamefont {C.}~\bibnamefont {Troessaert}},\ }\bibfield  {title} {\bibinfo {title} {{Aspects of the BMS/CFT correspondence}},\ }\href {https://doi.org/10.1007/JHEP05(2010)062} {\bibfield  {journal} {\bibinfo  {journal} {JHEP}\ }\textbf {\bibinfo {volume} {05}},\ \bibinfo {pages} {062}},\ \Eprint {https://arxiv.org/abs/1001.1541} {arXiv:1001.1541 [hep-th]} \BibitemShut {NoStop}%
\bibitem [{\citenamefont {Barnich}\ and\ \citenamefont {Troessaert}(2010{\natexlab{c}})}]{Barnich:2011ct}%
  \BibitemOpen
  \bibfield  {author} {\bibinfo {author} {\bibfnamefont {G.}~\bibnamefont {Barnich}}\ and\ \bibinfo {author} {\bibfnamefont {C.}~\bibnamefont {Troessaert}},\ }\bibfield  {title} {\bibinfo {title} {{Supertranslations call for superrotations}},\ }\bibfield  {booktitle} {\emph {\bibinfo {booktitle} {{Proceedings, Satellite Workshop on Non Commutative Field Theory and Gravity : 10th Hellenic School and Workshops on Elementary Particle Physics and Gravity (CORFU2010-NC): Corfu 2010, Greece, September 8-12, 2010}}},\ }\href {https://doi.org/10.22323/1.127.0010} {\bibfield  {journal} {\bibinfo  {journal} {PoS}\ }\textbf {\bibinfo {volume} {CNCFG2010}},\ \bibinfo {pages} {010} (\bibinfo {year} {2010}{\natexlab{c}})},\ \bibinfo {note} {[Ann. U. Craiova Phys.21,S11(2011)]},\ \Eprint {https://arxiv.org/abs/1102.4632} {arXiv:1102.4632 [gr-qc]} \BibitemShut {NoStop}%
\bibitem [{\citenamefont {Campiglia}\ and\ \citenamefont {Laddha}(2014)}]{Campiglia:2014yka}%
  \BibitemOpen
  \bibfield  {author} {\bibinfo {author} {\bibfnamefont {M.}~\bibnamefont {Campiglia}}\ and\ \bibinfo {author} {\bibfnamefont {A.}~\bibnamefont {Laddha}},\ }\bibfield  {title} {\bibinfo {title} {{Asymptotic symmetries and subleading soft graviton theorem}},\ }\href {https://doi.org/10.1103/PhysRevD.90.124028} {\bibfield  {journal} {\bibinfo  {journal} {Phys. Rev. D}\ }\textbf {\bibinfo {volume} {90}},\ \bibinfo {pages} {124028} (\bibinfo {year} {2014})},\ \Eprint {https://arxiv.org/abs/1408.2228} {arXiv:1408.2228 [hep-th]} \BibitemShut {NoStop}%
\bibitem [{\citenamefont {Campiglia}\ and\ \citenamefont {Laddha}(2015)}]{Campiglia:2015yka}%
  \BibitemOpen
  \bibfield  {author} {\bibinfo {author} {\bibfnamefont {M.}~\bibnamefont {Campiglia}}\ and\ \bibinfo {author} {\bibfnamefont {A.}~\bibnamefont {Laddha}},\ }\bibfield  {title} {\bibinfo {title} {{New symmetries for the Gravitational S-matrix}},\ }\href {https://doi.org/10.1007/JHEP04(2015)076} {\bibfield  {journal} {\bibinfo  {journal} {JHEP}\ }\textbf {\bibinfo {volume} {04}},\ \bibinfo {pages} {076}},\ \Eprint {https://arxiv.org/abs/1502.02318} {arXiv:1502.02318 [hep-th]} \BibitemShut {NoStop}%
\bibitem [{\citenamefont {Campiglia}\ and\ \citenamefont {Peraza}(2020)}]{Campiglia:2020qvc}%
  \BibitemOpen
  \bibfield  {author} {\bibinfo {author} {\bibfnamefont {M.}~\bibnamefont {Campiglia}}\ and\ \bibinfo {author} {\bibfnamefont {J.}~\bibnamefont {Peraza}},\ }\bibfield  {title} {\bibinfo {title} {{Generalized BMS charge algebra}},\ }\href {https://doi.org/10.1103/PhysRevD.101.104039} {\bibfield  {journal} {\bibinfo  {journal} {Phys. Rev. D}\ }\textbf {\bibinfo {volume} {101}},\ \bibinfo {pages} {104039} (\bibinfo {year} {2020})},\ \Eprint {https://arxiv.org/abs/2002.06691} {arXiv:2002.06691 [gr-qc]} \BibitemShut {NoStop}%
\bibitem [{\citenamefont {Freidel}\ \emph {et~al.}(2021)\citenamefont {Freidel}, \citenamefont {Oliveri}, \citenamefont {Pranzetti},\ and\ \citenamefont {Speziale}}]{Freidel:2021fxf}%
  \BibitemOpen
  \bibfield  {author} {\bibinfo {author} {\bibfnamefont {L.}~\bibnamefont {Freidel}}, \bibinfo {author} {\bibfnamefont {R.}~\bibnamefont {Oliveri}}, \bibinfo {author} {\bibfnamefont {D.}~\bibnamefont {Pranzetti}},\ and\ \bibinfo {author} {\bibfnamefont {S.}~\bibnamefont {Speziale}},\ }\bibfield  {title} {\bibinfo {title} {{The Weyl BMS group and Einstein\textquoteright{}s equations}},\ }\href {https://doi.org/10.1007/JHEP07(2021)170} {\bibfield  {journal} {\bibinfo  {journal} {JHEP}\ }\textbf {\bibinfo {volume} {07}},\ \bibinfo {pages} {170}},\ \Eprint {https://arxiv.org/abs/2104.05793} {arXiv:2104.05793 [hep-th]} \BibitemShut {NoStop}%
\bibitem [{\citenamefont {Compère}\ \emph {et~al.}(2018)\citenamefont {Compère}, \citenamefont {Fiorucci},\ and\ \citenamefont {Ruzziconi}}]{Compere:2018ylh}%
  \BibitemOpen
  \bibfield  {author} {\bibinfo {author} {\bibfnamefont {G.}~\bibnamefont {Compère}}, \bibinfo {author} {\bibfnamefont {A.}~\bibnamefont {Fiorucci}},\ and\ \bibinfo {author} {\bibfnamefont {R.}~\bibnamefont {Ruzziconi}},\ }\bibfield  {title} {\bibinfo {title} {{Superboost transitions, refraction memory and super-Lorentz charge algebra}},\ }\href {https://doi.org/10.1007/JHEP11(2018)200} {\bibfield  {journal} {\bibinfo  {journal} {JHEP}\ }\textbf {\bibinfo {volume} {11}},\ \bibinfo {pages} {200}},\ \bibinfo {note} {[Erratum: JHEP 04, 172 (2020)]},\ \Eprint {https://arxiv.org/abs/1810.00377} {arXiv:1810.00377 [hep-th]} \BibitemShut {NoStop}%
\bibitem [{\citenamefont {Hawking}(1976)}]{Hawking:1976ra}%
  \BibitemOpen
  \bibfield  {author} {\bibinfo {author} {\bibfnamefont {S.~W.}\ \bibnamefont {Hawking}},\ }\bibfield  {title} {\bibinfo {title} {{Breakdown of Predictability in Gravitational Collapse}},\ }\href {https://doi.org/10.1103/PhysRevD.14.2460} {\bibfield  {journal} {\bibinfo  {journal} {Phys. Rev. D}\ }\textbf {\bibinfo {volume} {14}},\ \bibinfo {pages} {2460} (\bibinfo {year} {1976})}\BibitemShut {NoStop}%
\bibitem [{\citenamefont {Vaidya}(1951)}]{Vaidya:1951zza}%
  \BibitemOpen
  \bibfield  {author} {\bibinfo {author} {\bibfnamefont {P.~C.}\ \bibnamefont {Vaidya}},\ }\bibfield  {title} {\bibinfo {title} {{Nonstatic Solutions of Einstein's Field Equations for Spheres of Fluids Radiating Energy}},\ }\href {https://doi.org/10.1103/PhysRev.83.10} {\bibfield  {journal} {\bibinfo  {journal} {Phys. Rev.}\ }\textbf {\bibinfo {volume} {83}},\ \bibinfo {pages} {10} (\bibinfo {year} {1951})}\BibitemShut {NoStop}%
\bibitem [{\citenamefont {Chu}\ and\ \citenamefont {Koyama}(2018)}]{Chu:2018tzu}%
  \BibitemOpen
  \bibfield  {author} {\bibinfo {author} {\bibfnamefont {C.-S.}\ \bibnamefont {Chu}}\ and\ \bibinfo {author} {\bibfnamefont {Y.}~\bibnamefont {Koyama}},\ }\bibfield  {title} {\bibinfo {title} {{Soft Hair of Dynamical Black Hole and Hawking Radiation}},\ }\href {https://doi.org/10.1007/JHEP04(2018)056} {\bibfield  {journal} {\bibinfo  {journal} {JHEP}\ }\textbf {\bibinfo {volume} {04}},\ \bibinfo {pages} {056}},\ \Eprint {https://arxiv.org/abs/1801.03658} {arXiv:1801.03658 [hep-th]} \BibitemShut {NoStop}%
\bibitem [{\citenamefont {Iofa}(2018)}]{Iofa:2017ukq}%
  \BibitemOpen
  \bibfield  {author} {\bibinfo {author} {\bibfnamefont {M.~Z.}\ \bibnamefont {Iofa}},\ }\bibfield  {title} {\bibinfo {title} {{Thermal Hawking radiation of black hole with supertranslation field}},\ }\href {https://doi.org/10.1007/JHEP01(2018)137} {\bibfield  {journal} {\bibinfo  {journal} {JHEP}\ }\textbf {\bibinfo {volume} {01}},\ \bibinfo {pages} {137}},\ \Eprint {https://arxiv.org/abs/1708.09169} {arXiv:1708.09169 [hep-th]} \BibitemShut {NoStop}%
\bibitem [{\citenamefont {Mirbabayi}\ and\ \citenamefont {Porrati}(2016)}]{Mirbabayi:2016axw}%
  \BibitemOpen
  \bibfield  {author} {\bibinfo {author} {\bibfnamefont {M.}~\bibnamefont {Mirbabayi}}\ and\ \bibinfo {author} {\bibfnamefont {M.}~\bibnamefont {Porrati}},\ }\bibfield  {title} {\bibinfo {title} {{Dressed Hard States and Black Hole Soft Hair}},\ }\href {https://doi.org/10.1103/PhysRevLett.117.211301} {\bibfield  {journal} {\bibinfo  {journal} {Phys. Rev. Lett.}\ }\textbf {\bibinfo {volume} {117}},\ \bibinfo {pages} {211301} (\bibinfo {year} {2016})},\ \Eprint {https://arxiv.org/abs/1607.03120} {arXiv:1607.03120 [hep-th]} \BibitemShut {NoStop}%
\bibitem [{\citenamefont {Bousso}\ and\ \citenamefont {Porrati}(2017{\natexlab{a}})}]{Bousso:2017dny}%
  \BibitemOpen
  \bibfield  {author} {\bibinfo {author} {\bibfnamefont {R.}~\bibnamefont {Bousso}}\ and\ \bibinfo {author} {\bibfnamefont {M.}~\bibnamefont {Porrati}},\ }\bibfield  {title} {\bibinfo {title} {{Soft Hair as a Soft Wig}},\ }\href {https://doi.org/10.1088/1361-6382/aa8be2} {\bibfield  {journal} {\bibinfo  {journal} {Class. Quant. Grav.}\ }\textbf {\bibinfo {volume} {34}},\ \bibinfo {pages} {204001} (\bibinfo {year} {2017}{\natexlab{a}})},\ \Eprint {https://arxiv.org/abs/1706.00436} {arXiv:1706.00436 [hep-th]} \BibitemShut {NoStop}%
\bibitem [{\citenamefont {Bousso}\ and\ \citenamefont {Porrati}(2017{\natexlab{b}})}]{Bousso:2017rsx}%
  \BibitemOpen
  \bibfield  {author} {\bibinfo {author} {\bibfnamefont {R.}~\bibnamefont {Bousso}}\ and\ \bibinfo {author} {\bibfnamefont {M.}~\bibnamefont {Porrati}},\ }\bibfield  {title} {\bibinfo {title} {{Observable Supertranslations}},\ }\href {https://doi.org/10.1103/PhysRevD.96.086016} {\bibfield  {journal} {\bibinfo  {journal} {Phys. Rev. D}\ }\textbf {\bibinfo {volume} {96}},\ \bibinfo {pages} {086016} (\bibinfo {year} {2017}{\natexlab{b}})},\ \Eprint {https://arxiv.org/abs/1706.09280} {arXiv:1706.09280 [hep-th]} \BibitemShut {NoStop}%
\bibitem [{\citenamefont {Marolf}(2017)}]{Marolf:2017jkr}%
  \BibitemOpen
  \bibfield  {author} {\bibinfo {author} {\bibfnamefont {D.}~\bibnamefont {Marolf}},\ }\bibfield  {title} {\bibinfo {title} {{The Black Hole information problem: past, present, and future}},\ }\href {https://doi.org/10.1088/1361-6633/aa77cc} {\bibfield  {journal} {\bibinfo  {journal} {Rept. Prog. Phys.}\ }\textbf {\bibinfo {volume} {80}},\ \bibinfo {pages} {092001} (\bibinfo {year} {2017})},\ \Eprint {https://arxiv.org/abs/1703.02143} {arXiv:1703.02143 [gr-qc]} \BibitemShut {NoStop}%
\bibitem [{\citenamefont {Comp{\`e}re}\ \emph {et~al.}(2019)\citenamefont {Comp{\`e}re}, \citenamefont {Long},\ and\ \citenamefont {Riegler}}]{Compere:2019rof}%
  \BibitemOpen
  \bibfield  {author} {\bibinfo {author} {\bibfnamefont {G.}~\bibnamefont {Comp{\`e}re}}, \bibinfo {author} {\bibfnamefont {J.}~\bibnamefont {Long}},\ and\ \bibinfo {author} {\bibfnamefont {M.}~\bibnamefont {Riegler}},\ }\bibfield  {title} {\bibinfo {title} {{Invariance of Unruh and Hawking radiation under matter-induced supertranslations}},\ }\href {https://doi.org/10.1007/JHEP05(2019)053} {\bibfield  {journal} {\bibinfo  {journal} {JHEP}\ }\textbf {\bibinfo {volume} {05}},\ \bibinfo {pages} {053}},\ \Eprint {https://arxiv.org/abs/1903.01812} {arXiv:1903.01812 [hep-th]} \BibitemShut {NoStop}%
\bibitem [{\citenamefont {Akiyama}\ \emph {et~al.}(2019{\natexlab{a}})\citenamefont {Akiyama} \emph {et~al.}}]{EventHorizonTelescope:2019dse}%
  \BibitemOpen
  \bibfield  {author} {\bibinfo {author} {\bibfnamefont {K.}~\bibnamefont {Akiyama}} \emph {et~al.} (\bibinfo {collaboration} {Event Horizon Telescope}),\ }\bibfield  {title} {\bibinfo {title} {{First M87 Event Horizon Telescope Results. I. The Shadow of the Supermassive Black Hole}},\ }\href {https://doi.org/10.3847/2041-8213/ab0ec7} {\bibfield  {journal} {\bibinfo  {journal} {Astrophys. J. Lett.}\ }\textbf {\bibinfo {volume} {875}},\ \bibinfo {pages} {L1} (\bibinfo {year} {2019}{\natexlab{a}})},\ \Eprint {https://arxiv.org/abs/1906.11238} {arXiv:1906.11238 [astro-ph.GA]} \BibitemShut {NoStop}%
\bibitem [{\citenamefont {Akiyama}\ \emph {et~al.}(2019{\natexlab{b}})\citenamefont {Akiyama} \emph {et~al.}}]{EventHorizonTelescope:2019uob}%
  \BibitemOpen
  \bibfield  {author} {\bibinfo {author} {\bibfnamefont {K.}~\bibnamefont {Akiyama}} \emph {et~al.} (\bibinfo {collaboration} {Event Horizon Telescope}),\ }\bibfield  {title} {\bibinfo {title} {{First M87 Event Horizon Telescope Results. II. Array and Instrumentation}},\ }\href {https://doi.org/10.3847/2041-8213/ab0c96} {\bibfield  {journal} {\bibinfo  {journal} {Astrophys. J. Lett.}\ }\textbf {\bibinfo {volume} {875}},\ \bibinfo {pages} {L2} (\bibinfo {year} {2019}{\natexlab{b}})},\ \Eprint {https://arxiv.org/abs/1906.11239} {arXiv:1906.11239 [astro-ph.IM]} \BibitemShut {NoStop}%
\bibitem [{\citenamefont {Akiyama}\ \emph {et~al.}(2019{\natexlab{c}})\citenamefont {Akiyama} \emph {et~al.}}]{EventHorizonTelescope:2019jan}%
  \BibitemOpen
  \bibfield  {author} {\bibinfo {author} {\bibfnamefont {K.}~\bibnamefont {Akiyama}} \emph {et~al.} (\bibinfo {collaboration} {Event Horizon Telescope}),\ }\bibfield  {title} {\bibinfo {title} {{First M87 Event Horizon Telescope Results. III. Data Processing and Calibration}},\ }\href {https://doi.org/10.3847/2041-8213/ab0c57} {\bibfield  {journal} {\bibinfo  {journal} {Astrophys. J. Lett.}\ }\textbf {\bibinfo {volume} {875}},\ \bibinfo {pages} {L3} (\bibinfo {year} {2019}{\natexlab{c}})},\ \Eprint {https://arxiv.org/abs/1906.11240} {arXiv:1906.11240 [astro-ph.GA]} \BibitemShut {NoStop}%
\bibitem [{\citenamefont {Akiyama}\ \emph {et~al.}(2019{\natexlab{d}})\citenamefont {Akiyama} \emph {et~al.}}]{EventHorizonTelescope:2019ths}%
  \BibitemOpen
  \bibfield  {author} {\bibinfo {author} {\bibfnamefont {K.}~\bibnamefont {Akiyama}} \emph {et~al.} (\bibinfo {collaboration} {Event Horizon Telescope}),\ }\bibfield  {title} {\bibinfo {title} {{First M87 Event Horizon Telescope Results. IV. Imaging the Central Supermassive Black Hole}},\ }\href {https://doi.org/10.3847/2041-8213/ab0e85} {\bibfield  {journal} {\bibinfo  {journal} {Astrophys. J. Lett.}\ }\textbf {\bibinfo {volume} {875}},\ \bibinfo {pages} {L4} (\bibinfo {year} {2019}{\natexlab{d}})},\ \Eprint {https://arxiv.org/abs/1906.11241} {arXiv:1906.11241 [astro-ph.GA]} \BibitemShut {NoStop}%
\bibitem [{\citenamefont {Akiyama}\ \emph {et~al.}(2019{\natexlab{e}})\citenamefont {Akiyama} \emph {et~al.}}]{EventHorizonTelescope:2019pgp}%
  \BibitemOpen
  \bibfield  {author} {\bibinfo {author} {\bibfnamefont {K.}~\bibnamefont {Akiyama}} \emph {et~al.} (\bibinfo {collaboration} {Event Horizon Telescope}),\ }\bibfield  {title} {\bibinfo {title} {{First M87 Event Horizon Telescope Results. V. Physical Origin of the Asymmetric Ring}},\ }\href {https://doi.org/10.3847/2041-8213/ab0f43} {\bibfield  {journal} {\bibinfo  {journal} {Astrophys. J. Lett.}\ }\textbf {\bibinfo {volume} {875}},\ \bibinfo {pages} {L5} (\bibinfo {year} {2019}{\natexlab{e}})},\ \Eprint {https://arxiv.org/abs/1906.11242} {arXiv:1906.11242 [astro-ph.GA]} \BibitemShut {NoStop}%
\bibitem [{\citenamefont {Akiyama}\ \emph {et~al.}(2019{\natexlab{f}})\citenamefont {Akiyama} \emph {et~al.}}]{EventHorizonTelescope:2019ggy}%
  \BibitemOpen
  \bibfield  {author} {\bibinfo {author} {\bibfnamefont {K.}~\bibnamefont {Akiyama}} \emph {et~al.} (\bibinfo {collaboration} {Event Horizon Telescope}),\ }\bibfield  {title} {\bibinfo {title} {{First M87 Event Horizon Telescope Results. VI. The Shadow and Mass of the Central Black Hole}},\ }\href {https://doi.org/10.3847/2041-8213/ab1141} {\bibfield  {journal} {\bibinfo  {journal} {Astrophys. J. Lett.}\ }\textbf {\bibinfo {volume} {875}},\ \bibinfo {pages} {L6} (\bibinfo {year} {2019}{\natexlab{f}})},\ \Eprint {https://arxiv.org/abs/1906.11243} {arXiv:1906.11243 [astro-ph.GA]} \BibitemShut {NoStop}%
\bibitem [{\citenamefont {Akiyama}\ \emph {et~al.}(2022{\natexlab{a}})\citenamefont {Akiyama} \emph {et~al.}}]{EventHorizonTelescope:2022wkp}%
  \BibitemOpen
  \bibfield  {author} {\bibinfo {author} {\bibfnamefont {K.}~\bibnamefont {Akiyama}} \emph {et~al.} (\bibinfo {collaboration} {Event Horizon Telescope}),\ }\bibfield  {title} {\bibinfo {title} {{First Sagittarius A* Event Horizon Telescope Results. I. The Shadow of the Supermassive Black Hole in the Center of the Milky Way}},\ }\href {https://doi.org/10.3847/2041-8213/ac6674} {\bibfield  {journal} {\bibinfo  {journal} {Astrophys. J. Lett.}\ }\textbf {\bibinfo {volume} {930}},\ \bibinfo {pages} {L12} (\bibinfo {year} {2022}{\natexlab{a}})},\ \Eprint {https://arxiv.org/abs/2311.08680} {arXiv:2311.08680 [astro-ph.HE]} \BibitemShut {NoStop}%
\bibitem [{\citenamefont {Akiyama}\ \emph {et~al.}(2022{\natexlab{b}})\citenamefont {Akiyama} \emph {et~al.}}]{EventHorizonTelescope:2022apq}%
  \BibitemOpen
  \bibfield  {author} {\bibinfo {author} {\bibfnamefont {K.}~\bibnamefont {Akiyama}} \emph {et~al.} (\bibinfo {collaboration} {Event Horizon Telescope}),\ }\bibfield  {title} {\bibinfo {title} {{First Sagittarius A* Event Horizon Telescope Results. II. EHT and Multiwavelength Observations, Data Processing, and Calibration}},\ }\href {https://doi.org/10.3847/2041-8213/ac6675} {\bibfield  {journal} {\bibinfo  {journal} {Astrophys. J. Lett.}\ }\textbf {\bibinfo {volume} {930}},\ \bibinfo {pages} {L13} (\bibinfo {year} {2022}{\natexlab{b}})},\ \Eprint {https://arxiv.org/abs/2311.08679} {arXiv:2311.08679 [astro-ph.HE]} \BibitemShut {NoStop}%
\bibitem [{\citenamefont {Akiyama}\ \emph {et~al.}(2022{\natexlab{c}})\citenamefont {Akiyama} \emph {et~al.}}]{EventHorizonTelescope:2022wok}%
  \BibitemOpen
  \bibfield  {author} {\bibinfo {author} {\bibfnamefont {K.}~\bibnamefont {Akiyama}} \emph {et~al.} (\bibinfo {collaboration} {Event Horizon Telescope}),\ }\bibfield  {title} {\bibinfo {title} {{First Sagittarius A* Event Horizon Telescope Results. III. Imaging of the Galactic Center Supermassive Black Hole}},\ }\href {https://doi.org/10.3847/2041-8213/ac6429} {\bibfield  {journal} {\bibinfo  {journal} {Astrophys. J. Lett.}\ }\textbf {\bibinfo {volume} {930}},\ \bibinfo {pages} {L14} (\bibinfo {year} {2022}{\natexlab{c}})},\ \Eprint {https://arxiv.org/abs/2311.09479} {arXiv:2311.09479 [astro-ph.HE]} \BibitemShut {NoStop}%
\bibitem [{\citenamefont {Akiyama}\ \emph {et~al.}(2022{\natexlab{d}})\citenamefont {Akiyama} \emph {et~al.}}]{EventHorizonTelescope:2022exc}%
  \BibitemOpen
  \bibfield  {author} {\bibinfo {author} {\bibfnamefont {K.}~\bibnamefont {Akiyama}} \emph {et~al.} (\bibinfo {collaboration} {Event Horizon Telescope}),\ }\bibfield  {title} {\bibinfo {title} {{First Sagittarius A* Event Horizon Telescope Results. IV. Variability, Morphology, and Black Hole Mass}},\ }\href {https://doi.org/10.3847/2041-8213/ac6736} {\bibfield  {journal} {\bibinfo  {journal} {Astrophys. J. Lett.}\ }\textbf {\bibinfo {volume} {930}},\ \bibinfo {pages} {L15} (\bibinfo {year} {2022}{\natexlab{d}})},\ \Eprint {https://arxiv.org/abs/2311.08697} {arXiv:2311.08697 [astro-ph.HE]} \BibitemShut {NoStop}%
\bibitem [{\citenamefont {Akiyama}\ \emph {et~al.}(2022{\natexlab{e}})\citenamefont {Akiyama} \emph {et~al.}}]{EventHorizonTelescope:2022urf}%
  \BibitemOpen
  \bibfield  {author} {\bibinfo {author} {\bibfnamefont {K.}~\bibnamefont {Akiyama}} \emph {et~al.} (\bibinfo {collaboration} {Event Horizon Telescope}),\ }\bibfield  {title} {\bibinfo {title} {{First Sagittarius A* Event Horizon Telescope Results. V. Testing Astrophysical Models of the Galactic Center Black Hole}},\ }\href {https://doi.org/10.3847/2041-8213/ac6672} {\bibfield  {journal} {\bibinfo  {journal} {Astrophys. J. Lett.}\ }\textbf {\bibinfo {volume} {930}},\ \bibinfo {pages} {L16} (\bibinfo {year} {2022}{\natexlab{e}})},\ \Eprint {https://arxiv.org/abs/2311.09478} {arXiv:2311.09478 [astro-ph.HE]} \BibitemShut {NoStop}%
\bibitem [{\citenamefont {Akiyama}\ \emph {et~al.}(2022{\natexlab{f}})\citenamefont {Akiyama} \emph {et~al.}}]{EventHorizonTelescope:2022xqj}%
  \BibitemOpen
  \bibfield  {author} {\bibinfo {author} {\bibfnamefont {K.}~\bibnamefont {Akiyama}} \emph {et~al.} (\bibinfo {collaboration} {Event Horizon Telescope}),\ }\bibfield  {title} {\bibinfo {title} {{First Sagittarius A* Event Horizon Telescope Results. VI. Testing the Black Hole Metric}},\ }\href {https://doi.org/10.3847/2041-8213/ac6756} {\bibfield  {journal} {\bibinfo  {journal} {Astrophys. J. Lett.}\ }\textbf {\bibinfo {volume} {930}},\ \bibinfo {pages} {L17} (\bibinfo {year} {2022}{\natexlab{f}})},\ \Eprint {https://arxiv.org/abs/2311.09484} {arXiv:2311.09484 [astro-ph.HE]} \BibitemShut {NoStop}%
\bibitem [{\citenamefont {Bambi}\ \emph {et~al.}(2019)\citenamefont {Bambi}, \citenamefont {Freese}, \citenamefont {Vagnozzi},\ and\ \citenamefont {Visinelli}}]{Bambi:2019tjh}%
  \BibitemOpen
  \bibfield  {author} {\bibinfo {author} {\bibfnamefont {C.}~\bibnamefont {Bambi}}, \bibinfo {author} {\bibfnamefont {K.}~\bibnamefont {Freese}}, \bibinfo {author} {\bibfnamefont {S.}~\bibnamefont {Vagnozzi}},\ and\ \bibinfo {author} {\bibfnamefont {L.}~\bibnamefont {Visinelli}},\ }\bibfield  {title} {\bibinfo {title} {{Testing the rotational nature of the supermassive object M87* from the circularity and size of its first image}},\ }\href {https://doi.org/10.1103/PhysRevD.100.044057} {\bibfield  {journal} {\bibinfo  {journal} {Phys. Rev. D}\ }\textbf {\bibinfo {volume} {100}},\ \bibinfo {pages} {044057} (\bibinfo {year} {2019})},\ \Eprint {https://arxiv.org/abs/1904.12983} {arXiv:1904.12983 [gr-qc]} \BibitemShut {NoStop}%
\bibitem [{\citenamefont {Vagnozzi}\ \emph {et~al.}(2023)\citenamefont {Vagnozzi} \emph {et~al.}}]{Vagnozzi:2022moj}%
  \BibitemOpen
  \bibfield  {author} {\bibinfo {author} {\bibfnamefont {S.}~\bibnamefont {Vagnozzi}} \emph {et~al.},\ }\bibfield  {title} {\bibinfo {title} {{Horizon-scale tests of gravity theories and fundamental physics from the Event Horizon Telescope image of Sagittarius A}},\ }\href {https://doi.org/10.1088/1361-6382/acd97b} {\bibfield  {journal} {\bibinfo  {journal} {Class. Quant. Grav.}\ }\textbf {\bibinfo {volume} {40}},\ \bibinfo {pages} {165007} (\bibinfo {year} {2023})},\ \Eprint {https://arxiv.org/abs/2205.07787} {arXiv:2205.07787 [gr-qc]} \BibitemShut {NoStop}%
\bibitem [{\citenamefont {Khodadi}\ \emph {et~al.}(2024)\citenamefont {Khodadi}, \citenamefont {Vagnozzi},\ and\ \citenamefont {Firouzjaee}}]{Khodadi:2024ubi}%
  \BibitemOpen
  \bibfield  {author} {\bibinfo {author} {\bibfnamefont {M.}~\bibnamefont {Khodadi}}, \bibinfo {author} {\bibfnamefont {S.}~\bibnamefont {Vagnozzi}},\ and\ \bibinfo {author} {\bibfnamefont {J.~T.}\ \bibnamefont {Firouzjaee}},\ }\bibfield  {title} {\bibinfo {title} {{Event Horizon Telescope observations exclude compact objects in baseline mimetic gravity}},\ }\href {https://doi.org/10.1038/s41598-024-78264-y} {\bibfield  {journal} {\bibinfo  {journal} {Sci. Rep.}\ }\textbf {\bibinfo {volume} {14}},\ \bibinfo {pages} {26932} (\bibinfo {year} {2024})},\ \Eprint {https://arxiv.org/abs/2408.03241} {arXiv:2408.03241 [gr-qc]} \BibitemShut {NoStop}%
\bibitem [{\citenamefont {Lin}\ \emph {et~al.}(2022)\citenamefont {Lin}, \citenamefont {Patel},\ and\ \citenamefont {Pu}}]{Lin:2022ksb}%
  \BibitemOpen
  \bibfield  {author} {\bibinfo {author} {\bibfnamefont {F.-L.}\ \bibnamefont {Lin}}, \bibinfo {author} {\bibfnamefont {A.}~\bibnamefont {Patel}},\ and\ \bibinfo {author} {\bibfnamefont {H.-Y.}\ \bibnamefont {Pu}},\ }\bibfield  {title} {\bibinfo {title} {{Black hole shadow with soft hairs}},\ }\href {https://doi.org/10.1007/JHEP09(2022)117} {\bibfield  {journal} {\bibinfo  {journal} {JHEP}\ }\textbf {\bibinfo {volume} {09}},\ \bibinfo {pages} {117}},\ \Eprint {https://arxiv.org/abs/2202.13559} {arXiv:2202.13559 [gr-qc]} \BibitemShut {NoStop}%
\bibitem [{\citenamefont {Zhu}\ \emph {et~al.}(2023)\citenamefont {Zhu}, \citenamefont {Han},\ and\ \citenamefont {Huang}}]{Zhu:2022shb}%
  \BibitemOpen
  \bibfield  {author} {\bibinfo {author} {\bibfnamefont {Q.-H.}\ \bibnamefont {Zhu}}, \bibinfo {author} {\bibfnamefont {Y.-X.}\ \bibnamefont {Han}},\ and\ \bibinfo {author} {\bibfnamefont {Q.-G.}\ \bibnamefont {Huang}},\ }\bibfield  {title} {\bibinfo {title} {{The shadow of supertranslated black hole}},\ }\href {https://doi.org/10.1140/epjc/s10052-023-11232-4} {\bibfield  {journal} {\bibinfo  {journal} {Eur. Phys. J. C}\ }\textbf {\bibinfo {volume} {83}},\ \bibinfo {pages} {88} (\bibinfo {year} {2023})},\ \Eprint {https://arxiv.org/abs/2205.14554} {arXiv:2205.14554 [gr-qc]} \BibitemShut {NoStop}%
\bibitem [{\citenamefont {Sarkar}\ \emph {et~al.}(2022)\citenamefont {Sarkar}, \citenamefont {Kumar},\ and\ \citenamefont {Bhattacharjee}}]{Sarkar:2021djs}%
  \BibitemOpen
  \bibfield  {author} {\bibinfo {author} {\bibfnamefont {S.}~\bibnamefont {Sarkar}}, \bibinfo {author} {\bibfnamefont {S.}~\bibnamefont {Kumar}},\ and\ \bibinfo {author} {\bibfnamefont {S.}~\bibnamefont {Bhattacharjee}},\ }\bibfield  {title} {\bibinfo {title} {{Can we detect a supertranslated black hole?}},\ }\href {https://doi.org/10.1103/PhysRevD.105.084001} {\bibfield  {journal} {\bibinfo  {journal} {Phys. Rev. D}\ }\textbf {\bibinfo {volume} {105}},\ \bibinfo {pages} {084001} (\bibinfo {year} {2022})},\ \Eprint {https://arxiv.org/abs/2110.03547} {arXiv:2110.03547 [gr-qc]} \BibitemShut {NoStop}%
\bibitem [{\citenamefont {Lin}\ \emph {et~al.}(2025)\citenamefont {Lin}, \citenamefont {Patel},\ and\ \citenamefont {Payne}}]{Lin:2024xzo}%
  \BibitemOpen
  \bibfield  {author} {\bibinfo {author} {\bibfnamefont {F.-L.}\ \bibnamefont {Lin}}, \bibinfo {author} {\bibfnamefont {A.}~\bibnamefont {Patel}},\ and\ \bibinfo {author} {\bibfnamefont {J.}~\bibnamefont {Payne}},\ }\bibfield  {title} {\bibinfo {title} {{Conic sections on the sky: Shadows of linearly superrotated black holes}},\ }\href {https://doi.org/10.1103/PhysRevD.111.024054} {\bibfield  {journal} {\bibinfo  {journal} {Phys. Rev. D}\ }\textbf {\bibinfo {volume} {111}},\ \bibinfo {pages} {024054} (\bibinfo {year} {2025})},\ \Eprint {https://arxiv.org/abs/2405.20181} {arXiv:2405.20181 [hep-th]} \BibitemShut {NoStop}%
\bibitem [{\citenamefont {Donnay}\ \emph {et~al.}(2016)\citenamefont {Donnay}, \citenamefont {Giribet}, \citenamefont {Gonzalez},\ and\ \citenamefont {Pino}}]{Donnay:2015abr}%
  \BibitemOpen
  \bibfield  {author} {\bibinfo {author} {\bibfnamefont {L.}~\bibnamefont {Donnay}}, \bibinfo {author} {\bibfnamefont {G.}~\bibnamefont {Giribet}}, \bibinfo {author} {\bibfnamefont {H.~A.}\ \bibnamefont {Gonzalez}},\ and\ \bibinfo {author} {\bibfnamefont {M.}~\bibnamefont {Pino}},\ }\bibfield  {title} {\bibinfo {title} {{Supertranslations and Superrotations at the Black Hole Horizon}},\ }\href {https://doi.org/10.1103/PhysRevLett.116.091101} {\bibfield  {journal} {\bibinfo  {journal} {Phys. Rev. Lett.}\ }\textbf {\bibinfo {volume} {116}},\ \bibinfo {pages} {091101} (\bibinfo {year} {2016})},\ \Eprint {https://arxiv.org/abs/1511.08687} {arXiv:1511.08687 [hep-th]} \BibitemShut {NoStop}%
\bibitem [{\citenamefont {Hou}\ \emph {et~al.}(2025{\natexlab{a}})\citenamefont {Hou}, \citenamefont {Lin},\ and\ \citenamefont {Zhu}}]{Hou:2024odb}%
  \BibitemOpen
  \bibfield  {author} {\bibinfo {author} {\bibfnamefont {S.}~\bibnamefont {Hou}}, \bibinfo {author} {\bibfnamefont {K.}~\bibnamefont {Lin}},\ and\ \bibinfo {author} {\bibfnamefont {Z.-H.}\ \bibnamefont {Zhu}},\ }\bibfield  {title} {\bibinfo {title} {{Finitely supertranslated Schwarzschild black hole and its perturbations}},\ }\href {https://doi.org/10.1103/PhysRevD.111.044029} {\bibfield  {journal} {\bibinfo  {journal} {Phys. Rev. D}\ }\textbf {\bibinfo {volume} {111}},\ \bibinfo {pages} {044029} (\bibinfo {year} {2025}{\natexlab{a}})},\ \Eprint {https://arxiv.org/abs/2410.12310} {arXiv:2410.12310 [gr-qc]} \BibitemShut {NoStop}%
\bibitem [{\citenamefont {Wald}(1984)}]{Wald:1984rg}%
  \BibitemOpen
  \bibfield  {author} {\bibinfo {author} {\bibfnamefont {R.~M.}\ \bibnamefont {Wald}},\ }\href {https://doi.org/10.7208/chicago/9780226870373.001.0001} {\emph {\bibinfo {title} {{General Relativity}}}}\ (\bibinfo  {publisher} {University of Chicago Press},\ \bibinfo {address} {Chicago, IL},\ \bibinfo {year} {1984})\BibitemShut {NoStop}%
\bibitem [{\citenamefont {Newman}\ and\ \citenamefont {Penrose}(1962)}]{Newman:1961qr}%
  \BibitemOpen
  \bibfield  {author} {\bibinfo {author} {\bibfnamefont {E.}~\bibnamefont {Newman}}\ and\ \bibinfo {author} {\bibfnamefont {R.}~\bibnamefont {Penrose}},\ }\bibfield  {title} {\bibinfo {title} {An approach to gravitational radiation by a method of spin coefficients},\ }\href {https://doi.org/10.1063/1.1724257} {\bibfield  {journal} {\bibinfo  {journal} {J. Math. Phys.}\ }\textbf {\bibinfo {volume} {3}},\ \bibinfo {pages} {566} (\bibinfo {year} {1962})},\ \bibinfo {note} {[Errata: J.Math.Phys.4,no.7,998(1963)]}\BibitemShut {NoStop}%
\bibitem [{\citenamefont {Newman}\ and\ \citenamefont {Unti}(1962)}]{Newman:1962cia}%
  \BibitemOpen
  \bibfield  {author} {\bibinfo {author} {\bibfnamefont {E.~T.}\ \bibnamefont {Newman}}\ and\ \bibinfo {author} {\bibfnamefont {T.~W.~J.}\ \bibnamefont {Unti}},\ }\bibfield  {title} {\bibinfo {title} {{Behavior of Asymptotically Flat Empty Spaces}},\ }\href {https://doi.org/10.1063/1.1724303} {\bibfield  {journal} {\bibinfo  {journal} {J. Math. Phys.}\ }\textbf {\bibinfo {volume} {3}},\ \bibinfo {pages} {891} (\bibinfo {year} {1962})}\BibitemShut {NoStop}%
\bibitem [{\citenamefont {Newman}\ and\ \citenamefont {Penrose}(1968)}]{Newman:1968uj}%
  \BibitemOpen
  \bibfield  {author} {\bibinfo {author} {\bibfnamefont {E.~T.}\ \bibnamefont {Newman}}\ and\ \bibinfo {author} {\bibfnamefont {R.}~\bibnamefont {Penrose}},\ }\bibfield  {title} {\bibinfo {title} {{New conservation laws for zero rest-mass fields in asymptotically flat space-time}},\ }\href {https://doi.org/10.1098/rspa.1968.0112} {\bibfield  {journal} {\bibinfo  {journal} {Proc. Roy. Soc. Lond. A}\ }\textbf {\bibinfo {volume} {305}},\ \bibinfo {pages} {175} (\bibinfo {year} {1968})}\BibitemShut {NoStop}%
\bibitem [{\citenamefont {Adamo}\ \emph {et~al.}(2009)\citenamefont {Adamo}, \citenamefont {Kozameh},\ and\ \citenamefont {Newman}}]{Adamo:2009vu}%
  \BibitemOpen
  \bibfield  {author} {\bibinfo {author} {\bibfnamefont {T.~M.}\ \bibnamefont {Adamo}}, \bibinfo {author} {\bibfnamefont {C.~N.}\ \bibnamefont {Kozameh}},\ and\ \bibinfo {author} {\bibfnamefont {E.~T.}\ \bibnamefont {Newman}},\ }\bibfield  {title} {\bibinfo {title} {{Null Geodesic Congruences, Asymptotically Flat Space-Times and Their Physical Interpretation}},\ }\href {https://doi.org/10.12942/lrr-2009-6} {\bibfield  {journal} {\bibinfo  {journal} {Living Rev. Rel.}\ }\textbf {\bibinfo {volume} {12}},\ \bibinfo {pages} {6} (\bibinfo {year} {2009})},\ \bibinfo {note} {[Living Rev. Rel.15,1(2012)]},\ \Eprint {https://arxiv.org/abs/0906.2155} {arXiv:0906.2155 [gr-qc]} \BibitemShut {NoStop}%
\bibitem [{\citenamefont {Freidel}\ and\ \citenamefont {Pranzetti}(2022)}]{Freidel:2021qpz}%
  \BibitemOpen
  \bibfield  {author} {\bibinfo {author} {\bibfnamefont {L.}~\bibnamefont {Freidel}}\ and\ \bibinfo {author} {\bibfnamefont {D.}~\bibnamefont {Pranzetti}},\ }\bibfield  {title} {\bibinfo {title} {{Gravity from symmetry: duality and impulsive waves}},\ }\href {https://doi.org/10.1007/JHEP04(2022)125} {\bibfield  {journal} {\bibinfo  {journal} {JHEP}\ }\textbf {\bibinfo {volume} {04}},\ \bibinfo {pages} {125}},\ \Eprint {https://arxiv.org/abs/2109.06342} {arXiv:2109.06342 [hep-th]} \BibitemShut {NoStop}%
\bibitem [{\citenamefont {Campoleoni}\ \emph {et~al.}(2023)\citenamefont {Campoleoni}, \citenamefont {Delfante}, \citenamefont {Pekar}, \citenamefont {Petropoulos}, \citenamefont {Rivera-Betancour},\ and\ \citenamefont {Vilatte}}]{Campoleoni:2023fug}%
  \BibitemOpen
  \bibfield  {author} {\bibinfo {author} {\bibfnamefont {A.}~\bibnamefont {Campoleoni}}, \bibinfo {author} {\bibfnamefont {A.}~\bibnamefont {Delfante}}, \bibinfo {author} {\bibfnamefont {S.}~\bibnamefont {Pekar}}, \bibinfo {author} {\bibfnamefont {P.~M.}\ \bibnamefont {Petropoulos}}, \bibinfo {author} {\bibfnamefont {D.}~\bibnamefont {Rivera-Betancour}},\ and\ \bibinfo {author} {\bibfnamefont {M.}~\bibnamefont {Vilatte}},\ }\bibfield  {title} {\bibinfo {title} {{Flat from anti de Sitter}},\ }\href {https://doi.org/10.1007/JHEP12(2023)078} {\bibfield  {journal} {\bibinfo  {journal} {JHEP}\ }\textbf {\bibinfo {volume} {12}},\ \bibinfo {pages} {078}},\ \Eprint {https://arxiv.org/abs/2309.15182} {arXiv:2309.15182 [hep-th]} \BibitemShut {NoStop}%
\bibitem [{\citenamefont {Schwarzschild}(1916)}]{Schwarzschild:1916uq}%
  \BibitemOpen
  \bibfield  {author} {\bibinfo {author} {\bibfnamefont {K.}~\bibnamefont {Schwarzschild}},\ }\bibfield  {title} {\bibinfo {title} {{On the gravitational field of a mass point according to Einstein's theory}},\ }\href@noop {} {\bibfield  {journal} {\bibinfo  {journal} {Sitzungsber. Preuss. Akad. Wiss. Berlin (Math. Phys. )}\ }\textbf {\bibinfo {volume} {1916}},\ \bibinfo {pages} {189} (\bibinfo {year} {1916})},\ \Eprint {https://arxiv.org/abs/physics/9905030} {arXiv:physics/9905030} \BibitemShut {NoStop}%
\bibitem [{\citenamefont {Kerr}(1963)}]{Kerr:1963ud}%
  \BibitemOpen
  \bibfield  {author} {\bibinfo {author} {\bibfnamefont {R.~P.}\ \bibnamefont {Kerr}},\ }\bibfield  {title} {\bibinfo {title} {{Gravitational field of a spinning mass as an example of algebraically special metrics}},\ }\href {https://doi.org/10.1103/PhysRevLett.11.237} {\bibfield  {journal} {\bibinfo  {journal} {Phys. Rev. Lett.}\ }\textbf {\bibinfo {volume} {11}},\ \bibinfo {pages} {237} (\bibinfo {year} {1963})}\BibitemShut {NoStop}%
\bibitem [{\citenamefont {Comp\`ere}\ and\ \citenamefont {Long}(2016)}]{Compere:2016hzt}%
  \BibitemOpen
  \bibfield  {author} {\bibinfo {author} {\bibfnamefont {G.}~\bibnamefont {Comp\`ere}}\ and\ \bibinfo {author} {\bibfnamefont {J.}~\bibnamefont {Long}},\ }\bibfield  {title} {\bibinfo {title} {{Classical static final state of collapse with supertranslation memory}},\ }\href {https://doi.org/10.1088/0264-9381/33/19/195001} {\bibfield  {journal} {\bibinfo  {journal} {Class. Quant. Grav.}\ }\textbf {\bibinfo {volume} {33}},\ \bibinfo {pages} {195001} (\bibinfo {year} {2016})},\ \Eprint {https://arxiv.org/abs/1602.05197} {arXiv:1602.05197 [gr-qc]} \BibitemShut {NoStop}%
\bibitem [{\citenamefont {Comp{\`e}re}\ \emph {et~al.}(2020)\citenamefont {Comp{\`e}re}, \citenamefont {Fiorucci},\ and\ \citenamefont {Ruzziconi}}]{Compere:2020lrt}%
  \BibitemOpen
  \bibfield  {author} {\bibinfo {author} {\bibfnamefont {G.}~\bibnamefont {Comp{\`e}re}}, \bibinfo {author} {\bibfnamefont {A.}~\bibnamefont {Fiorucci}},\ and\ \bibinfo {author} {\bibfnamefont {R.}~\bibnamefont {Ruzziconi}},\ }\bibfield  {title} {\bibinfo {title} {{The $\Lambda$-BMS$_4$ charge algebra}},\ }\href {https://doi.org/10.1007/JHEP10(2020)205} {\bibfield  {journal} {\bibinfo  {journal} {JHEP}\ }\textbf {\bibinfo {volume} {10}},\ \bibinfo {pages} {205}},\ \Eprint {https://arxiv.org/abs/2004.10769} {arXiv:2004.10769 [hep-th]} \BibitemShut {NoStop}%
\bibitem [{\citenamefont {Chru\'sciel}\ \emph {et~al.}(1993)\citenamefont {Chru\'sciel}, \citenamefont {MacCallum},\ and\ \citenamefont {Singleton}}]{Chrusciel:1993hx}%
  \BibitemOpen
  \bibfield  {author} {\bibinfo {author} {\bibfnamefont {P.~T.}\ \bibnamefont {Chru\'sciel}}, \bibinfo {author} {\bibfnamefont {M.~A.~H.}\ \bibnamefont {MacCallum}},\ and\ \bibinfo {author} {\bibfnamefont {D.~B.}\ \bibnamefont {Singleton}},\ }\bibfield  {title} {\bibinfo {title} {{Gravitational waves in general relativity XIV. Bondi expansions and the polyhomogeneity of $\mathscr I$}},\ }\href {https://doi.org/10.1098/rsta.1995.0004} {\bibfield  {journal} {\bibinfo  {journal} {Phil. Trans. Royal Soc. of London}\ }\textbf {\bibinfo {volume} {A350}},\ \bibinfo {pages} {113} (\bibinfo {year} {1993})},\ \Eprint {https://arxiv.org/abs/gr-qc/9305021} {arXiv:gr-qc/9305021} \BibitemShut {NoStop}%
\bibitem [{\citenamefont {Flanagan}\ and\ \citenamefont {Nichols}(2024)}]{Flanagan:2023jio}%
  \BibitemOpen
  \bibfield  {author} {\bibinfo {author} {\bibfnamefont {E.~E.}\ \bibnamefont {Flanagan}}\ and\ \bibinfo {author} {\bibfnamefont {D.~A.}\ \bibnamefont {Nichols}},\ }\bibfield  {title} {\bibinfo {title} {{Fully nonlinear transformations of the Weyl-Bondi-Metzner-Sachs asymptotic symmetry group}},\ }\href {https://doi.org/10.1007/JHEP03(2024)120} {\bibfield  {journal} {\bibinfo  {journal} {JHEP}\ }\textbf {\bibinfo {volume} {03}},\ \bibinfo {pages} {120}},\ \Eprint {https://arxiv.org/abs/2311.03130} {arXiv:2311.03130 [gr-qc]} \BibitemShut {NoStop}%
\bibitem [{Note1()}]{Note1}%
  \BibitemOpen
  \bibinfo {note} {For other ways of parameterizing the vacuum, please refer to Refs.~\cite {Compere:2018ylh,Freidel:2021fxf,Flanagan:2023jio}.}\BibitemShut {Stop}%
\bibitem [{Note2()}]{Note2}%
  \BibitemOpen
  \bibinfo {note} {For the complete specification of a nontrivial flat vacuum, please refer to Refs.~\cite {Compere:2016jwb,Compere:2018ylh}.}\BibitemShut {Stop}%
\bibitem [{Note3()}]{Note3}%
  \BibitemOpen
  \bibinfo {note} {Since one starts with a particular vacuum~\protect \eqref {eq-def-fvac}, the Poincar\'e transformation can be well defined.}\BibitemShut {Stop}%
\bibitem [{\citenamefont {Grenzebach}\ \emph {et~al.}(2014)\citenamefont {Grenzebach}, \citenamefont {Perlick},\ and\ \citenamefont {L{\"a}mmerzahl}}]{Grenzebach:2014fha}%
  \BibitemOpen
  \bibfield  {author} {\bibinfo {author} {\bibfnamefont {A.}~\bibnamefont {Grenzebach}}, \bibinfo {author} {\bibfnamefont {V.}~\bibnamefont {Perlick}},\ and\ \bibinfo {author} {\bibfnamefont {C.}~\bibnamefont {L{\"a}mmerzahl}},\ }\bibfield  {title} {\bibinfo {title} {{Photon Regions and Shadows of Kerr-Newman-NUT Black Holes with a Cosmological Constant}},\ }\href {https://doi.org/10.1103/PhysRevD.89.124004} {\bibfield  {journal} {\bibinfo  {journal} {Phys. Rev. D}\ }\textbf {\bibinfo {volume} {89}},\ \bibinfo {pages} {124004} (\bibinfo {year} {2014})},\ \Eprint {https://arxiv.org/abs/1403.5234} {arXiv:1403.5234 [gr-qc]} \BibitemShut {NoStop}%
\bibitem [{\citenamefont {Gralla}\ \emph {et~al.}(2018)\citenamefont {Gralla}, \citenamefont {Lupsasca},\ and\ \citenamefont {Strominger}}]{Gralla:2017ufe}%
  \BibitemOpen
  \bibfield  {author} {\bibinfo {author} {\bibfnamefont {S.~E.}\ \bibnamefont {Gralla}}, \bibinfo {author} {\bibfnamefont {A.}~\bibnamefont {Lupsasca}},\ and\ \bibinfo {author} {\bibfnamefont {A.}~\bibnamefont {Strominger}},\ }\bibfield  {title} {\bibinfo {title} {{Observational Signature of High Spin at the Event Horizon Telescope}},\ }\href {https://doi.org/10.1093/mnras/sty039} {\bibfield  {journal} {\bibinfo  {journal} {Mon. Not. Roy. Astron. Soc.}\ }\textbf {\bibinfo {volume} {475}},\ \bibinfo {pages} {3829} (\bibinfo {year} {2018})},\ \Eprint {https://arxiv.org/abs/1710.11112} {arXiv:1710.11112 [astro-ph.HE]} \BibitemShut {NoStop}%
\bibitem [{\citenamefont {Perlick}\ and\ \citenamefont {Tsupko}(2022)}]{Perlick:2021aok}%
  \BibitemOpen
  \bibfield  {author} {\bibinfo {author} {\bibfnamefont {V.}~\bibnamefont {Perlick}}\ and\ \bibinfo {author} {\bibfnamefont {O.~Y.}\ \bibnamefont {Tsupko}},\ }\bibfield  {title} {\bibinfo {title} {{Calculating black hole shadows: Review of analytical studies}},\ }\href {https://doi.org/10.1016/j.physrep.2021.10.004} {\bibfield  {journal} {\bibinfo  {journal} {Phys. Rept.}\ }\textbf {\bibinfo {volume} {947}},\ \bibinfo {pages} {1} (\bibinfo {year} {2022})},\ \Eprint {https://arxiv.org/abs/2105.07101} {arXiv:2105.07101 [gr-qc]} \BibitemShut {NoStop}%
\bibitem [{\citenamefont {Walker}\ and\ \citenamefont {Penrose}(1970)}]{Walker:1970un}%
  \BibitemOpen
  \bibfield  {author} {\bibinfo {author} {\bibfnamefont {M.}~\bibnamefont {Walker}}\ and\ \bibinfo {author} {\bibfnamefont {R.}~\bibnamefont {Penrose}},\ }\bibfield  {title} {\bibinfo {title} {{On quadratic first integrals of the geodesic equations for type [22] spacetimes}},\ }\href {https://doi.org/10.1007/BF01649445} {\bibfield  {journal} {\bibinfo  {journal} {Commun. Math. Phys.}\ }\textbf {\bibinfo {volume} {18}},\ \bibinfo {pages} {265} (\bibinfo {year} {1970})}\BibitemShut {NoStop}%
\bibitem [{\citenamefont {Carter}(1968)}]{Carter:1968rr}%
  \BibitemOpen
  \bibfield  {author} {\bibinfo {author} {\bibfnamefont {B.}~\bibnamefont {Carter}},\ }\bibfield  {title} {\bibinfo {title} {{Global structure of the Kerr family of gravitational fields}},\ }\href {https://doi.org/10.1103/PhysRev.174.1559} {\bibfield  {journal} {\bibinfo  {journal} {Phys. Rev.}\ }\textbf {\bibinfo {volume} {174}},\ \bibinfo {pages} {1559} (\bibinfo {year} {1968})}\BibitemShut {NoStop}%
\bibitem [{\citenamefont {{Fletcher}}\ and\ \citenamefont {{Lun}}(2003)}]{Fletcher2003bs}%
  \BibitemOpen
  \bibfield  {author} {\bibinfo {author} {\bibfnamefont {S.~J.}\ \bibnamefont {{Fletcher}}}\ and\ \bibinfo {author} {\bibfnamefont {A.~W.~C.}\ \bibnamefont {{Lun}}},\ }\bibfield  {title} {\bibinfo {title} {{The Kerr spacetime in generalized Bondi-Sachs coordinates}},\ }\href {https://doi.org/10.1088/0264-9381/20/19/302} {\bibfield  {journal} {\bibinfo  {journal} {Class. Quant. Grav.}\ }\textbf {\bibinfo {volume} {20}},\ \bibinfo {pages} {4153} (\bibinfo {year} {2003})}\BibitemShut {NoStop}%
\bibitem [{\citenamefont {Bishop}\ and\ \citenamefont {Venter}(2006{\natexlab{a}})}]{Bishop:2006kerr}%
  \BibitemOpen
  \bibfield  {author} {\bibinfo {author} {\bibfnamefont {N.~T.}\ \bibnamefont {Bishop}}\ and\ \bibinfo {author} {\bibfnamefont {L.~R.}\ \bibnamefont {Venter}},\ }\bibfield  {title} {\bibinfo {title} {Kerr metric in bondi-sachs form},\ }\href {https://doi.org/10.1103/PhysRevD.73.084023} {\bibfield  {journal} {\bibinfo  {journal} {Phys. Rev. D}\ }\textbf {\bibinfo {volume} {73}},\ \bibinfo {pages} {084023} (\bibinfo {year} {2006}{\natexlab{a}})}\BibitemShut {NoStop}%
\bibitem [{\citenamefont {Bai}\ \emph {et~al.}(2007)\citenamefont {Bai}, \citenamefont {Cao}, \citenamefont {Gong}, \citenamefont {Shang}, \citenamefont {Wu},\ and\ \citenamefont {Lau}}]{Bai:2007rs}%
  \BibitemOpen
  \bibfield  {author} {\bibinfo {author} {\bibfnamefont {S.}~\bibnamefont {Bai}}, \bibinfo {author} {\bibfnamefont {Z.-J.}\ \bibnamefont {Cao}}, \bibinfo {author} {\bibfnamefont {X.-F.}\ \bibnamefont {Gong}}, \bibinfo {author} {\bibfnamefont {Y.}~\bibnamefont {Shang}}, \bibinfo {author} {\bibfnamefont {X.-N.}\ \bibnamefont {Wu}},\ and\ \bibinfo {author} {\bibfnamefont {Y.~K.}\ \bibnamefont {Lau}},\ }\bibfield  {title} {\bibinfo {title} {{Light Cone Structure near Null Infinity of the Kerr Metric}},\ }\href {https://doi.org/10.1103/PhysRevD.75.044003} {\bibfield  {journal} {\bibinfo  {journal} {Phys. Rev. D}\ }\textbf {\bibinfo {volume} {75}},\ \bibinfo {pages} {044003} (\bibinfo {year} {2007})},\ \Eprint {https://arxiv.org/abs/gr-qc/0701171} {arXiv:gr-qc/0701171} \BibitemShut {NoStop}%
\bibitem [{\citenamefont {Wu}\ and\ \citenamefont {Bai}(2008)}]{Wu:2008yi}%
  \BibitemOpen
  \bibfield  {author} {\bibinfo {author} {\bibfnamefont {X.}~\bibnamefont {Wu}}\ and\ \bibinfo {author} {\bibfnamefont {S.}~\bibnamefont {Bai}},\ }\bibfield  {title} {\bibinfo {title} {{On Uniqueness of Kerr Space-time near null infinity}},\ }\href {https://doi.org/10.1103/PhysRevD.78.124009} {\bibfield  {journal} {\bibinfo  {journal} {Phys. Rev. D}\ }\textbf {\bibinfo {volume} {78}},\ \bibinfo {pages} {124009} (\bibinfo {year} {2008})},\ \Eprint {https://arxiv.org/abs/0811.3794} {arXiv:0811.3794 [gr-qc]} \BibitemShut {NoStop}%
\bibitem [{Note4()}]{Note4}%
  \BibitemOpen
  \bibinfo {note} {There is a closed form \cite {PhysRevD.73.084023}. Unfortunately, that is too complicated for our purpose.}\BibitemShut {Stop}%
\bibitem [{\citenamefont {Chandrasekhar}(1998)}]{Chandrasekhar:1985kt}%
  \BibitemOpen
  \bibfield  {author} {\bibinfo {author} {\bibfnamefont {S.}~\bibnamefont {Chandrasekhar}},\ }\href {https://global.oup.com/academic/product/the-mathematical-theory-of-black-holes-9780198503705?q=the%20mathematical%20theory%20of%20black%20holes&lang=en&cc=cn#} {\emph {\bibinfo {title} {{The mathematical theory of black holes}}}},\ Oxford classic texts in the physical sciences\ (\bibinfo  {publisher} {Oxford University Press},\ \bibinfo {address} {Oxford},\ \bibinfo {year} {1998})\BibitemShut {NoStop}%
\bibitem [{\citenamefont {Jackson}(1998)}]{Jackson:1998nia}%
  \BibitemOpen
  \bibfield  {author} {\bibinfo {author} {\bibfnamefont {J.~D.}\ \bibnamefont {Jackson}},\ }\href@noop {} {\emph {\bibinfo {title} {{Classical Electrodynamics}}}}\ (\bibinfo  {publisher} {Wiley},\ \bibinfo {year} {1998})\BibitemShut {NoStop}%
\bibitem [{\citenamefont {{Balbus}}\ and\ \citenamefont {{Hawley}}(1991)}]{1991ApJ...376..214B}%
  \BibitemOpen
  \bibfield  {author} {\bibinfo {author} {\bibfnamefont {S.~A.}\ \bibnamefont {{Balbus}}}\ and\ \bibinfo {author} {\bibfnamefont {J.~F.}\ \bibnamefont {{Hawley}}},\ }\bibfield  {title} {\bibinfo {title} {{A Powerful Local Shear Instability in Weakly Magnetized Disks. I. Linear Analysis}},\ }\href {https://doi.org/10.1086/170270} {\bibfield  {journal} {\bibinfo  {journal} {Astrophys. J.}\ }\textbf {\bibinfo {volume} {376}},\ \bibinfo {pages} {214} (\bibinfo {year} {1991})}\BibitemShut {NoStop}%
\bibitem [{\citenamefont {Balbus}\ and\ \citenamefont {Hawley}(1998)}]{RevModPhys.70.1}%
  \BibitemOpen
  \bibfield  {author} {\bibinfo {author} {\bibfnamefont {S.~A.}\ \bibnamefont {Balbus}}\ and\ \bibinfo {author} {\bibfnamefont {J.~F.}\ \bibnamefont {Hawley}},\ }\bibfield  {title} {\bibinfo {title} {Instability, turbulence, and enhanced transport in accretion disks},\ }\href {https://doi.org/10.1103/RevModPhys.70.1} {\bibfield  {journal} {\bibinfo  {journal} {Rev. Mod. Phys.}\ }\textbf {\bibinfo {volume} {70}},\ \bibinfo {pages} {1} (\bibinfo {year} {1998})}\BibitemShut {NoStop}%
\bibitem [{\citenamefont {{Curtis}}(1918)}]{1918PLicO..13....9C}%
  \BibitemOpen
  \bibfield  {author} {\bibinfo {author} {\bibfnamefont {H.~D.}\ \bibnamefont {{Curtis}}},\ }\bibfield  {title} {\bibinfo {title} {{Descriptions of 762 Nebulae and Clusters Photographed with the Crossley Reflector}},\ }\href@noop {} {\bibfield  {journal} {\bibinfo  {journal} {Publications of Lick Observatory}\ }\textbf {\bibinfo {volume} {13}},\ \bibinfo {pages} {9} (\bibinfo {year} {1918})}\BibitemShut {NoStop}%
\bibitem [{\citenamefont {{Contopoulos}}\ \emph {et~al.}(2015)\citenamefont {{Contopoulos}}, \citenamefont {{Gabuzda}},\ and\ \citenamefont {{Kylafis}}}]{2015ASSL..414.....C}%
  \BibitemOpen
  \bibinfo {editor} {\bibfnamefont {I.}~\bibnamefont {{Contopoulos}}}, \bibinfo {editor} {\bibfnamefont {D.}~\bibnamefont {{Gabuzda}}},\ and\ \bibinfo {editor} {\bibfnamefont {N.}~\bibnamefont {{Kylafis}}},\ eds.,\ \href {https://doi.org/10.1007/978-3-319-10356-3} {\emph {\bibinfo {title} {The Formation and Disruption of Black Hole Jets}}},\ \bibinfo {series} {Astrophysics and Space Science Library}, Vol.\ \bibinfo {volume} {414}\ (\bibinfo {year} {2015})\BibitemShut {NoStop}%
\bibitem [{\citenamefont {{Hills}}(1975)}]{1975Natur.254..295H}%
  \BibitemOpen
  \bibfield  {author} {\bibinfo {author} {\bibfnamefont {J.~G.}\ \bibnamefont {{Hills}}},\ }\bibfield  {title} {\bibinfo {title} {{Possible power source of Seyfert galaxies and QSOs}},\ }\href {https://doi.org/10.1038/254295a0} {\bibfield  {journal} {\bibinfo  {journal} {Nature}\ }\textbf {\bibinfo {volume} {254}},\ \bibinfo {pages} {295} (\bibinfo {year} {1975})}\BibitemShut {NoStop}%
\bibitem [{\citenamefont {Rees}(1988)}]{Rees:1988bf}%
  \BibitemOpen
  \bibfield  {author} {\bibinfo {author} {\bibfnamefont {M.~J.}\ \bibnamefont {Rees}},\ }\bibfield  {title} {\bibinfo {title} {{Tidal disruption of stars by black holes of 10 to the 6th-10 to the 8th solar masses in nearby galaxies}},\ }\href {https://doi.org/10.1038/333523a0} {\bibfield  {journal} {\bibinfo  {journal} {Nature}\ }\textbf {\bibinfo {volume} {333}},\ \bibinfo {pages} {523} (\bibinfo {year} {1988})}\BibitemShut {NoStop}%
\bibitem [{\citenamefont {Kozai}(1962)}]{Kozai:1962zz}%
  \BibitemOpen
  \bibfield  {author} {\bibinfo {author} {\bibfnamefont {Y.}~\bibnamefont {Kozai}},\ }\bibfield  {title} {\bibinfo {title} {{Secular perturbations of asteroids with high inclination and eccentricity}},\ }\href {https://doi.org/10.1086/108790} {\bibfield  {journal} {\bibinfo  {journal} {Astron. J.}\ }\textbf {\bibinfo {volume} {67}},\ \bibinfo {pages} {591} (\bibinfo {year} {1962})}\BibitemShut {NoStop}%
\bibitem [{\citenamefont {Lidov}(1962)}]{Lidov:1962wjn}%
  \BibitemOpen
  \bibfield  {author} {\bibinfo {author} {\bibfnamefont {M.~L.}\ \bibnamefont {Lidov}},\ }\bibfield  {title} {\bibinfo {title} {{The evolution of orbits of artificial satellites of planets under the action of gravitational perturbations of external bodies}},\ }\href {https://doi.org/10.1016/0032-0633(62)90129-0} {\bibfield  {journal} {\bibinfo  {journal} {Planet. Space Sci.}\ }\textbf {\bibinfo {volume} {9}},\ \bibinfo {pages} {719} (\bibinfo {year} {1962})}\BibitemShut {NoStop}%
\bibitem [{\citenamefont {Will}(2021)}]{Will:2020tri}%
  \BibitemOpen
  \bibfield  {author} {\bibinfo {author} {\bibfnamefont {C.~M.}\ \bibnamefont {Will}},\ }\bibfield  {title} {\bibinfo {title} {{Higher-order effects in the dynamics of hierarchical triple systems. Quadrupole-squared terms}},\ }\href {https://doi.org/10.1103/PhysRevD.103.063003} {\bibfield  {journal} {\bibinfo  {journal} {Phys. Rev. D}\ }\textbf {\bibinfo {volume} {103}},\ \bibinfo {pages} {063003} (\bibinfo {year} {2021})},\ \Eprint {https://arxiv.org/abs/2011.13286} {arXiv:2011.13286 [astro-ph.EP]} \BibitemShut {NoStop}%
\bibitem [{\citenamefont {Conway}\ and\ \citenamefont {Will}(2024)}]{Conway:2024azg}%
  \BibitemOpen
  \bibfield  {author} {\bibinfo {author} {\bibfnamefont {L.}~\bibnamefont {Conway}}\ and\ \bibinfo {author} {\bibfnamefont {C.~M.}\ \bibnamefont {Will}},\ }\bibfield  {title} {\bibinfo {title} {{Higher-order effects in the dynamics of hierarchical triple systems. II. Second-order and dotriacontapole-order effects}},\ }\href {https://doi.org/10.1103/PhysRevD.110.083022} {\bibfield  {journal} {\bibinfo  {journal} {Phys. Rev. D}\ }\textbf {\bibinfo {volume} {110}},\ \bibinfo {pages} {083022} (\bibinfo {year} {2024})},\ \Eprint {https://arxiv.org/abs/2408.04411} {arXiv:2408.04411 [astro-ph.EP]} \BibitemShut {NoStop}%
\bibitem [{\citenamefont {Conway}\ and\ \citenamefont {Will}(2025)}]{Conway:2025ixt}%
  \BibitemOpen
  \bibfield  {author} {\bibinfo {author} {\bibfnamefont {L.}~\bibnamefont {Conway}}\ and\ \bibinfo {author} {\bibfnamefont {C.~M.}\ \bibnamefont {Will}},\ }\bibfield  {title} {\bibinfo {title} {{Higher-order effects in the dynamics of hierarchical triple systems. III. Astrophysical implications of second-order and dotriacontapole terms}},\ }\href {https://doi.org/10.1103/PhysRevD.111.083052} {\bibfield  {journal} {\bibinfo  {journal} {Phys. Rev. D}\ }\textbf {\bibinfo {volume} {111}},\ \bibinfo {pages} {083052} (\bibinfo {year} {2025})},\ \Eprint {https://arxiv.org/abs/2501.11187} {arXiv:2501.11187 [astro-ph.EP]} \BibitemShut {NoStop}%
\bibitem [{\citenamefont {Gupta}\ \emph {et~al.}(2020)\citenamefont {Gupta}, \citenamefont {Suzuki}, \citenamefont {Okawa},\ and\ \citenamefont {Maeda}}]{Gupta:2019unn}%
  \BibitemOpen
  \bibfield  {author} {\bibinfo {author} {\bibfnamefont {P.}~\bibnamefont {Gupta}}, \bibinfo {author} {\bibfnamefont {H.}~\bibnamefont {Suzuki}}, \bibinfo {author} {\bibfnamefont {H.}~\bibnamefont {Okawa}},\ and\ \bibinfo {author} {\bibfnamefont {K.-i.}\ \bibnamefont {Maeda}},\ }\bibfield  {title} {\bibinfo {title} {{Gravitational Waves from Hierarchical Triple Systems with Kozai-Lidov Oscillation}},\ }\href {https://doi.org/10.1103/PhysRevD.101.104053} {\bibfield  {journal} {\bibinfo  {journal} {Phys. Rev. D}\ }\textbf {\bibinfo {volume} {101}},\ \bibinfo {pages} {104053} (\bibinfo {year} {2020})},\ \Eprint {https://arxiv.org/abs/1911.11318} {arXiv:1911.11318 [gr-qc]} \BibitemShut {NoStop}%
\bibitem [{\citenamefont {Toscani}\ \emph {et~al.}(2022)\citenamefont {Toscani}, \citenamefont {Lodato}, \citenamefont {Price},\ and\ \citenamefont {Liptai}}]{Toscani:2021bzr}%
  \BibitemOpen
  \bibfield  {author} {\bibinfo {author} {\bibfnamefont {M.}~\bibnamefont {Toscani}}, \bibinfo {author} {\bibfnamefont {G.}~\bibnamefont {Lodato}}, \bibinfo {author} {\bibfnamefont {D.~J.}\ \bibnamefont {Price}},\ and\ \bibinfo {author} {\bibfnamefont {D.}~\bibnamefont {Liptai}},\ }\bibfield  {title} {\bibinfo {title} {{Gravitational waves from tidal disruption events: an open and comprehensive catalog}},\ }\href {https://doi.org/10.1093/mnras/stab3384} {\bibfield  {journal} {\bibinfo  {journal} {Mon. Not. Roy. Astron. Soc.}\ }\textbf {\bibinfo {volume} {510}},\ \bibinfo {pages} {992} (\bibinfo {year} {2022})},\ \Eprint {https://arxiv.org/abs/2111.05145} {arXiv:2111.05145 [astro-ph.HE]} \BibitemShut {NoStop}%
\bibitem [{\citenamefont {Pfister}\ \emph {et~al.}(2022)\citenamefont {Pfister}, \citenamefont {Toscani}, \citenamefont {Wong}, \citenamefont {Dai}, \citenamefont {Lodato},\ and\ \citenamefont {Rossi}}]{Pfister:2021ton}%
  \BibitemOpen
  \bibfield  {author} {\bibinfo {author} {\bibfnamefont {H.}~\bibnamefont {Pfister}}, \bibinfo {author} {\bibfnamefont {M.}~\bibnamefont {Toscani}}, \bibinfo {author} {\bibfnamefont {T.~H.~T.}\ \bibnamefont {Wong}}, \bibinfo {author} {\bibfnamefont {J.~L.}\ \bibnamefont {Dai}}, \bibinfo {author} {\bibfnamefont {G.}~\bibnamefont {Lodato}},\ and\ \bibinfo {author} {\bibfnamefont {E.~M.}\ \bibnamefont {Rossi}},\ }\bibfield  {title} {\bibinfo {title} {{Observable gravitational waves from tidal disruption events and their electromagnetic counterpart}},\ }\href {https://doi.org/10.1093/mnras/stab3387} {\bibfield  {journal} {\bibinfo  {journal} {Mon. Not. Roy. Astron. Soc.}\ }\textbf {\bibinfo {volume} {510}},\ \bibinfo {pages} {2025} (\bibinfo {year} {2022})},\ \Eprint {https://arxiv.org/abs/2103.05883} {arXiv:2103.05883 [astro-ph.HE]} \BibitemShut {NoStop}%
\bibitem [{\citenamefont {Yuan}\ \emph {et~al.}(2025)\citenamefont {Yuan}, \citenamefont {Cardoso}, \citenamefont {Duque},\ and\ \citenamefont {Younsi}}]{Yuan:2025fde}%
  \BibitemOpen
  \bibfield  {author} {\bibinfo {author} {\bibfnamefont {C.}~\bibnamefont {Yuan}}, \bibinfo {author} {\bibfnamefont {V.}~\bibnamefont {Cardoso}}, \bibinfo {author} {\bibfnamefont {F.}~\bibnamefont {Duque}},\ and\ \bibinfo {author} {\bibfnamefont {Z.}~\bibnamefont {Younsi}},\ }\bibfield  {title} {\bibinfo {title} {{Gravitational waves from accretion disks: Turbulence, mode excitation, and prospects for future detectors}},\ }\href {https://doi.org/10.1103/PhysRevD.111.063048} {\bibfield  {journal} {\bibinfo  {journal} {Phys. Rev. D}\ }\textbf {\bibinfo {volume} {111}},\ \bibinfo {pages} {063048} (\bibinfo {year} {2025})},\ \Eprint {https://arxiv.org/abs/2502.07871} {arXiv:2502.07871 [gr-qc]} \BibitemShut {NoStop}%
\bibitem [{\citenamefont {Zel'dovich}\ and\ \citenamefont {Polnarev}(1974)}]{Zeldovich:1974gvh}%
  \BibitemOpen
  \bibfield  {author} {\bibinfo {author} {\bibfnamefont {Y.~B.}\ \bibnamefont {Zel'dovich}}\ and\ \bibinfo {author} {\bibfnamefont {A.~G.}\ \bibnamefont {Polnarev}},\ }\bibfield  {title} {\bibinfo {title} {{Radiation of gravitational waves by a cluster of superdense stars}},\ }\href@noop {} {\bibfield  {journal} {\bibinfo  {journal} {Sov. Astron.}\ }\textbf {\bibinfo {volume} {18}},\ \bibinfo {pages} {17} (\bibinfo {year} {1974})}\BibitemShut {NoStop}%
\bibitem [{\citenamefont {Braginsky}\ and\ \citenamefont {Grishchuk}(1985)}]{Braginsky:1986ia}%
  \BibitemOpen
  \bibfield  {author} {\bibinfo {author} {\bibfnamefont {V.}~\bibnamefont {Braginsky}}\ and\ \bibinfo {author} {\bibfnamefont {L.}~\bibnamefont {Grishchuk}},\ }\bibfield  {title} {\bibinfo {title} {{Kinematic Resonance and Memory Effect in Free Mass Gravitational Antennas}},\ }\href@noop {} {\bibfield  {journal} {\bibinfo  {journal} {Sov. Phys. JETP}\ }\textbf {\bibinfo {volume} {62}},\ \bibinfo {pages} {427} (\bibinfo {year} {1985})}\BibitemShut {NoStop}%
\bibitem [{\citenamefont {Christodoulou}(1991)}]{Christodoulou1991}%
  \BibitemOpen
  \bibfield  {author} {\bibinfo {author} {\bibfnamefont {D.}~\bibnamefont {Christodoulou}},\ }\bibfield  {title} {\bibinfo {title} {Nonlinear nature of gravitation and gravitational-wave experiments},\ }\href {https://doi.org/10.1103/PhysRevLett.67.1486} {\bibfield  {journal} {\bibinfo  {journal} {Phys. Rev. Lett.}\ }\textbf {\bibinfo {volume} {67}},\ \bibinfo {pages} {1486} (\bibinfo {year} {1991})}\BibitemShut {NoStop}%
\bibitem [{\citenamefont {Thorne}(1992)}]{Thorne:1992sdb}%
  \BibitemOpen
  \bibfield  {author} {\bibinfo {author} {\bibfnamefont {K.~S.}\ \bibnamefont {Thorne}},\ }\bibfield  {title} {\bibinfo {title} {{Gravitational-wave bursts with memory: The Christodoulou effect}},\ }\href {https://doi.org/10.1103/PhysRevD.45.520} {\bibfield  {journal} {\bibinfo  {journal} {Phys. Rev. D}\ }\textbf {\bibinfo {volume} {45}},\ \bibinfo {pages} {520} (\bibinfo {year} {1992})}\BibitemShut {NoStop}%
\bibitem [{\citenamefont {De~Luca}\ \emph {et~al.}(2025{\natexlab{a}})\citenamefont {De~Luca}, \citenamefont {Khoury},\ and\ \citenamefont {Wong}}]{DeLuca:2024cjl}%
  \BibitemOpen
  \bibfield  {author} {\bibinfo {author} {\bibfnamefont {V.}~\bibnamefont {De~Luca}}, \bibinfo {author} {\bibfnamefont {J.}~\bibnamefont {Khoury}},\ and\ \bibinfo {author} {\bibfnamefont {S.~S.~C.}\ \bibnamefont {Wong}},\ }\bibfield  {title} {\bibinfo {title} {{Gravitational memory and soft theorems: The local perspective}},\ }\href {https://doi.org/10.1103/gbg1-mz49} {\bibfield  {journal} {\bibinfo  {journal} {Phys. Rev. D}\ }\textbf {\bibinfo {volume} {112}},\ \bibinfo {pages} {L021502} (\bibinfo {year} {2025}{\natexlab{a}})},\ \Eprint {https://arxiv.org/abs/2412.01910} {arXiv:2412.01910 [gr-qc]} \BibitemShut {NoStop}%
\bibitem [{\citenamefont {De~Luca}\ \emph {et~al.}(2025{\natexlab{b}})\citenamefont {De~Luca}, \citenamefont {Khoury},\ and\ \citenamefont {Wong}}]{DeLuca:2024asq}%
  \BibitemOpen
  \bibfield  {author} {\bibinfo {author} {\bibfnamefont {V.}~\bibnamefont {De~Luca}}, \bibinfo {author} {\bibfnamefont {J.}~\bibnamefont {Khoury}},\ and\ \bibinfo {author} {\bibfnamefont {S.~S.~C.}\ \bibnamefont {Wong}},\ }\bibfield  {title} {\bibinfo {title} {{Gravitational memory and Ward identities in the local detector frame}},\ }\href {https://doi.org/10.1103/PhysRevD.112.024032} {\bibfield  {journal} {\bibinfo  {journal} {Phys. Rev. D}\ }\textbf {\bibinfo {volume} {112}},\ \bibinfo {pages} {024032} (\bibinfo {year} {2025}{\natexlab{b}})},\ \Eprint {https://arxiv.org/abs/2412.12273} {arXiv:2412.12273 [gr-qc]} \BibitemShut {NoStop}%
\bibitem [{\citenamefont {H\"ubner}\ \emph {et~al.}(2020)\citenamefont {H\"ubner}, \citenamefont {Talbot}, \citenamefont {Lasky},\ and\ \citenamefont {Thrane}}]{Hubner:2019sly}%
  \BibitemOpen
  \bibfield  {author} {\bibinfo {author} {\bibfnamefont {M.}~\bibnamefont {H\"ubner}}, \bibinfo {author} {\bibfnamefont {C.}~\bibnamefont {Talbot}}, \bibinfo {author} {\bibfnamefont {P.~D.}\ \bibnamefont {Lasky}},\ and\ \bibinfo {author} {\bibfnamefont {E.}~\bibnamefont {Thrane}},\ }\bibfield  {title} {\bibinfo {title} {{Measuring gravitational-wave memory in the first LIGO/Virgo gravitational-wave transient catalog}},\ }\href {https://doi.org/10.1103/PhysRevD.101.023011} {\bibfield  {journal} {\bibinfo  {journal} {Phys. Rev. D}\ }\textbf {\bibinfo {volume} {101}},\ \bibinfo {pages} {023011} (\bibinfo {year} {2020})},\ \Eprint {https://arxiv.org/abs/1911.12496} {arXiv:1911.12496 [astro-ph.HE]} \BibitemShut {NoStop}%
\bibitem [{\citenamefont {Grant}\ and\ \citenamefont {Nichols}(2023)}]{Grant:2022bla}%
  \BibitemOpen
  \bibfield  {author} {\bibinfo {author} {\bibfnamefont {A.~M.}\ \bibnamefont {Grant}}\ and\ \bibinfo {author} {\bibfnamefont {D.~A.}\ \bibnamefont {Nichols}},\ }\bibfield  {title} {\bibinfo {title} {{Outlook for detecting the gravitational-wave displacement and spin memory effects with current and future gravitational-wave detectors}},\ }\href {https://doi.org/10.1103/PhysRevD.107.064056} {\bibfield  {journal} {\bibinfo  {journal} {Phys. Rev. D}\ }\textbf {\bibinfo {volume} {107}},\ \bibinfo {pages} {064056} (\bibinfo {year} {2023})},\ \bibinfo {note} {[Erratum: Phys.Rev.D 108, 029901 (2023)]},\ \Eprint {https://arxiv.org/abs/2210.16266} {arXiv:2210.16266 [gr-qc]} \BibitemShut {NoStop}%
\bibitem [{\citenamefont {Gasparotto}\ \emph {et~al.}(2023)\citenamefont {Gasparotto}, \citenamefont {Vicente}, \citenamefont {Blas}, \citenamefont {Jenkins},\ and\ \citenamefont {Barausse}}]{Gasparotto:2023fcg}%
  \BibitemOpen
  \bibfield  {author} {\bibinfo {author} {\bibfnamefont {S.}~\bibnamefont {Gasparotto}}, \bibinfo {author} {\bibfnamefont {R.}~\bibnamefont {Vicente}}, \bibinfo {author} {\bibfnamefont {D.}~\bibnamefont {Blas}}, \bibinfo {author} {\bibfnamefont {A.~C.}\ \bibnamefont {Jenkins}},\ and\ \bibinfo {author} {\bibfnamefont {E.}~\bibnamefont {Barausse}},\ }\bibfield  {title} {\bibinfo {title} {{Can gravitational-wave memory help constrain binary black-hole parameters? A LISA case study}},\ }\href {https://doi.org/10.1103/PhysRevD.107.124033} {\bibfield  {journal} {\bibinfo  {journal} {Phys. Rev. D}\ }\textbf {\bibinfo {volume} {107}},\ \bibinfo {pages} {124033} (\bibinfo {year} {2023})},\ \Eprint {https://arxiv.org/abs/2301.13228} {arXiv:2301.13228 [gr-qc]} \BibitemShut {NoStop}%
\bibitem [{\citenamefont {Inchausp{\'e}}\ \emph {et~al.}(2025)\citenamefont {Inchausp{\'e}}, \citenamefont {Gasparotto}, \citenamefont {Blas}, \citenamefont {Heisenberg}, \citenamefont {Zosso},\ and\ \citenamefont {Tiwari}}]{Inchauspe:2024ibs}%
  \BibitemOpen
  \bibfield  {author} {\bibinfo {author} {\bibfnamefont {H.}~\bibnamefont {Inchausp{\'e}}}, \bibinfo {author} {\bibfnamefont {S.}~\bibnamefont {Gasparotto}}, \bibinfo {author} {\bibfnamefont {D.}~\bibnamefont {Blas}}, \bibinfo {author} {\bibfnamefont {L.}~\bibnamefont {Heisenberg}}, \bibinfo {author} {\bibfnamefont {J.}~\bibnamefont {Zosso}},\ and\ \bibinfo {author} {\bibfnamefont {S.}~\bibnamefont {Tiwari}},\ }\bibfield  {title} {\bibinfo {title} {{Measuring gravitational wave memory with LISA}},\ }\href {https://doi.org/10.1103/PhysRevD.111.044044} {\bibfield  {journal} {\bibinfo  {journal} {Phys. Rev. D}\ }\textbf {\bibinfo {volume} {111}},\ \bibinfo {pages} {044044} (\bibinfo {year} {2025})},\ \Eprint {https://arxiv.org/abs/2406.09228} {arXiv:2406.09228 [gr-qc]} \BibitemShut {NoStop}%
\bibitem [{\citenamefont {Hou}\ \emph {et~al.}(2025{\natexlab{b}})\citenamefont {Hou}, \citenamefont {Zhao}, \citenamefont {Cao},\ and\ \citenamefont {Zhu}}]{Hou:2024rgo}%
  \BibitemOpen
  \bibfield  {author} {\bibinfo {author} {\bibfnamefont {S.}~\bibnamefont {Hou}}, \bibinfo {author} {\bibfnamefont {Z.-C.}\ \bibnamefont {Zhao}}, \bibinfo {author} {\bibfnamefont {Z.}~\bibnamefont {Cao}},\ and\ \bibinfo {author} {\bibfnamefont {Z.-H.}\ \bibnamefont {Zhu}},\ }\bibfield  {title} {\bibinfo {title} {{Space-Borne Interferometers to Detect Thousands of Memory Signals Emitted by Stellar-Mass Binary Black Holes}},\ }\href {https://doi.org/10.1088/0256-307X/42/10/101101} {\bibfield  {journal} {\bibinfo  {journal} {Chin. Phys. Lett.}\ }\textbf {\bibinfo {volume} {42}},\ \bibinfo {pages} {101101} (\bibinfo {year} {2025}{\natexlab{b}})},\ \Eprint {https://arxiv.org/abs/2411.18053} {arXiv:2411.18053 [gr-qc]} \BibitemShut {NoStop}%
\bibitem [{\citenamefont {Hou}\ and\ \citenamefont {Zhu}(2021)}]{Hou:2020tnd}%
  \BibitemOpen
  \bibfield  {author} {\bibinfo {author} {\bibfnamefont {S.}~\bibnamefont {Hou}}\ and\ \bibinfo {author} {\bibfnamefont {Z.-H.}\ \bibnamefont {Zhu}},\ }\bibfield  {title} {\bibinfo {title} {{Gravitational memory effects and Bondi-Metzner-Sachs symmetries in scalar-tensor theories}},\ }\href {https://doi.org/10.1007/JHEP01(2021)083} {\bibfield  {journal} {\bibinfo  {journal} {JHEP}\ }\textbf {\bibinfo {volume} {01}},\ \bibinfo {pages} {083}},\ \Eprint {https://arxiv.org/abs/2005.01310} {arXiv:2005.01310 [gr-qc]} \BibitemShut {NoStop}%
\bibitem [{\citenamefont {Tahura}\ \emph {et~al.}(2021)\citenamefont {Tahura}, \citenamefont {Nichols}, \citenamefont {Saffer}, \citenamefont {Stein},\ and\ \citenamefont {Yagi}}]{Tahura:2020vsa}%
  \BibitemOpen
  \bibfield  {author} {\bibinfo {author} {\bibfnamefont {S.}~\bibnamefont {Tahura}}, \bibinfo {author} {\bibfnamefont {D.~A.}\ \bibnamefont {Nichols}}, \bibinfo {author} {\bibfnamefont {A.}~\bibnamefont {Saffer}}, \bibinfo {author} {\bibfnamefont {L.~C.}\ \bibnamefont {Stein}},\ and\ \bibinfo {author} {\bibfnamefont {K.}~\bibnamefont {Yagi}},\ }\bibfield  {title} {\bibinfo {title} {{Brans-Dicke theory in Bondi-Sachs form: Asymptotically flat solutions, asymptotic symmetries and gravitational-wave memory effects}},\ }\href {https://doi.org/10.1103/PhysRevD.103.104026} {\bibfield  {journal} {\bibinfo  {journal} {Phys. Rev. D}\ }\textbf {\bibinfo {volume} {103}},\ \bibinfo {pages} {104026} (\bibinfo {year} {2021})},\ \Eprint {https://arxiv.org/abs/2007.13799} {arXiv:2007.13799 [gr-qc]} \BibitemShut {NoStop}%
\bibitem [{\citenamefont {Seraj}(2021)}]{Seraj:2021qja}%
  \BibitemOpen
  \bibfield  {author} {\bibinfo {author} {\bibfnamefont {A.}~\bibnamefont {Seraj}},\ }\bibfield  {title} {\bibinfo {title} {{Gravitational breathing memory and dual symmetries}},\ }\href {https://doi.org/10.1007/JHEP05(2021)283} {\bibfield  {journal} {\bibinfo  {journal} {JHEP}\ }\textbf {\bibinfo {volume} {05}},\ \bibinfo {pages} {283}},\ \Eprint {https://arxiv.org/abs/2103.12185} {arXiv:2103.12185 [hep-th]} \BibitemShut {NoStop}%
\bibitem [{\citenamefont {Hou}\ \emph {et~al.}(2022)\citenamefont {Hou}, \citenamefont {Zhu},\ and\ \citenamefont {Zhu}}]{Hou:2021oxe}%
  \BibitemOpen
  \bibfield  {author} {\bibinfo {author} {\bibfnamefont {S.}~\bibnamefont {Hou}}, \bibinfo {author} {\bibfnamefont {T.}~\bibnamefont {Zhu}},\ and\ \bibinfo {author} {\bibfnamefont {Z.-H.}\ \bibnamefont {Zhu}},\ }\bibfield  {title} {\bibinfo {title} {{Asymptotic analysis of Chern-Simons modified gravity and its memory effects}},\ }\href {https://doi.org/10.1103/PhysRevD.105.024025} {\bibfield  {journal} {\bibinfo  {journal} {Phys. Rev. D}\ }\textbf {\bibinfo {volume} {105}},\ \bibinfo {pages} {024025} (\bibinfo {year} {2022})},\ \Eprint {https://arxiv.org/abs/2109.04238} {arXiv:2109.04238 [gr-qc]} \BibitemShut {NoStop}%
\bibitem [{\citenamefont {Hou}\ \emph {et~al.}(2024{\natexlab{a}})\citenamefont {Hou}, \citenamefont {Wang},\ and\ \citenamefont {Zhu}}]{Hou:2023pfz}%
  \BibitemOpen
  \bibfield  {author} {\bibinfo {author} {\bibfnamefont {S.}~\bibnamefont {Hou}}, \bibinfo {author} {\bibfnamefont {A.}~\bibnamefont {Wang}},\ and\ \bibinfo {author} {\bibfnamefont {Z.-H.}\ \bibnamefont {Zhu}},\ }\bibfield  {title} {\bibinfo {title} {{Asymptotic analysis of Einstein-\AE{}ther theory and its memory effects: The linearized case}},\ }\href {https://doi.org/10.1103/PhysRevD.109.044025} {\bibfield  {journal} {\bibinfo  {journal} {Phys. Rev. D}\ }\textbf {\bibinfo {volume} {109}},\ \bibinfo {pages} {044025} (\bibinfo {year} {2024}{\natexlab{a}})},\ \Eprint {https://arxiv.org/abs/2309.01165} {arXiv:2309.01165 [gr-qc]} \BibitemShut {NoStop}%
\bibitem [{\citenamefont {Hou}\ \emph {et~al.}(2024{\natexlab{b}})\citenamefont {Hou}, \citenamefont {Fan}, \citenamefont {Zhu},\ and\ \citenamefont {Zhu}}]{Hou:2024xbv}%
  \BibitemOpen
  \bibfield  {author} {\bibinfo {author} {\bibfnamefont {S.}~\bibnamefont {Hou}}, \bibinfo {author} {\bibfnamefont {X.-L.}\ \bibnamefont {Fan}}, \bibinfo {author} {\bibfnamefont {T.}~\bibnamefont {Zhu}},\ and\ \bibinfo {author} {\bibfnamefont {Z.-H.}\ \bibnamefont {Zhu}},\ }\bibfield  {title} {\bibinfo {title} {{Nontensorial gravitational wave polarizations from the tensorial degrees of freedom: Linearized Lorentz-violating theory of gravity}},\ }\href {https://doi.org/10.1103/PhysRevD.109.084011} {\bibfield  {journal} {\bibinfo  {journal} {Phys. Rev. D}\ }\textbf {\bibinfo {volume} {109}},\ \bibinfo {pages} {084011} (\bibinfo {year} {2024}{\natexlab{b}})},\ \Eprint {https://arxiv.org/abs/2401.03474} {arXiv:2401.03474 [gr-qc]} \BibitemShut {NoStop}%
\bibitem [{\citenamefont {Mitman}\ \emph {et~al.}(2020)\citenamefont {Mitman}, \citenamefont {Moxon}, \citenamefont {Scheel}, \citenamefont {Teukolsky}, \citenamefont {Boyle}, \citenamefont {Deppe}, \citenamefont {Kidder},\ and\ \citenamefont {Throwe}}]{Mitman:2020pbt}%
  \BibitemOpen
  \bibfield  {author} {\bibinfo {author} {\bibfnamefont {K.}~\bibnamefont {Mitman}}, \bibinfo {author} {\bibfnamefont {J.}~\bibnamefont {Moxon}}, \bibinfo {author} {\bibfnamefont {M.~A.}\ \bibnamefont {Scheel}}, \bibinfo {author} {\bibfnamefont {S.~A.}\ \bibnamefont {Teukolsky}}, \bibinfo {author} {\bibfnamefont {M.}~\bibnamefont {Boyle}}, \bibinfo {author} {\bibfnamefont {N.}~\bibnamefont {Deppe}}, \bibinfo {author} {\bibfnamefont {L.~E.}\ \bibnamefont {Kidder}},\ and\ \bibinfo {author} {\bibfnamefont {W.}~\bibnamefont {Throwe}},\ }\bibfield  {title} {\bibinfo {title} {{Computation of displacement and spin gravitational memory in numerical relativity}},\ }\href {https://doi.org/10.1103/PhysRevD.102.104007} {\bibfield  {journal} {\bibinfo  {journal} {Phys. Rev. D}\ }\textbf {\bibinfo {volume} {102}},\ \bibinfo {pages} {104007} (\bibinfo {year} {2020})},\ \Eprint {https://arxiv.org/abs/2007.11562} {arXiv:2007.11562 [gr-qc]} \BibitemShut {NoStop}%
\bibitem [{\citenamefont {Bieri}\ and\ \citenamefont {Garfinkle}(2014)}]{Bieri:2013ada}%
  \BibitemOpen
  \bibfield  {author} {\bibinfo {author} {\bibfnamefont {L.}~\bibnamefont {Bieri}}\ and\ \bibinfo {author} {\bibfnamefont {D.}~\bibnamefont {Garfinkle}},\ }\bibfield  {title} {\bibinfo {title} {{Perturbative and gauge invariant treatment of gravitational wave memory}},\ }\href {https://doi.org/10.1103/PhysRevD.89.084039} {\bibfield  {journal} {\bibinfo  {journal} {Phys. Rev. D}\ }\textbf {\bibinfo {volume} {89}},\ \bibinfo {pages} {084039} (\bibinfo {year} {2014})},\ \Eprint {https://arxiv.org/abs/1312.6871} {arXiv:1312.6871 [gr-qc]} \BibitemShut {NoStop}%
\bibitem [{\citenamefont {Islam}\ \emph {et~al.}(2022)\citenamefont {Islam}, \citenamefont {Field}, \citenamefont {Hughes}, \citenamefont {Khanna}, \citenamefont {Varma}, \citenamefont {Giesler}, \citenamefont {Scheel}, \citenamefont {Kidder},\ and\ \citenamefont {Pfeiffer}}]{islam2022surrogate}%
  \BibitemOpen
  \bibfield  {author} {\bibinfo {author} {\bibfnamefont {T.}~\bibnamefont {Islam}}, \bibinfo {author} {\bibfnamefont {S.~E.}\ \bibnamefont {Field}}, \bibinfo {author} {\bibfnamefont {S.~A.}\ \bibnamefont {Hughes}}, \bibinfo {author} {\bibfnamefont {G.}~\bibnamefont {Khanna}}, \bibinfo {author} {\bibfnamefont {V.}~\bibnamefont {Varma}}, \bibinfo {author} {\bibfnamefont {M.}~\bibnamefont {Giesler}}, \bibinfo {author} {\bibfnamefont {M.~A.}\ \bibnamefont {Scheel}}, \bibinfo {author} {\bibfnamefont {L.~E.}\ \bibnamefont {Kidder}},\ and\ \bibinfo {author} {\bibfnamefont {H.~P.}\ \bibnamefont {Pfeiffer}},\ }\bibfield  {title} {\bibinfo {title} {{Surrogate model for gravitational wave signals from nonspinning, comparable-to large-mass-ratio black hole binaries built on black hole perturbation theory waveforms calibrated to numerical relativity}},\ }\href {https://doi.org/10.1103/PhysRevD.106.104025} {\bibfield  {journal} {\bibinfo  {journal} {Phys. Rev. D}\ }\textbf {\bibinfo {volume} {106}},\ \bibinfo {pages} {104025} (\bibinfo {year} {2022})},\ \Eprint {https://arxiv.org/abs/2204.01972} {arXiv:2204.01972 [gr-qc]} \BibitemShut {NoStop}%
\bibitem [{\citenamefont {Rink}\ \emph {et~al.}(2024)\citenamefont {Rink}, \citenamefont {Bachhar}, \citenamefont {Islam}, \citenamefont {Rifat}, \citenamefont {Gonzalez-Quesada}, \citenamefont {Field}, \citenamefont {Khanna}, \citenamefont {Hughes},\ and\ \citenamefont {Varma}}]{Rink:2024swg}%
  \BibitemOpen
  \bibfield  {author} {\bibinfo {author} {\bibfnamefont {K.}~\bibnamefont {Rink}}, \bibinfo {author} {\bibfnamefont {R.}~\bibnamefont {Bachhar}}, \bibinfo {author} {\bibfnamefont {T.}~\bibnamefont {Islam}}, \bibinfo {author} {\bibfnamefont {N.~E.~M.}\ \bibnamefont {Rifat}}, \bibinfo {author} {\bibfnamefont {K.}~\bibnamefont {Gonzalez-Quesada}}, \bibinfo {author} {\bibfnamefont {S.~E.}\ \bibnamefont {Field}}, \bibinfo {author} {\bibfnamefont {G.}~\bibnamefont {Khanna}}, \bibinfo {author} {\bibfnamefont {S.~A.}\ \bibnamefont {Hughes}},\ and\ \bibinfo {author} {\bibfnamefont {V.}~\bibnamefont {Varma}},\ }\bibfield  {title} {\bibinfo {title} {{Gravitational wave surrogate model for spinning, intermediate mass ratio binaries based on perturbation theory and numerical relativity}},\ }\href {https://doi.org/10.1103/PhysRevD.110.124069} {\bibfield  {journal} {\bibinfo  {journal} {Phys. Rev. D}\ }\textbf {\bibinfo {volume} {110}},\ \bibinfo {pages} {124069} (\bibinfo {year} {2024})},\ \Eprint {https://arxiv.org/abs/2407.18319} {arXiv:2407.18319 [gr-qc]} \BibitemShut {NoStop}%
\bibitem [{\citenamefont {Carr}\ \emph {et~al.}(2021)\citenamefont {Carr}, \citenamefont {Kuhnel},\ and\ \citenamefont {Visinelli}}]{Carr:2020erq}%
  \BibitemOpen
  \bibfield  {author} {\bibinfo {author} {\bibfnamefont {B.}~\bibnamefont {Carr}}, \bibinfo {author} {\bibfnamefont {F.}~\bibnamefont {Kuhnel}},\ and\ \bibinfo {author} {\bibfnamefont {L.}~\bibnamefont {Visinelli}},\ }\bibfield  {title} {\bibinfo {title} {{Constraints on Stupendously Large Black Holes}},\ }\href {https://doi.org/10.1093/mnras/staa3651} {\bibfield  {journal} {\bibinfo  {journal} {Mon. Not. Roy. Astron. Soc.}\ }\textbf {\bibinfo {volume} {501}},\ \bibinfo {pages} {2029} (\bibinfo {year} {2021})},\ \Eprint {https://arxiv.org/abs/2008.08077} {arXiv:2008.08077 [astro-ph.CO]} \BibitemShut {NoStop}%
\bibitem [{\citenamefont {Carr}\ \emph {et~al.}(2024)\citenamefont {Carr}, \citenamefont {Clesse}, \citenamefont {Garcia-Bellido}, \citenamefont {Hawkins},\ and\ \citenamefont {Kuhnel}}]{Carr:2023tpt}%
  \BibitemOpen
  \bibfield  {author} {\bibinfo {author} {\bibfnamefont {B.}~\bibnamefont {Carr}}, \bibinfo {author} {\bibfnamefont {S.}~\bibnamefont {Clesse}}, \bibinfo {author} {\bibfnamefont {J.}~\bibnamefont {Garcia-Bellido}}, \bibinfo {author} {\bibfnamefont {M.}~\bibnamefont {Hawkins}},\ and\ \bibinfo {author} {\bibfnamefont {F.}~\bibnamefont {Kuhnel}},\ }\bibfield  {title} {\bibinfo {title} {{Observational evidence for primordial black holes: A positivist perspective}},\ }\href {https://doi.org/10.1016/j.physrep.2023.11.005} {\bibfield  {journal} {\bibinfo  {journal} {Phys. Rept.}\ }\textbf {\bibinfo {volume} {1054}},\ \bibinfo {pages} {1} (\bibinfo {year} {2024})},\ \Eprint {https://arxiv.org/abs/2306.03903} {arXiv:2306.03903 [astro-ph.CO]} \BibitemShut {NoStop}%
\bibitem [{\citenamefont {Ayzenberg}\ \emph {et~al.}(2025)\citenamefont {Ayzenberg} \emph {et~al.}}]{Ayzenberg:2023hfw}%
  \BibitemOpen
  \bibfield  {author} {\bibinfo {author} {\bibfnamefont {D.}~\bibnamefont {Ayzenberg}} \emph {et~al.},\ }\bibfield  {title} {\bibinfo {title} {{Fundamental physics opportunities with future ground-based mm/sub-mm VLBI arrays}},\ }\href {https://doi.org/10.1007/s41114-025-00057-0} {\bibfield  {journal} {\bibinfo  {journal} {Living Rev. Rel.}\ }\textbf {\bibinfo {volume} {28}},\ \bibinfo {pages} {4} (\bibinfo {year} {2025})},\ \bibinfo {note} {[Erratum: Living Rev.Rel. 28, 7 (2025)]},\ \Eprint {https://arxiv.org/abs/2312.02130} {arXiv:2312.02130 [astro-ph.HE]} \BibitemShut {NoStop}%
\bibitem [{\citenamefont {{Andrianov}}\ \emph {et~al.}(2021)\citenamefont {{Andrianov}}, \citenamefont {{Baryshev}}, \citenamefont {{Falcke}}, \citenamefont {{Girin}}, \citenamefont {{de Graauw}}, \citenamefont {{Kostenko}}, \citenamefont {{Kudriashov}}, \citenamefont {{Ladygin}}, \citenamefont {{Likhachev}}, \citenamefont {{Roelofs}}, \citenamefont {{Rudnitskiy}}, \citenamefont {{Shaykhutdinov}}, \citenamefont {{Shchekinov}},\ and\ \citenamefont {{Shchurov}}}]{2021MNRAS.500.4866A}%
  \BibitemOpen
  \bibfield  {author} {\bibinfo {author} {\bibfnamefont {A.~S.}\ \bibnamefont {{Andrianov}}}, \bibinfo {author} {\bibfnamefont {A.~M.}\ \bibnamefont {{Baryshev}}}, \bibinfo {author} {\bibfnamefont {H.}~\bibnamefont {{Falcke}}}, \bibinfo {author} {\bibfnamefont {I.~A.}\ \bibnamefont {{Girin}}}, \bibinfo {author} {\bibfnamefont {T.}~\bibnamefont {{de Graauw}}}, \bibinfo {author} {\bibfnamefont {V.~I.}\ \bibnamefont {{Kostenko}}}, \bibinfo {author} {\bibfnamefont {V.}~\bibnamefont {{Kudriashov}}}, \bibinfo {author} {\bibfnamefont {V.~A.}\ \bibnamefont {{Ladygin}}}, \bibinfo {author} {\bibfnamefont {S.~F.}\ \bibnamefont {{Likhachev}}}, \bibinfo {author} {\bibfnamefont {F.}~\bibnamefont {{Roelofs}}}, \bibinfo {author} {\bibfnamefont {A.~G.}\ \bibnamefont {{Rudnitskiy}}}, \bibinfo {author} {\bibfnamefont {A.~R.}\ \bibnamefont {{Shaykhutdinov}}}, \bibinfo {author} {\bibfnamefont {Y.~A.}\ \bibnamefont {{Shchekinov}}},\ and\ \bibinfo {author} {\bibfnamefont {M.~A.}\ \bibnamefont {{Shchurov}}},\ }\bibfield  {title} {\bibinfo {title} {{Simulations of M87 and Sgr A* imaging with the Millimetron Space Observatory on near-Earth orbits}},\ }\href {https://doi.org/10.1093/mnras/staa2709} {\bibfield  {journal} {\bibinfo  {journal} {Mon. Not. Roy. Astron. Soc.}\ }\textbf {\bibinfo {volume} {500}},\ \bibinfo {pages} {4866} (\bibinfo {year} {2021})},\ \Eprint {https://arxiv.org/abs/2006.10120} {arXiv:2006.10120 [astro-ph.GA]} \BibitemShut {NoStop}%
\bibitem [{\citenamefont {Gurvits}\ \emph {et~al.}(2022)\citenamefont {Gurvits} \emph {et~al.}}]{Gurvits:2022wgm}%
  \BibitemOpen
  \bibfield  {author} {\bibinfo {author} {\bibfnamefont {L.~I.}\ \bibnamefont {Gurvits}} \emph {et~al.},\ }\bibfield  {title} {\bibinfo {title} {{The science case and challenges of space-borne sub-millimeter interferometry}},\ }\href {https://doi.org/10.1016/j.actaastro.2022.04.020} {\bibfield  {journal} {\bibinfo  {journal} {Acta Astronaut.}\ }\textbf {\bibinfo {volume} {196}},\ \bibinfo {pages} {314} (\bibinfo {year} {2022})},\ \Eprint {https://arxiv.org/abs/2204.09144} {arXiv:2204.09144 [astro-ph.IM]} \BibitemShut {NoStop}%
\bibitem [{\citenamefont {Hudson}\ \emph {et~al.}(2023)\citenamefont {Hudson}, \citenamefont {Gurvits}, \citenamefont {Wielgus}, \citenamefont {Paragi}, \citenamefont {Liu},\ and\ \citenamefont {Zheng}}]{Hudson:2023tuy}%
  \BibitemOpen
  \bibfield  {author} {\bibinfo {author} {\bibfnamefont {B.}~\bibnamefont {Hudson}}, \bibinfo {author} {\bibfnamefont {L.~I.}\ \bibnamefont {Gurvits}}, \bibinfo {author} {\bibfnamefont {M.}~\bibnamefont {Wielgus}}, \bibinfo {author} {\bibfnamefont {Z.}~\bibnamefont {Paragi}}, \bibinfo {author} {\bibfnamefont {L.}~\bibnamefont {Liu}},\ and\ \bibinfo {author} {\bibfnamefont {W.}~\bibnamefont {Zheng}},\ }\bibfield  {title} {\bibinfo {title} {{Orbital configurations of spaceborne interferometers for studying photon rings of supermassive black holes}},\ }\href {https://doi.org/10.1016/j.actaastro.2023.09.035} {\bibfield  {journal} {\bibinfo  {journal} {Acta Astronaut.}\ }\textbf {\bibinfo {volume} {213}},\ \bibinfo {pages} {681} (\bibinfo {year} {2023})},\ \Eprint {https://arxiv.org/abs/2309.17127} {arXiv:2309.17127 [astro-ph.IM]} \BibitemShut {NoStop}%
\bibitem [{\citenamefont {{Coyle}}\ \emph {et~al.}(2024)\citenamefont {{Coyle}}, \citenamefont {{Matsuura}},\ and\ \citenamefont {{Perrin}}}]{2024SPIE13092E....C}%
  \BibitemOpen
  \bibinfo {editor} {\bibfnamefont {L.~E.}\ \bibnamefont {{Coyle}}}, \bibinfo {editor} {\bibfnamefont {S.}~\bibnamefont {{Matsuura}}},\ and\ \bibinfo {editor} {\bibfnamefont {M.~D.}\ \bibnamefont {{Perrin}}},\ eds.,\ \href {https://doi.org/10.1117/12.3049198} {\emph {\bibinfo {title} {Space Telescopes and Instrumentation 2024: Optical, Infrared, and Millimeter Wave}}},\ \bibinfo {series} {Society of Photo-Optical Instrumentation Engineers (SPIE) Conference Series}, Vol.\ \bibinfo {volume} {13092}\ (\bibinfo {year} {2024})\BibitemShut {NoStop}%
\bibitem [{\citenamefont {{Kurdubov}}\ \emph {et~al.}(2019)\citenamefont {{Kurdubov}}, \citenamefont {{Pavlov}}, \citenamefont {{Mironova}},\ and\ \citenamefont {{Kaplev}}}]{2019MNRAS.486..815K}%
  \BibitemOpen
  \bibfield  {author} {\bibinfo {author} {\bibfnamefont {S.~L.}\ \bibnamefont {{Kurdubov}}}, \bibinfo {author} {\bibfnamefont {D.~A.}\ \bibnamefont {{Pavlov}}}, \bibinfo {author} {\bibfnamefont {S.~M.}\ \bibnamefont {{Mironova}}},\ and\ \bibinfo {author} {\bibfnamefont {S.~A.}\ \bibnamefont {{Kaplev}}},\ }\bibfield  {title} {\bibinfo {title} {{Earth-Moon very-long-baseline interferometry project: modelling of the scientific outcome}},\ }\href {https://doi.org/10.1093/mnras/stz827} {\bibfield  {journal} {\bibinfo  {journal} {Mon. Not. Roy. Astron. Soc.}\ }\textbf {\bibinfo {volume} {486}},\ \bibinfo {pages} {815} (\bibinfo {year} {2019})}\BibitemShut {NoStop}%
\bibitem [{\citenamefont {{Likhachev}}\ \emph {et~al.}(2022)\citenamefont {{Likhachev}}, \citenamefont {{Rudnitskiy}}, \citenamefont {{Shchurov}}, \citenamefont {{Andrianov}}, \citenamefont {{Baryshev}}, \citenamefont {{Chernov}},\ and\ \citenamefont {{Kostenko}}}]{2022MNRAS.511..668L}%
  \BibitemOpen
  \bibfield  {author} {\bibinfo {author} {\bibfnamefont {S.~F.}\ \bibnamefont {{Likhachev}}}, \bibinfo {author} {\bibfnamefont {A.~G.}\ \bibnamefont {{Rudnitskiy}}}, \bibinfo {author} {\bibfnamefont {M.~A.}\ \bibnamefont {{Shchurov}}}, \bibinfo {author} {\bibfnamefont {A.~S.}\ \bibnamefont {{Andrianov}}}, \bibinfo {author} {\bibfnamefont {A.~M.}\ \bibnamefont {{Baryshev}}}, \bibinfo {author} {\bibfnamefont {S.~V.}\ \bibnamefont {{Chernov}}},\ and\ \bibinfo {author} {\bibfnamefont {V.~I.}\ \bibnamefont {{Kostenko}}},\ }\bibfield  {title} {\bibinfo {title} {{High-resolution imaging of a black hole shadow with Millimetron orbit around lagrange point l2}},\ }\href {https://doi.org/10.1093/mnras/stac079} {\bibfield  {journal} {\bibinfo  {journal} {Mon. Not. Roy. Astron. Soc.}\ }\textbf {\bibinfo {volume} {511}},\ \bibinfo {pages} {668} (\bibinfo {year} {2022})},\ \Eprint {https://arxiv.org/abs/2108.03077} {arXiv:2108.03077 [astro-ph.GA]} \BibitemShut {NoStop}%
\bibitem [{\citenamefont {{Hudson}}\ \emph {et~al.}(2025)\citenamefont {{Hudson}}, \citenamefont {{Gurvits}}, \citenamefont {{Palumbo}}, \citenamefont {{Issaoun}},\ and\ \citenamefont {{Rana}}}]{2025AcAau.232..564H}%
  \BibitemOpen
  \bibfield  {author} {\bibinfo {author} {\bibfnamefont {B.}~\bibnamefont {{Hudson}}}, \bibinfo {author} {\bibfnamefont {L.~I.}\ \bibnamefont {{Gurvits}}}, \bibinfo {author} {\bibfnamefont {D.}~\bibnamefont {{Palumbo}}}, \bibinfo {author} {\bibfnamefont {S.}~\bibnamefont {{Issaoun}}},\ and\ \bibinfo {author} {\bibfnamefont {H.}~\bibnamefont {{Rana}}},\ }\bibfield  {title} {\bibinfo {title} {{Toward optimisation of a sub-Terahertz spaceborne VLBI mission}},\ }\href {https://doi.org/10.1016/j.actaastro.2025.03.022} {\bibfield  {journal} {\bibinfo  {journal} {Acta Astronautica}\ }\textbf {\bibinfo {volume} {232}},\ \bibinfo {pages} {564} (\bibinfo {year} {2025})},\ \Eprint {https://arxiv.org/abs/2503.20312} {arXiv:2503.20312 [astro-ph.IM]} \BibitemShut {NoStop}%
\bibitem [{\citenamefont {{Hong}}\ \emph {et~al.}(2025)\citenamefont {{Hong}}, \citenamefont {{Wu}}, \citenamefont {{Liu}}, \citenamefont {{Yu}}, \citenamefont {{Wang}}, \citenamefont {{Shuai}}, \citenamefont {{Zhong}}, \citenamefont {{Zhu}}, \citenamefont {{Xie}}, \citenamefont {{Zhang}}, \citenamefont {{Xiong}}, \citenamefont {{Tang}}, \citenamefont {{Zou}}, \citenamefont {{Li}}, \citenamefont {{Wang}}, \citenamefont {{Xie}}, \citenamefont {{Xue}}, \citenamefont {{Geng}}, \citenamefont {{Zhang}}, \citenamefont {{Wu}}, \citenamefont {{Huang}}, \citenamefont {{Zheng}}, \citenamefont {{Liu}}, \citenamefont {{Wu}}, \citenamefont {{Zhang}}, \citenamefont {{An}}, \citenamefont {{Yang}}, \citenamefont {{Tong}}, \citenamefont {{Gurvits}}, \citenamefont {{Zheng}}, \citenamefont {{Gu}}, \citenamefont {{Ma}}, \citenamefont {{Li}}, \citenamefont {{Li}}, \citenamefont {{Zhao}}, \citenamefont {{Rui}}, \citenamefont {{Chen}}, \citenamefont {{Chen}}, \citenamefont {{Li}}, \citenamefont {{Zhang}}, \citenamefont {{Liu}}, \citenamefont {{Jiang}}, \citenamefont {{Wang}}, \citenamefont {{Wang}}, \citenamefont {{Sun}}, \citenamefont {{Hao}}, \citenamefont {{Cui}}, \citenamefont {{Jiang}}, \citenamefont {{Qian}},\ and\ \citenamefont {{Ye}}}]{2025SCPMA..6919511H}%
  \BibitemOpen
  \bibfield  {author} {\bibinfo {author} {\bibfnamefont {X.}~\bibnamefont {{Hong}}}, \bibinfo {author} {\bibfnamefont {W.}~\bibnamefont {{Wu}}}, \bibinfo {author} {\bibfnamefont {Q.}~\bibnamefont {{Liu}}}, \bibinfo {author} {\bibfnamefont {D.}~\bibnamefont {{Yu}}}, \bibinfo {author} {\bibfnamefont {C.}~\bibnamefont {{Wang}}}, \bibinfo {author} {\bibfnamefont {T.}~\bibnamefont {{Shuai}}}, \bibinfo {author} {\bibfnamefont {W.}~\bibnamefont {{Zhong}}}, \bibinfo {author} {\bibfnamefont {R.}~\bibnamefont {{Zhu}}}, \bibinfo {author} {\bibfnamefont {Y.}~\bibnamefont {{Xie}}}, \bibinfo {author} {\bibfnamefont {L.}~\bibnamefont {{Zhang}}}, \bibinfo {author} {\bibfnamefont {L.}~\bibnamefont {{Xiong}}}, \bibinfo {author} {\bibfnamefont {Y.}~\bibnamefont {{Tang}}}, \bibinfo {author} {\bibfnamefont {Y.}~\bibnamefont {{Zou}}}, \bibinfo {author} {\bibfnamefont {H.}~\bibnamefont {{Li}}}, \bibinfo {author} {\bibfnamefont {G.}~\bibnamefont {{Wang}}}, \bibinfo {author} {\bibfnamefont {J.}~\bibnamefont {{Xie}}}, \bibinfo {author} {\bibfnamefont {C.}~\bibnamefont {{Xue}}}, \bibinfo {author} {\bibfnamefont {H.}~\bibnamefont {{Geng}}}, \bibinfo {author} {\bibfnamefont {J.}~\bibnamefont {{Zhang}}}, \bibinfo {author} {\bibfnamefont {X.}~\bibnamefont {{Wu}}}, \bibinfo {author} {\bibfnamefont {Y.}~\bibnamefont {{Huang}}}, \bibinfo {author} {\bibfnamefont {W.}~\bibnamefont {{Zheng}}}, \bibinfo {author} {\bibfnamefont {L.}~\bibnamefont {{Liu}}}, \bibinfo {author} {\bibfnamefont {F.}~\bibnamefont {{Wu}}}, \bibinfo {author} {\bibfnamefont {X.}~\bibnamefont {{Zhang}}}, \bibinfo {author} {\bibfnamefont {T.}~\bibnamefont {{An}}}, \bibinfo {author} {\bibfnamefont {X.}~\bibnamefont {{Yang}}}, \bibinfo {author} {\bibfnamefont {F.}~\bibnamefont {{Tong}}}, \bibinfo {author} {\bibfnamefont {L.~I.}\ \bibnamefont {{Gurvits}}}, \bibinfo {author} {\bibfnamefont {Y.}~\bibnamefont {{Zheng}}}, \bibinfo {author} {\bibfnamefont {M.}~\bibnamefont {{Gu}}}, \bibinfo {author} {\bibfnamefont {X.}~\bibnamefont {{Ma}}}, \bibinfo {author} {\bibfnamefont {L.}~\bibnamefont {{Li}}}, \bibinfo {author} {\bibfnamefont {P.}~\bibnamefont {{Li}}}, \bibinfo {author} {\bibfnamefont {S.}~\bibnamefont {{Zhao}}}, \bibinfo {author} {\bibfnamefont {P.}~\bibnamefont {{Rui}}}, \bibinfo {author} {\bibfnamefont {L.}~\bibnamefont {{Chen}}}, \bibinfo {author} {\bibfnamefont {G.}~\bibnamefont {{Chen}}}, \bibinfo {author} {\bibfnamefont {K.}~\bibnamefont {{Li}}}, \bibinfo {author} {\bibfnamefont {C.}~\bibnamefont {{Zhang}}}, \bibinfo {author} {\bibfnamefont {Y.}~\bibnamefont {{Liu}}}, \bibinfo {author} {\bibfnamefont {Y.}~\bibnamefont {{Jiang}}}, \bibinfo {author} {\bibfnamefont {J.}~\bibnamefont {{Wang}}}, \bibinfo {author} {\bibfnamefont {W.}~\bibnamefont {{Wang}}}, \bibinfo {author} {\bibfnamefont {Y.}~\bibnamefont {{Sun}}}, \bibinfo {author} {\bibfnamefont {L.}~\bibnamefont {{Hao}}}, \bibinfo {author} {\bibfnamefont {L.}~\bibnamefont {{Cui}}}, \bibinfo {author} {\bibfnamefont {D.}~\bibnamefont {{Jiang}}}, \bibinfo {author} {\bibfnamefont {Z.}~\bibnamefont {{Qian}}},\ and\ \bibinfo {author} {\bibfnamefont {S.}~\bibnamefont {{Ye}}},\ }\bibfield  {title} {\bibinfo {title} {{Lunar Orbital VLBI Experiment: Motivation, scientific purposes and status}},\ }\href {https://doi.org/10.1007/s11433-025-2751-2} {\bibfield  {journal} {\bibinfo  {journal} {Science China Physics, Mechanics, and Astronomy}\ }\textbf {\bibinfo {volume} {69}},\ \bibinfo {eid} {219511} (\bibinfo {year} {2025})},\ \Eprint {https://arxiv.org/abs/2507.16317} {arXiv:2507.16317 [astro-ph.IM]} \BibitemShut {NoStop}%
\bibitem [{\citenamefont {{Liu}}\ \emph {et~al.}(2022)\citenamefont {{Liu}}, \citenamefont {{Zheng}}, \citenamefont {{Fu}},\ and\ \citenamefont {{Xu}}}]{2022AJ....164...67L}%
  \BibitemOpen
  \bibfield  {author} {\bibinfo {author} {\bibfnamefont {L.}~\bibnamefont {{Liu}}}, \bibinfo {author} {\bibfnamefont {W.}~\bibnamefont {{Zheng}}}, \bibinfo {author} {\bibfnamefont {J.}~\bibnamefont {{Fu}}},\ and\ \bibinfo {author} {\bibfnamefont {Z.}~\bibnamefont {{Xu}}},\ }\bibfield  {title} {\bibinfo {title} {{OmniUV: A Multipurpose Simulation Toolkit for VLBI Observation}},\ }\href {https://doi.org/10.3847/1538-3881/ac77f0} {\bibfield  {journal} {\bibinfo  {journal} {{Astron. J.}}\ }\textbf {\bibinfo {volume} {164}},\ \bibinfo {eid} {67} (\bibinfo {year} {2022})},\ \Eprint {https://arxiv.org/abs/2201.03797} {arXiv:2201.03797 [astro-ph.IM]} \BibitemShut {NoStop}%
\bibitem [{\citenamefont {Perlick}\ \emph {et~al.}(2018)\citenamefont {Perlick}, \citenamefont {Tsupko},\ and\ \citenamefont {Bisnovatyi-Kogan}}]{Perlick:2018iye}%
  \BibitemOpen
  \bibfield  {author} {\bibinfo {author} {\bibfnamefont {V.}~\bibnamefont {Perlick}}, \bibinfo {author} {\bibfnamefont {O.~Y.}\ \bibnamefont {Tsupko}},\ and\ \bibinfo {author} {\bibfnamefont {G.~S.}\ \bibnamefont {Bisnovatyi-Kogan}},\ }\bibfield  {title} {\bibinfo {title} {{Black hole shadow in an expanding universe with a cosmological constant}},\ }\href {https://doi.org/10.1103/PhysRevD.97.104062} {\bibfield  {journal} {\bibinfo  {journal} {Phys. Rev. D}\ }\textbf {\bibinfo {volume} {97}},\ \bibinfo {pages} {104062} (\bibinfo {year} {2018})},\ \Eprint {https://arxiv.org/abs/1804.04898} {arXiv:1804.04898 [gr-qc]} \BibitemShut {NoStop}%
\bibitem [{\citenamefont {Bisnovatyi-Kogan}\ and\ \citenamefont {Tsupko}(2018)}]{Bisnovatyi-Kogan:2018vxl}%
  \BibitemOpen
  \bibfield  {author} {\bibinfo {author} {\bibfnamefont {G.~S.}\ \bibnamefont {Bisnovatyi-Kogan}}\ and\ \bibinfo {author} {\bibfnamefont {O.~Y.}\ \bibnamefont {Tsupko}},\ }\bibfield  {title} {\bibinfo {title} {{Shadow of a black hole at cosmological distances}},\ }\href {https://doi.org/10.1103/PhysRevD.98.084020} {\bibfield  {journal} {\bibinfo  {journal} {Phys. Rev. D}\ }\textbf {\bibinfo {volume} {98}},\ \bibinfo {pages} {084020} (\bibinfo {year} {2018})},\ \Eprint {https://arxiv.org/abs/1805.03311} {arXiv:1805.03311 [gr-qc]} \BibitemShut {NoStop}%
\bibitem [{\citenamefont {Tsupko}\ and\ \citenamefont {Bisnovatyi-Kogan}(2020)}]{Tsupko:2019mfo}%
  \BibitemOpen
  \bibfield  {author} {\bibinfo {author} {\bibfnamefont {O.~Y.}\ \bibnamefont {Tsupko}}\ and\ \bibinfo {author} {\bibfnamefont {G.~S.}\ \bibnamefont {Bisnovatyi-Kogan}},\ }\bibfield  {title} {\bibinfo {title} {{First analytical calculation of black hole shadow in McVittie metric}},\ }\href {https://doi.org/10.1142/S0218271820500625} {\bibfield  {journal} {\bibinfo  {journal} {Int. J. Mod. Phys. D}\ }\textbf {\bibinfo {volume} {29}},\ \bibinfo {pages} {2050062} (\bibinfo {year} {2020})},\ \Eprint {https://arxiv.org/abs/1912.07495} {arXiv:1912.07495 [gr-qc]} \BibitemShut {NoStop}%
\bibitem [{\citenamefont {Li}\ \emph {et~al.}(2020)\citenamefont {Li}, \citenamefont {Guo},\ and\ \citenamefont {Chen}}]{Li:2020drn}%
  \BibitemOpen
  \bibfield  {author} {\bibinfo {author} {\bibfnamefont {P.-C.}\ \bibnamefont {Li}}, \bibinfo {author} {\bibfnamefont {M.}~\bibnamefont {Guo}},\ and\ \bibinfo {author} {\bibfnamefont {B.}~\bibnamefont {Chen}},\ }\bibfield  {title} {\bibinfo {title} {{Shadow of a Spinning Black Hole in an Expanding Universe}},\ }\href {https://doi.org/10.1103/PhysRevD.101.084041} {\bibfield  {journal} {\bibinfo  {journal} {Phys. Rev. D}\ }\textbf {\bibinfo {volume} {101}},\ \bibinfo {pages} {084041} (\bibinfo {year} {2020})},\ \Eprint {https://arxiv.org/abs/2001.04231} {arXiv:2001.04231 [gr-qc]} \BibitemShut {NoStop}%
\bibitem [{\citenamefont {Mukhanov}(2005)}]{Mukhanov:2005sc}%
  \BibitemOpen
  \bibfield  {author} {\bibinfo {author} {\bibfnamefont {V.}~\bibnamefont {Mukhanov}},\ }\href {http://www-spires.fnal.gov/spires/find/books/www?cl=QB981.M89::2005} {\emph {\bibinfo {title} {{Physical Foundations of Cosmology}}}}\ (\bibinfo  {publisher} {Cambridge University Press},\ \bibinfo {address} {Oxford},\ \bibinfo {year} {2005})\BibitemShut {NoStop}%
\bibitem [{\citenamefont {Comp{\`e}re}\ and\ \citenamefont {Long}(2016)}]{Compere:2016jwb}%
  \BibitemOpen
  \bibfield  {author} {\bibinfo {author} {\bibfnamefont {G.}~\bibnamefont {Comp{\`e}re}}\ and\ \bibinfo {author} {\bibfnamefont {J.}~\bibnamefont {Long}},\ }\bibfield  {title} {\bibinfo {title} {{Vacua of the gravitational field}},\ }\href {https://doi.org/10.1007/JHEP07(2016)137} {\bibfield  {journal} {\bibinfo  {journal} {JHEP}\ }\textbf {\bibinfo {volume} {07}},\ \bibinfo {pages} {137}},\ \Eprint {https://arxiv.org/abs/1601.04958} {arXiv:1601.04958 [hep-th]} \BibitemShut {NoStop}%
\bibitem [{\citenamefont {Bishop}\ and\ \citenamefont {Venter}(2006{\natexlab{b}})}]{PhysRevD.73.084023}%
  \BibitemOpen
  \bibfield  {author} {\bibinfo {author} {\bibfnamefont {N.~T.}\ \bibnamefont {Bishop}}\ and\ \bibinfo {author} {\bibfnamefont {L.~R.}\ \bibnamefont {Venter}},\ }\bibfield  {title} {\bibinfo {title} {Kerr metric in bondi-sachs form},\ }\href {https://doi.org/10.1103/PhysRevD.73.084023} {\bibfield  {journal} {\bibinfo  {journal} {Phys. Rev. D}\ }\textbf {\bibinfo {volume} {73}},\ \bibinfo {pages} {084023} (\bibinfo {year} {2006}{\natexlab{b}})}\BibitemShut {NoStop}%
\end{thebibliography}%

\end{document}